\PassOptionsToPackage{table,xcdraw}{xcolor}
\documentclass{article}

\usepackage{microtype}
\usepackage{graphicx}
\usepackage{booktabs} 

\usepackage{hyperref}

\usepackage[accepted]{icml2025}

\usepackage{amsmath}
\usepackage{amssymb}
\usepackage{mathtools}
\usepackage{amsthm}

\usepackage[capitalize,noabbrev]{cleveref}

\usepackage{makecell}
\usepackage{subcaption}
\usepackage{alltt}
\usepackage{tikz}
\usepackage{fontawesome5}
\usetikzlibrary{arrows.meta,positioning}

\theoremstyle{plain}

\theoremstyle{definition}

\theoremstyle{remark}

\usepackage[textsize=tiny]{todonotes}

\begin{document}

\title{Tests as Prompt: A Test-Driven-Development Benchmark for LLM Code Generation}

\author{
    Yi Cui \\
    ONEKQ Lab, USA \\
    \texttt{yi@onekq.ai}
}

\date{}

\twocolumn[
\maketitle
\vspace{-0.5cm}
]




\begin{abstract}
We introduce WebApp1K, a novel benchmark for evaluating large language models (LLMs) in test-driven development (TDD) tasks, where test cases serve as both prompt and verification for code generation. Unlike traditional approaches relying on natural language prompts, our benchmark emphasizes the ability of LLMs to interpret and implement functionality directly from test cases, reflecting real-world software development practices. Comprising 1000 diverse challenges across 20 application domains, the benchmark evaluates LLMs on their ability to generate compact, functional code under the constraints of context length and multi-feature complexity. Our findings highlight instruction following and in-context learning as critical capabilities for TDD success, surpassing the importance of general coding proficiency or pretraining knowledge. Through comprehensive evaluation of 19 frontier models, we reveal performance bottlenecks, such as instruction loss in long prompts, and provide a detailed error analysis spanning multiple root causes. This work underscores the practical value of TDD-specific benchmarks and lays the foundation for advancing LLM capabilities in rigorous, application-driven coding scenarios.
\end{abstract}

\section{Introduction}

Code generation is a classical genre of LLM (large language model) tasks and one of its most important applications. The focus of current research and benchmarks include algorithms\cite{mbpp, codex}, tool use\cite{bfcl}, debugging\cite{swebench}, etc. Here, a coding task consists of two parts: \textbf{prompt} and \textbf{verification}. The prompt is the task description in natural language. The verification is a collection of tests to run against the code generated by LLM. The task succeeds if all tests pass.

In this paper, we explore a new avenue where tests are \textbf{both} prompt and verification. We call this a TDD (test-driven development) task, and for the evaluation purpose, an essemble of TDD tasks a TDD benchmark. Tab.~\ref{tab:tdd-tld} illustrates a TDD task, in comparison to a natural-language-prompted coding task which we term as TLD (test-last development) task.
\begin{table*}[htbp] 
\caption{A Comparison of TDD task vs TLD task}
\centering
\begin{tabular}{|l|p{7cm}|p{7cm}|}
\hline
 & \textbf{TDD: Test-Driven Development}                         & \textbf{TLD: Test-Last Development}                          \\ \hline
\textbf{Prompt}         & Write \texttt{\small parseJson} to pass the following tests:\newline
\texttt{\small testParseSingleKeyValue() \{...\}} \newline
\texttt{\small testParseEmptyObject() \{...\}} \newline
\texttt{\small testParseArrayOfNumbers() \{...\}} \newline
\texttt{\small testParseNestedObject() \{...\}} \newline
\texttt{\small testParseArrayOfObjects() \{...\}} & Write \texttt{\small parseJson} function which converts a string into a JSON object.\\ \hline
\textbf{Verification}   & The task succeeds if all following tests pass:\newline
\texttt{\small testParseSingleKeyValue}, \newline
\texttt{\small testParseEmptyObject}, \newline
\texttt{\small testParseArrayOfNumbers}, \newline
\texttt{\small testParseNestedObject}, \newline
\texttt{\small testParseArrayOfObjects}. & The task succeeds if all following tests pass:\newline
\texttt{\small testParseSingleKeyValue}, \newline
\texttt{\small testParseEmptyObject}, \newline
\texttt{\small testParseArrayOfNumbers}, \newline
\texttt{\small testParseNestedObject}, \newline
\texttt{\small testParseArrayOfObjects}. \\ \hline
\end{tabular}
\label{tab:tdd-tld}
\end{table*}

For human developers, TLD is a common and appropriate approach for protoyping projects with vague scope and short lifecycle. But for mature projects of large stakeholder values, TDD\cite{tdd-book} must be followed for its unambiguous scope and clear contract. Here, tests are the de facto system specifications. Using the example in Tab.~\ref{tab:tdd-tld}, an enterprise platform is very likely to fork its own JSON parser for nuanced business needs, e.g. toleranting certain faults from its proprietary data stream instead of raising exceptions. A non-compliant implementation would not only fail the guarding tests, but also cause irrevocable loss if introduced into production.

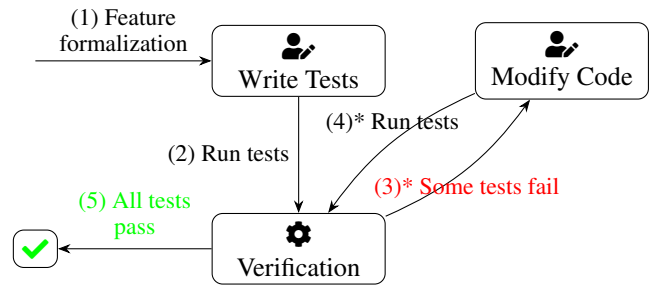
\begin{figure}[htbp]
\centering
\begin{tikzpicture}[
    ->,              
    >=Stealth,       
    on grid          
]


\node (write) [
    draw, rectangle, rounded corners,
    minimum width=2.3cm,
    inner xsep=3pt, inner ysep=3pt,
    align=center
] at (0,0) {\faUserEdit \\ Write Tests};

\node (start) [coordinate] at (-3.5,0) {};

\node (tdd) [
    draw, rectangle, rounded corners,
    minimum width=2.3cm,
    inner xsep=3pt, inner ysep=3pt,
    align=center
] at (3.5,0) {\faUserEdit \\Modify Code};

\node (ver) [
    draw, rectangle, rounded corners,
    minimum width=2.3cm,
    inner xsep=3pt, inner ysep=3pt,
    align=center
] at (0,-2.5) {\faCog \\Verification};

\node (end) [
    draw, rectangle, rounded corners
] at (-3.5,-2.5) {\textcolor{green}{\faCheck}};


\draw (start) 
  -- node[font=\footnotesize, above, align=center]{(1) Feature\\formalization} (write);

\draw (write) 
  -- node[font=\footnotesize, left]{(2) Run tests} (ver);

\draw[bend right=15] (ver) 
  to node[font=\footnotesize, below, text=red]{(3)* Some tests fail} (tdd);

\draw[bend right=15] (tdd) 
  to node[font=\footnotesize, above]{(4)* Run tests} (ver);

\draw (ver) 
  -- node[font=\footnotesize, above, align=center, text=green]{(5) All tests\\pass} (end);
\end{tikzpicture}
\caption{Incremental TDD by Human}
\label{fig:tdd-human}
\end{figure}

TDD is time-consuming and expensive. Shown in Fig.~\ref{fig:tdd-human}, human developers approach TDD incrementally. For example, the five tests in Tab.~\ref{tab:tdd-tld} are gradually added. During each iteration, the developer uses test failure as the main feedback to modify existing code, until all tests pass.

\begin{figure}[htbp]
\centering
\begin{tikzpicture}[
    ->,              
    >=Stealth,       
    on grid          
]


\node (write) [
    draw, rectangle, rounded corners,
    minimum width=2.3cm,
    inner xsep=3pt, inner ysep=3pt,
    align=center
] at (0,0) {\faUserEdit \\ Write Tests};

\node (start) [coordinate] at (-3.5,0) {};

\node (tdd) [
    draw, rectangle, rounded corners,
    minimum width=2.3cm,
    inner xsep=3pt, inner ysep=3pt,
    align=center
] at (3.5,0) {\faRobot \\Generate Code};

\node (ver) [
    draw, rectangle, rounded corners,
    minimum width=2.3cm,
    inner xsep=3pt, inner ysep=3pt,
    align=center
] at (0,-2.5) {\faCog \\Verification};

\node (end1) [
    draw, rectangle, rounded corners
] at (-3.5,-2.5) {\textcolor{green}{\faCheck}};

\node (end2) [
    draw, rectangle, rounded corners
] at (4,-2.5) {\textcolor{red}{\faTimes}};


\draw (start) 
  -- node[font=\footnotesize, above, align=center]{(1) Feature\\formalization} (write);

\draw (write) 
  -- node[font=\footnotesize, above, align=center]{(2) Tests as\\prompt} (tdd);

\draw (tdd) 
  to node[font=\footnotesize, above]{(3) Run tests} (ver);

\draw (ver) 
  -- node[font=\footnotesize, above, align=center, text=green]{(4a) All tests\\pass} (end1);

\draw (ver) 
  -- node[font=\footnotesize, above, align=center, text=red]{(4b) Some tests\\fail} (end2);
\end{tikzpicture}
\caption{Transactional TDD by LLM}
\label{fig:tdd-llm}
\end{figure}
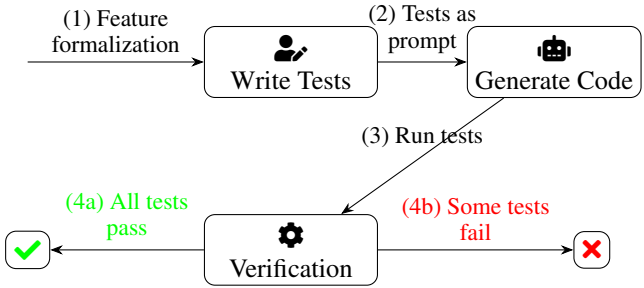

On the other hand, LLM approaches TDD in a transactional manner (Fig.~\ref{fig:tdd-llm}). A strong LLM should be able to accept all five tests in Tab.~\ref{fig:tdd-llm} in one single prompt, and write code to pass all tests with high success rate. Another difference is that LLMs do no need test failures in their prompt, as test themselves are sufficient. Therefore, improving TDD success rate of LLMs will yield tremendous values for the software industry, saving both time and cost.

The goal of this paper is to identify key LLM capabilities for TDD task success, in other words, what cause LLMs to fail at the task. To get these insights, we need a greenfield benchmark, with following considerations.
\begin{itemize}
\item \textbf{Untapped Semantic Space}: The key for an LLM to succeed on TDD is to capture and understand the instructions from tests, instead of reciting from pretraining knowledge. Unfortunately, tooling (JSON parsing) and algorithmic (LeetCode) challenges are too well-represented, hence cannot be trusted. In comparison, application layer is a much larger space for us to construct nuanced challenges unseen by LLMs.
\item \textbf{Rich Tool Ecosystem}: We need LLMs to focus on addressing instructions defined by tests, instead of reinventing wheels. If JSON parsing is needed to solve a problem, LLMs should not write a parser from scratch, but assume that a high-quality library is ready at its disposal.
\item \textbf{Compact Code}: Context length is a major performance bottleneck and highly heterogeneous among LLMs. With these constraints in mind, the benchmark should be designed to challenge LLMs to implement as many features in as few tokens as possible. 
\end{itemize}

To this end, we present WebApp1K\cite{webapp1k}, a TDD benchmark containing 1000 challenges covering a wide range of application scenarios. In each challenge, the LLM is instructed to build a mini web application supporting a singular or a combination of features. Human designers conceive each scenario, then instruct GPT to implement it into tests. Our original contributions are as follows.
\begin{itemize}
\item \textbf{New Task and New Benchmark}: We propose the TDD task as a new category of coding tasks. We also build to our knowledge the first TDD benchmark. This effort reveals the following critical insights.
\item \textbf{Critical Abilities as LLM Differentiators}: We identify instruction following and in-context learning as differentiating abilities for TDD success. Proficiency in algorithms and programming are neither critical or sufficient. We prove this point by showing that LLMs of low TDD success rate have high success rate on sibling TLD tasks.
\item \textbf{Scaling Challenge to All LLMs}: We discover the input context length to be the main bottleneck to TDD success rate impacting all LLMs. We suspect the root cause to be attention decay\cite{lostinthemiddle}.
\end{itemize}

Lastly, we fully acknowledge the pivotal role tests play in all coding benchmarks. Tab.~\ref{tab:benchmarks} compares them in three categories. Specifically related to tests, algorithmic benchmarks follow the TLD approach. Problem-solving benchmarks partially follow TDD, where tests are only a slice of the context (along with logs, source code, and the issue description).

\begin{table*}[htbp]
\caption{Comparison of coding benchmarks}
\centering
\begin{tabular}{|p{2.6cm}|p{3.6cm}|p{4.4cm}|p{4.4cm}|}
\hline
 & \textbf{Algorithmic Benchmarks} & \textbf{TDD Benchmarks} & \textbf{Problem-Solving Benchmarks} \\ \hline
\textbf{Example} & Palindrome Check in HumanEval \cite{codex} and LeetCode & JSON parsing in Tab.~\ref{tab:tdd-tld} & scikit-learn-14520 issue in SWE-bench (examplified in \cite{swebench-verified}) \\ \hline
\textbf{Relationship with Tests} & TLD & TDD & Partial TDD \\ \hline
\textbf{Scope} & Function & Module & Application \\ \hline
\textbf{Context Length} & Short & Medium & Long \\ \hline
\textbf{Running Overhead} & Low (programming language dependencies) & Medium (programming language and framework dependencies) & High (containerized dependencies) \\ \hline
\textbf{Primary Applications} & Coding interview and brainstorming & Pre-launch feature development & Post-launch patch and bugfix \\ \hline
\end{tabular}
\label{tab:benchmarks}
\end{table*}

The rest of this paper is organized as follows. In Sec.~\ref{sec:benchmark}, we introduce how WebApp1K is built and run. In Sec.~\ref{sec:evaluation}, we show LLM performance and analyze their errors. In Sec.~\ref{sec:duo}, we discuss the degraded LLM performance under more test cases, and potential root causes. Finally, we discuss related works in Sec.~\ref{sec:related} and conclude the paper in Sec.~\ref{sec:conclude}.

\section{Benchmark}\label{sec:benchmark}
\subsection{Rationales on Technology Stack}
Our benchmark is based on JavaScript React\cite{react} framework following the rationales below.

We believe an application-layer benchmark has the largest and most untapped semantic space to evaluate TDD capabilities of LLMs. Among different application pillars, e.g. web, mobile, enterprise, desktop, we choose web apps for its broadness and ubiquity. On broadness, practically any application scenario can be implemented in a web app. On ubiquity, the web elements are already rooted in other pillars, such as the PWA (progressive web app) approach in mobile app development.

Next on code compactness, web apps run on verbose HTML code embedded with style files and Javascript functions, which easily saturate an LLM's output context window. On the other hand, modern web app frameworks abstract away  boiler plates and low level constructs, and allow LLMs to build a lot more functionalities with the same number of tokens. We choose web app frameworks over raw HTML code.

Finally, we want LLMs to solve new problems using a familiar tool. This means the chosen framework must be well represented in pretraining corpora. However, there are many popular open source frameworks backed by major languages (JavaScript, Python, Java, PHP, etc.). Also each of them has rich ecosystem and testing support to help the benchmark build solid and reproducible evaluation. 

The tie-breaking choice here hinges on SPA (single page application), a design pattern well aligned with the present LLM APIs. SPA allows a developer (human or LLM) to implement multiple features in a single code file, allowing the benchmark to easily and reliablely hook LLM output to evaluation. As such, our final choice lands on React, the top SPA framework.

\begin{table}[h!]
	\caption{Template of React-based solution}
	\centering
	\begin{tabular}{|l|}
		\hline
		\begin{minipage}{\dimexpr\columnwidth-2\fboxsep-2\fboxrule}
			\vspace{2mm}
			\begingroup
			\renewcommand{\ttdefault}{pcr}
			\scriptsize
			\begin{verbatim}
// Import Statements
...
import React from 'react';

// Main component of the application
function App() {
	...
	// Business logics to handle user actions
	const functionA = (...) -> {
		...
	};
	const functionB = (...) -> {
		...
	};

	// JSX-based UI layout
	return (
		<div>
			// UI events are wired to the calling of
			// functionA and functionB
		</div>
	);
};

// Export Statement
export default App;
			\end{verbatim}
			\endgroup
		\end{minipage} \\
		\hline
	\end{tabular}
	\label{tab:template}
\end{table}

\subsection{Task Prompt}\label{sec:benchmark.formulation}
Since the prompt of a TDD task primarily consists of verbatim test code, we use a sample web app scenario to explain.

Consider a blogging website, in which a user adds comment to an existing blog post. This user journey is simulated by the unit test in Table \ref{tab:success}. Here, \texttt{fetchMock.post} is a lightweight setup to mock a successful API response without running any additional software components. The following \texttt{await} lines simulate user actions (text input, mouse click etc.). Finally, \texttt{expect} lines examine the expected outcome, i.e. the mocked API should be invoked exactly once and the system response of success should appear on the updated webpage. Similarly, the pairing failure case is shown in Table \ref{tab:failure}, where a mocked API failure is expected to lead to error message on the updated webpage.
\begin{table}[h!]
	\caption{Success case for adding a comment to a blog post}
	\centering
	\begin{tabular}{|l|}
		\hline
		\begin{minipage}{\dimexpr\columnwidth-2\fboxsep-2\fboxrule}
			\vspace{2mm}
			\begingroup
			\renewcommand{\ttdefault}{pcr} 
			\scriptsize
			\begin{verbatim}
test("successfully add comment to a post", async () => {
	fetchMock.post("/api/comments", 200);

	await act(async () => { 
		render(<MemoryRouter><App /></MemoryRouter>); 
	});
	await act(async () => { 
		fireEvent.change(screen.getByPlaceholderText(
			/Add a comment/i), 
			{ target: { value: "Great post!" } }); 
	});
	await act(async () => { 
		fireEvent.click(screen.getByText(/Submit/i)); 
	});

	expect(fetchMock.calls("/api/comments").length).toBe(1);
	expect(screen.getByText(
		/Comment added successfully/i))
		.toBeInTheDocument();
}, 10000);
			\end{verbatim}
			\endgroup
		\end{minipage} \\
		\hline
	\end{tabular}
	\label{tab:success}
\end{table}

\begin{table}[h!]
	\centering
	\caption{Failure case for adding a comment to a blog post}
	\begin{tabular}{|l|}
		\hline
		\begin{minipage}{\dimexpr\columnwidth-2\fboxsep-2\fboxrule}
			\vspace{2mm}
			\begingroup
			\renewcommand{\ttdefault}{pcr} 
			\scriptsize
			\begin{verbatim}
test("fails to add comment to a post", async () => {
	fetchMock.post("/api/comments", 500);

	// Lines identical to the success case are ignored.
	
	expect(screen.getByText(
		/Failed to add comment/i))
		.toBeInTheDocument();
}, 10000);
			\end{verbatim}
			\endgroup
		\end{minipage} \\
		\hline
	\end{tabular}
	\label{tab:failure}
\end{table}

The prompt is straightforward: we feed test files to the LLM, expecting it to generate code passing these tests. The token length of the prompt is around 0.5K.
\begin{align}
	&\text{Generate App.js to pass the tests below: } \label{eq:prompt} &\\
	&\{Tab.~\ref{tab:success}\}\{Tab.~\ref{tab:failure}\}.\text{ RETURN CODE ONLY.} \nonumber &
\end{align}

The benchmark consists of 1000 such tasks. Each task uses a success case and failure case to describe the scenario. These 1000 tasks are aggregated under 20 application domains, e.g. blogging, e-commerce, traveling. More details can be found in Appendix \ref{sec:construction}.

\subsection{Task Verification}
To succeed at the task defined in Tab.~\ref{tab:success} and \ref{tab:failure}, an LLM is expected to output code following the template in Tab.~\ref{tab:template}. The code generates a single webpage decorated with a form-like UI element allowing the test-simulated user to add comment. If all expectations in Tab.~\ref{tab:success} and \ref{tab:failure} are met, the tests pass, and the task succeeds.

We use $pass@k$, a metric defined in \cite{codex} and commonly accepted by subsequent works. Due to budget and rate limit constraints, each task is evaluated at most 10 times, i.e. $n=10$. Since $k$ must be no larger than $n$, we measure $pass@1$, $pass@5$, and $pass@10$. More details on the experiment setup can be found in Appendix \ref{sec:experiment}.

\section{Evaluation Results}\label{sec:evaluation}
\subsection{LLM Performances}
Tab.~\ref{tab:leaderboard} summarizes the $pass@k$ results of 19 frontier LLMs. We only measure $pass@1$ for top reasoning models primarily due to their inference cost. But since the value of $pass@k$ asymptotically increases with $k$, there is no doubt that the top reasoning models lead other LLMs by an obvious gap.
\begin{table}[h!]
\caption{pass@k results for frontier LLMs}
\centering
\begin{tabular}{|l|c|c|c|}
\hline
\textbf{Model} & \textbf{pass@1} & \textbf{pass@5} & \textbf{pass@10} \\ \hline
o1-preview & 0.952 & N/A & N/A \\ \hline
o1-mini & 0.939 & N/A & N/A \\ \hline
deepseek-r1&0.927 & N/A & N/A\\ \hline
gpt-4o-2024-08-06 & 0.885 & 0.9047 & 0.909 \\ \hline
claude-3.5-sonnet & 0.8808 & 0.8845 & 0.886 \\ \hline
deepseek-v3 & 0.8723 & 0.8968 & 0.902 \\ \hline
gemini-2.0-thinking & 0.859 & 0.879 & 0.884\\ \hline
deepseek-v2.5 & 0.834 & 0.8595 & 0.869 \\ \hline
gpt-4o-mini & 0.8271 & 0.8534 & 0.858 \\ \hline
gemini-2.0-flash & 0.822 & 0.848 & 0.852 \\ \hline
mistral-large-2 & 0.7804 & 0.8191 & 0.831 \\ \hline
qwen2.5-coder-32b & 0.7002 & 0.8009 & 0.827 \\ \hline
mixtral-8x22b & 0.3074 & 0.4821 & 0.533 \\ \hline
llama-v3-70b & 0.3323 & 0.4462 & 0.489 \\ \hline
llama-v3p1-405b & 0.302 & 0.4053 & 0.437 \\ \hline
llama-v3p1-8b & 0.2512 & 0.3941 & 0.432 \\ \hline
llama-v3p1-70b & 0.1027 & 0.1848 & 0.246 \\ \hline
mixtral-8x7b & 0.1269 & 0.196 & 0.218 \\ \hline
llama-v3-8b & 0.0679 & 0.1183 & 0.139 \\ \hline
\end{tabular}
\label{tab:leaderboard}
\end{table}

\subsection{Benchmark Difficulty}
\begin{figure}[htbp]
    \centering
    \includegraphics[width=1.05\columnwidth]{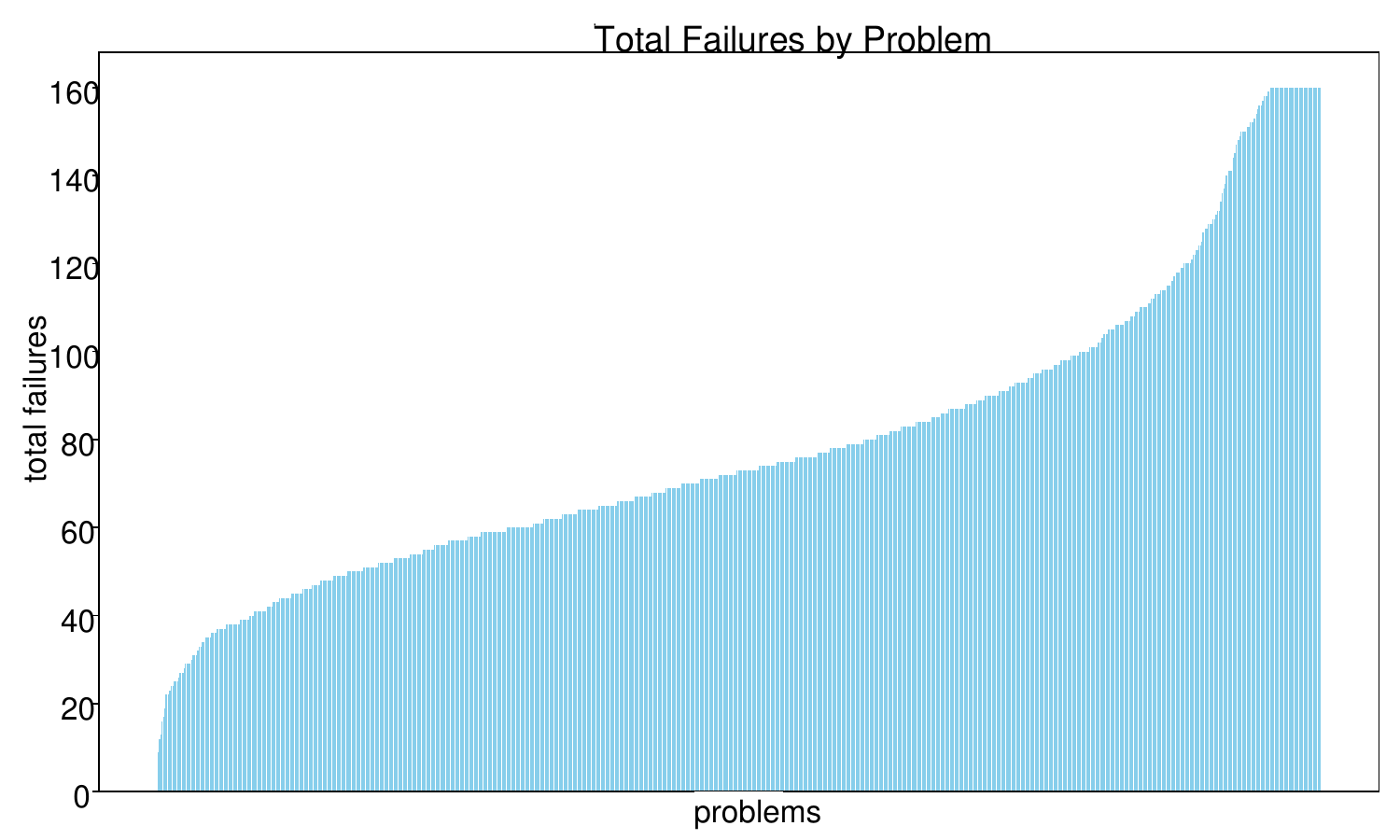}
    \caption{Failures per problem}
    \label{fig:failures}
\end{figure}

Since each LLM solves each task for at most 10 times, this gives us 160 solutions per task\footnote{We exclude reasoning models because they are only evaluated once per task. Also given their high success rates, they leave very small impact to the distribution.}. Fig.~\ref{fig:failures} shows number of failures per task. The more failures a task collects, the more difficult it is.

As indicated by the figure, the majority of the tasks have low failure rates, i.e. they are relatively easy for LLMs to solve. Conversely, a small cluster of problems on the far right exhibit extremely high failure rates, some remain unsolved by any LLM. Appendix \ref{sec:o1} will reveal more insights on why they are difficult.

\subsection{Error Types}
We study error logs and find LLMs make seven types of errors, coded to A through G. They are summarized in Tab.~\ref{tab:errors}.
\begin{table*}[h!]
\caption{Error table}
\centering
\begin{tabular}{|c|p{2.6cm}|p{2.6cm}|p{4.8cm}|p{3.2cm}|}
\hline
\textbf{Error Code} & \textbf{Name} & \textbf{Verbatim Error} & \textbf{Root Cause} & \textbf{Model Capability}\\ \hline
\cellcolor[HTML]{FDB515} A & Version Mismatch & TypeError & Deprecated framework functions are used & Preference Alignment\\ \hline
\cellcolor[HTML]{F4711E} B & Text Mismatch & TestingLibrary ElementError & Attributes or texts of HTML tags do not match test expectations & In-context Learning \\ \hline
\cellcolor[HTML]{C51E3A}C & API Call Mismatch & expect(received) & Mock APIs are called less or more than expected  & In-context Learning\\ \hline
\cellcolor[HTML]{FF6FFF} D & Uninstalled Module & Cannot find module & Imported module is not installed  & Instruction Following\\ \hline
\cellcolor[HTML]{09D0EF} E & Invalid API Call & fetch-mock &The call signature does not match the test expectation  & In-context Learning\\ \hline
\cellcolor[HTML]{20B2AA} F & Scope Violation & ReferenceError & An out-of-scope call is made to a locally-defined function & Pretraining knowledge\\ \hline
\cellcolor[HTML]{33B864} G & Missing UI Element & Element type is invalid & No UI element is defined in the code & Instruction Following \\ \hline
\end{tabular}
\label{tab:errors}
\end{table*}

The verbatim errors are the original error messages or codes captured by the log. Each of them is broadly scoped to contain a wide array of behaviors. However, in the context of our benchmark, we find all verbatim errors are projected to a narrowband of behaviors attributed to the same root causes. 

Based on the root causes, we further conjecture their connections to model capabilities.
\begin{itemize}
\item \textbf{Preference Alignment}: violating unspecified user preference, i.e. the latest stable version

\item \textbf{In-context Learning}: mismatching string or integer values specified in the model input 

\item \textbf{Instruction Following}: misunderstanding or missing the feature requested in test cases

\item \textbf{Pretraining Knowledge}: violating scoping rule of the programming language
\end{itemize}

\subsection{Singular and Twin Errors}
\begin{figure}[h!]
    \centering
    \includegraphics[width=1.05\columnwidth]{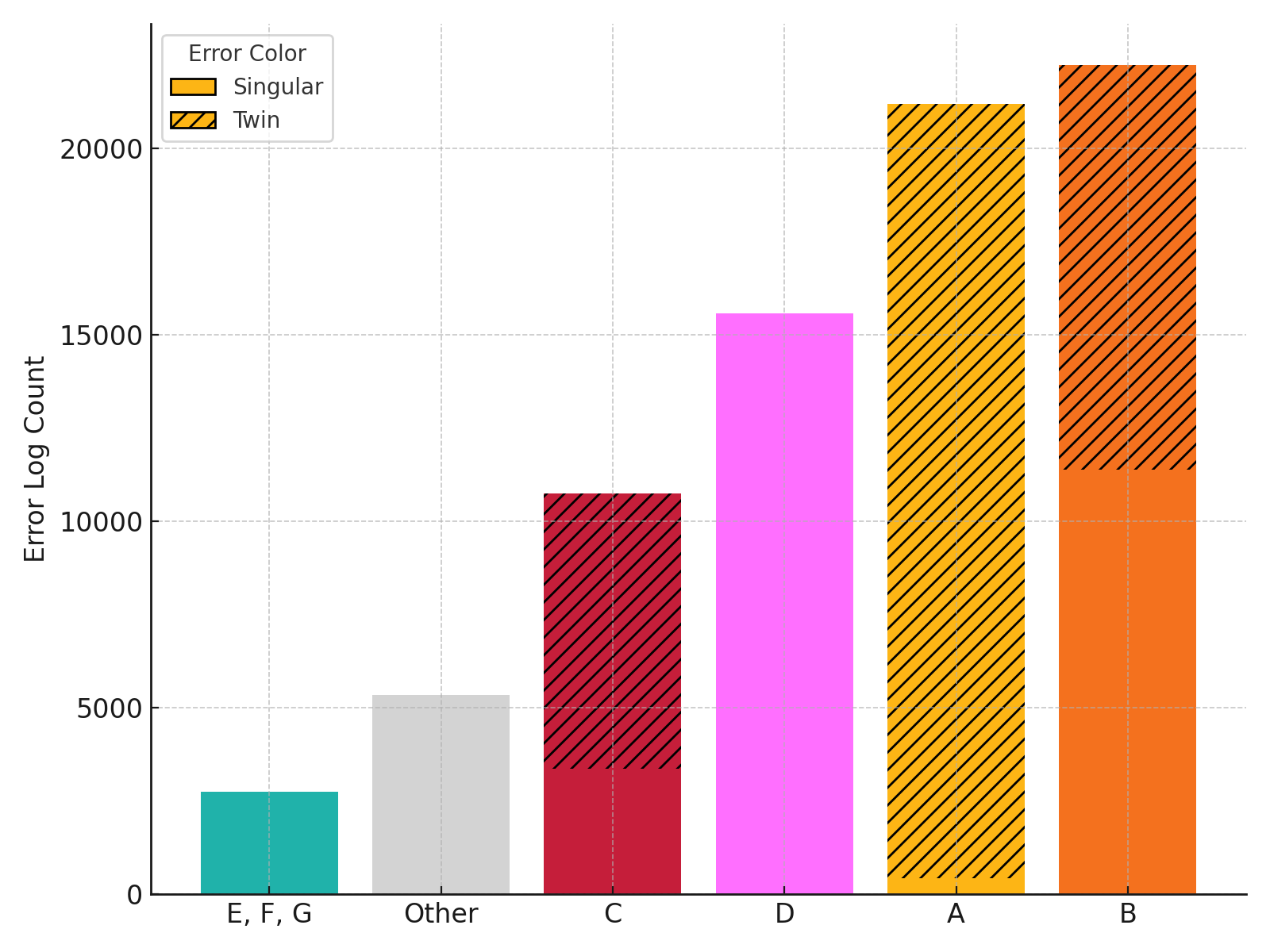}
    \caption{Distribution of singular and twin errors}
    \label{fig:error_logs}
\end{figure}

An error log can contain a combination of many error types, indicating the code is poorly implemented. But this is not the dominant pattern. 93\% of error logs contain either a singular error or twin errors. Fig.~\ref{fig:error_logs} shows the distribution of singular and twin errors.

Singular error means the log contains only one error pointing to a single line. Twin errors are two errors of the same type, preeminently pointing to the same error line. Since the code needs to pass two unit tests, often times the same bug offends both tests. This means that even upon failures, all LLMs produce quality code, but with only one error.

\subsection{Error Distribution by Models}
\begin{figure}[h!]
    \centering
    \includegraphics[width=\columnwidth]{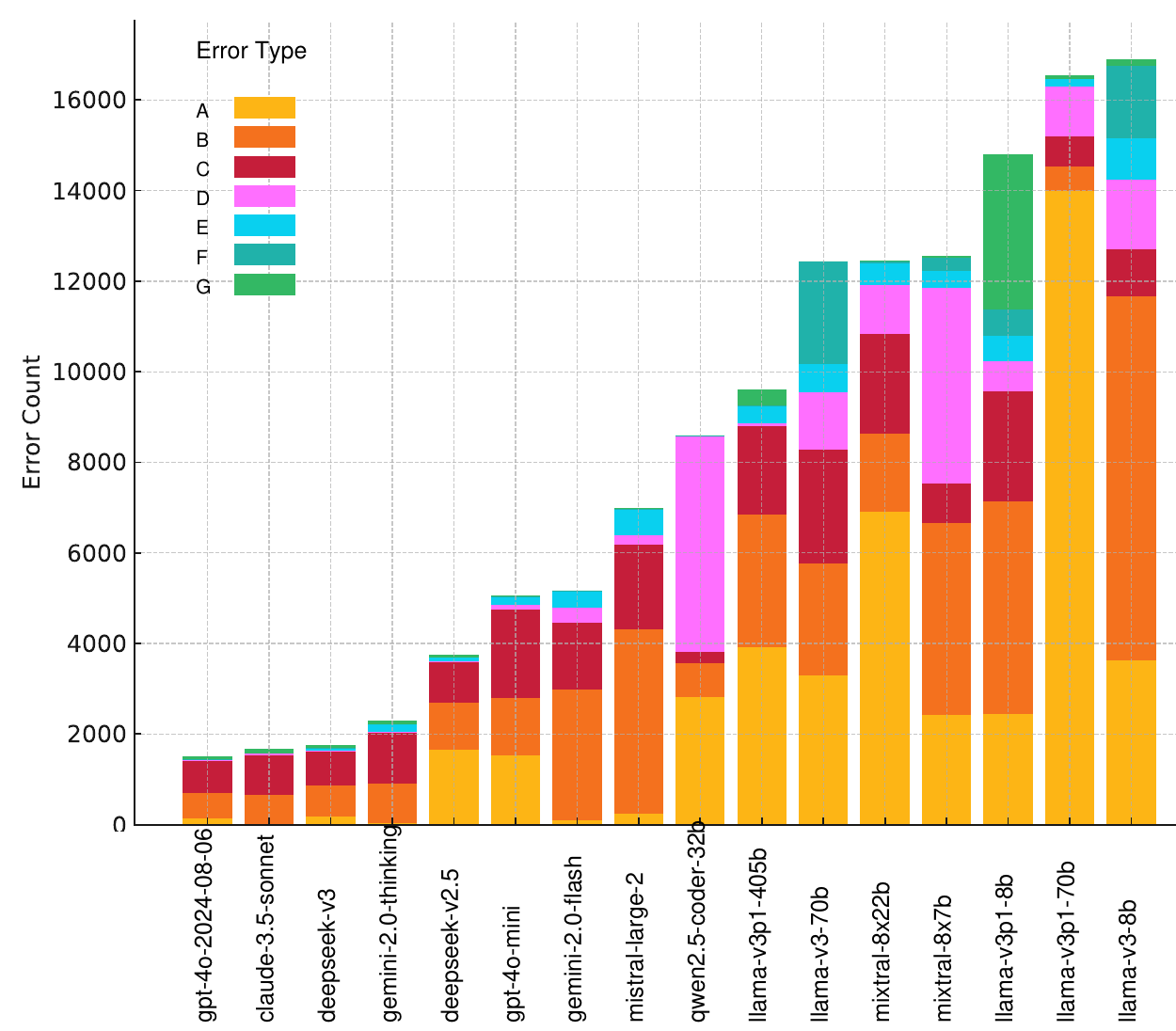}
    \caption{Error distribution by models}
    \label{fig:errors_models}
\end{figure}

In Fig.~\ref{fig:errors_models}, we show the error distribution separately for each LLM\footnote{Reasoning models are excluded because their sample sizes are too small (1 run per task instead of 10). They still make the same types of errors as other LLMs.}. The most important finding here is that no model is immune to any of the seven error types, even when the raw error counts differ by one order of magnitude bewteen LLMs with the highest and lowest success rates. 

This means that all LLMs possess the same knowledge and capabilities to write high-quality code delivering features desired by the task, and same inherent vulnerabilities resulting in the same types of errors. The key differentiator here is that top LLMs meet test instructions, where others fail instructions.

\subsection{TLD Experiment}
\begin{figure}
\centering
\begin{tikzpicture}[
    ->,              
    >=Stealth,       
    on grid          
]


\node (write) [
    draw, rectangle, rounded corners,
    minimum width=2.3cm,
    inner xsep=3pt, inner ysep=3pt,
    align=center
] at (0,0) {\faRobot \\ Generate Code};

\node (start) [coordinate] at (-3.5,0) {};

\node (tld) [
    draw, rectangle, rounded corners,
    minimum width=2.3cm,
    inner xsep=3pt, inner ysep=3pt,
    align=center
] at (3.5,0) {\faUserEdit \\Modify Tests};

\node (ver) [
    draw, rectangle, rounded corners,
    minimum width=2.3cm,
    inner xsep=3pt, inner ysep=3pt,
    align=center
] at (0,-2.5) {\faCog \\Verification};

\node (end1) [
    draw, rectangle, rounded corners
] at (-3.5,-2.5) {\textcolor{green}{\faCheck}};

\node (end2) [
    draw, rectangle, rounded corners
] at (4,-2.5) {\textcolor{red}{\faTimes}};


\draw (start) 
  -- node[font=\footnotesize, above, align=center]{(1) Feature\\formalization} (write);

\draw (write) 
  -- node[font=\footnotesize, above, align=center]{(2) Write tests\\based on\\generated\\code} (tld);

\draw (tld) 
  to node[font=\footnotesize, above]{(3) Run tests} (ver);

\draw (ver) 
  -- node[font=\footnotesize, above, align=center, text=green]{(4a) All tests\\pass} (end1);

\draw (ver) 
  -- node[font=\footnotesize, above, align=center, text=red]{(4b) Some tests\\fail} (end2);
\end{tikzpicture}
\caption{TLD by LLM}
\label{fig:tld-llm}
\end{figure}
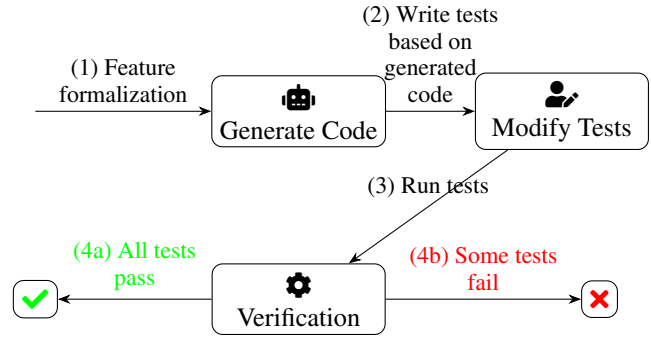

Of the total 160,000 solutions included in Fig.~\ref{fig:failures}, only 172 have syntax errors, i.e. the build failure rate is 0.1\%. In particular, the solutions by reasoning models, Claude 3.5, and Mistral Large 2 have no syntax errors. Of all error types in Tab.~\ref{tab:errors}, type A, B, and D account for overwhelming share among LLMs with weaker performances (Fig.~\ref{fig:errors_models}). On the other hand, these errors do not indicate the code is dysfunctional, only violating the tests. In light of this counter argument, we conducte a TLD (test-last development) experiment (Fig.~\ref{fig:tld-llm}), where we modify the violated tests to accommodate the verbtaim code output.
\begin{itemize}
\item \textbf{Type A Error:} Rollback to an older version of React if the code uses functions therein
\item \textbf{Type B Error:} Retrofit attribute or text property expectations to match the code
\item \textbf{Type D Error:} Refactor mock statements to accommodate the module referenced in the code
\end{itemize}

To prevent test semantic drifts, we ensure that the test code structure is unmodified, and restrict each of the above actions to the scope of single statement. As shown in Tab.~\ref{tab:tld}, all LLMs demonstrate significant $pass@1$ lift after test modification.
\begin{table}[h!]
\caption{TLD experiment: pass@1 results}
\centering
\begin{tabular}{|l|c|c|}
\hline
\textbf{Model}                 & \textbf{TDD pass@1} & \textbf{TLD pass@1} \\ \hline
llama-v3-70b          & 0.3323                   & 0.6400             \\ \hline
mixtral-8x22b         & 0.3074                   & 0.8000             \\ \hline
llama-v3p1-405b       & 0.3020                   & 0.8850             \\ \hline
llama-v3p1-8b         & 0.2512                   & 0.7550             \\ \hline
mixtral-8x7b          & 0.1269                   & 0.7300             \\ \hline
llama-v3p1-70b        & 0.1027                   & 0.7900             \\ \hline
llama-v3-8b           & 0.0679                   & 0.6500             \\ \hline
\end{tabular}
\label{tab:tld}
\end{table}

Note that TLD is a popular approach for experimental and prototyping projects, but is widely considered unfit for tablestake projects. Also TLD bears an implicit cost, since test modification itself is time-consuming and hard to automate.
\section{Duo-Feature Upgrade}\label{sec:duo}
To make the benchmark more challenging, we merge two singular tasks into a duo-feature task. Under this upgradedd benchmark, each task consists of four test cases: two successes and two failures. Accordingly, the prompt length is doubled to around 1K tokens.

Given the SPA flexibility of React, the output template (Tab.~\ref{tab:template}) remains a single webpage decorated with multiple UI elements to support two features.
\subsection{LLM Performnaces}
As shown in Tab.~\ref{tab:leaderboard_duo}, longer input context with more test cases cause $pass@1$ of all LLMs to decrease significantly. Also the SOTA is owned by Claude 3.5.
\begin{table}[h!]
\caption{Duo-feature pass@1 for selected LLMs}
\centering
\begin{tabular}{|l|c|}
\hline
\textbf{Model} & \textbf{pass@1} \\ \hline
claude-3.5-sonnet & 0.75 \\ \hline
deepseek-r1 & 0.687 \\ \hline
o1-mini & 0.667 \\ \hline
o1-preview & 0.652 \\ \hline
deepseek-v3 & 0.585 \\ \hline
gemini-2.0-thinking & 0.58 \\ \hline
gemini-2.0-flash & 0.578 \\ \hline
gpt-4o-2024-08-06 & 0.531 \\ \hline
deepseek-v2.5 & 0.49 \\ \hline
mistral-large-2 & 0.449 \\ \hline
\end{tabular}
\label{tab:leaderboard_duo}
\end{table}

Meanwhile, the LLM behaviors remain largely the same on other aspects described in Sec.~\ref{sec:evaluation}. The output code is functional with occasional build failures, and make the same errors more frequently.

\subsection{Instruction Loss}
To illustrate the essentiality of instruction following, we demonstrate a task solved by Claude 3.5, but failed by o1-preview despite its advanced reasoning capabilities. As shown in Tab.~\ref{tab:duo}, this task requires the duo feature of adding comment and retrieving blog posts in a single webpage.
\begin{table}[h!]
    \caption{A duo-feature TDD task: add comment and retrieve all blog posts}
    \centering
    \begin{tabular}{|l|}
        \hline
        \begin{minipage}{\dimexpr\columnwidth-2\fboxsep-2\fboxrule\relax}
            \vspace{2mm}
            \begingroup
            \renewcommand{\ttdefault}{pcr}
            \scriptsize
            \begin{alltt}
import App from './addComment_retrieveAllBlogPosts';
...
test("successfully add comment to a post",
	async () => \{ ...
\}

test("fails to add comment to a post",
	async () => \{ ...
\}

test("successfully get all blog posts",
	async () => \{ ... 
\}

test("fails to get all blog posts with server error",
	async () => \{
	fetchMock.get(
		'/api/posts',
		\{status:500, body:\{error: 'Internal Server Error'\}\});
	...
	expect(fetchMock.calls()).toHaveLength(1);
	expect(\textbf{screen.getByText('Internal Server Error')})
		.toBeInTheDocument();
\}, 10000);
            \end{alltt}
            \endgroup
        \end{minipage} \\
        \hline
    \end{tabular}
    \label{tab:duo}
\end{table}

Here, o1-preview passes all tests but the last one. The output code neither attempts to catch the 500 error nor prints out the \textbf{Internal Server Error} string. The reasoning chain is normal, and no step specifically mentions the need to catch internal server errors. 
\begin{flushleft} 
\hspace{1.5cm}Crafting the component $\longrightarrow$ \\
\hspace{1.5cm}Laying out the requirements $\longrightarrow$ \\
\hspace{1.5cm}Importing dependencies $\longrightarrow$ \\
\hspace{1.5cm}Breaking down the code $\longrightarrow$ \\
\hspace{1.5cm}Setting up the app $\longrightarrow$ \\
\hspace{1.5cm}Testing a post functionality $\longrightarrow$ \\
\hspace{1.5cm}Testing API integration
\end{flushleft}

The o1-preview's inherent coding ability is solid, because it solves both tasks separately under the single-feature benchmark. To this end, we argue the root cause to be instruction loss. It remains unknown whether the instruction is never picked up from the model input, or lost during an early reasoning stage. What we are sure of is the necessity of full instruction set as the foundation for reasoning, without which any LLM will fail the task.

\section{Related Works}\label{sec:related}
\subsection{Coding-Related Tasks and Benchmarks}
Prompt-driven coding has become mainstream since the introduction of Codex\cite{codex}. The evolution of benchmarks reflect the scaled-up challenges posed to LLMs, from algorithms\cite{mbpp}, to data science problems\cite{ds1000}, object-oriented coding\cite{classeval}, code execution\cite{codereval}, function calling\cite{bfcl}, SQL queries\cite{bird}, project-level resolution\cite{swebench}, etc. These benchmarks all rely on test suite of different sizes to verify task success. On the other hand, the prompt is becoming longer and harder to specify, resulting in misalignment with its verification counterpart, which can be only addressed by human calibration\cite{swebench-verified}.

TDD benchmarks avoid such misalignment by unifying task prompt and verification, meanwhile introducing other challenges to LLMs.

\subsection{Instruction Following and In-Context Learning}
Instruction following and in-context learning are two of the most desired LLM abilities to ace TDD tasks. Both topics have been extensively researched\cite{survey-icl, survey-instructions}, and their close relations revealed by several empirical or mechanistic studies\cite{zero-shot, implicit-icl, implicit-bayesian, no-instruction-tuning, induction-head}. Several well-known benchmarks\cite{instructeval, followbench, infobench} were also introduced to measure LLM progress on these abilities.

However, majority of the existing works focus on natural language instructions. Given the practical values of TDD tasks, we would like to see more interests developed over code-based instructions. Our evaluation demonstrates LLMs' remarkable ability to follow coded instructions. But it also revealed their vulnerabilities when coded instructions grow longer. This is related to another stream of works which try to scale natural language instructions\cite{multi-task, batch-prompting}. We will track closely the development of these two work streams.

\subsection{Reinforcement Learning and Reasoning}
The recent advancement of reasoning models leverage many seminal works on reinforcement learning. Works on the learning side include self-play\cite{sp-survey}, self-taught\cite{star, quiet-star}, learning from running environment\cite{alphazero}, etc. Works on the inference side include process modeling\cite{verifystepbystep}, inductive reasoning\cite{hypothesissearch}, tree search\cite{fast-slow}, etc.

Aside from general reasoning models, many works have applied reinforcement learning to coding-specific problems, including code generation\cite{rl-code-generation}, test generation\cite{rl-test-generation}, error repair\cite{rl-repair, rl-security-repair}, etc.

The values of reasoning and self-improvement techniques to TDD tasks are best showcased by the exciting SOTA lift to our benchmark. Unfortunately, we also observe the negative impact of instruction loss to reasoning model performances. We think it is worthwhile to incorporate nuanced and complex model input into future reasoning model development.

\subsection{Coding LLMs}
Recently, we have witnesssed the flourishing of many affordably-trained coding LLMs or SLMs (small language models)\cite{starcoder, starcoder2, qwen25, opencoder}. Unlike frontier LLMs, these coding models should be specialized. As such, their base models can be more specially fine-tuned to instructoin sets aligned with TDD tasks, e.g. long context with multiple expectations, dominated by code. We look forward to impressive performance by these specialized LLMs on our benchmark.

\subsection{TDD in LLM Coding}
Much similar to this paper, some recent works introduced TDD to coding task prompt, and studied best practice and performance impact\cite{tdd-function, tdd-prompt, llm4tdd}. But to our knowledge, this is the first paper focusing on TDD benchmarking.

Finally, one may argue that it is easy to repurpose classical coding benchmarks to evaluate TDD tasks by simply appending their test cases to the prompt. But we argue the benefits and necessity to have dedicated benchmarks to this cause. Just as TDD is the norm in application development emphasizing on business logic, knowledge on input instructions is the most critical factor to task success, overshadowing pretraining knowledge\footnote{This is a comparative argument relevant to other tasks akin to algorithms and data structures. A TDD task cannot succeed without a strong coding LLM.}. We think benchmarks crafted along this line of thinking can appropriately evaluate and challenge LLMs to keep improving on TDD tasks.

\section{Conclusions}\label{sec:conclude}
This paper focuses on the TDD aspect of LLM code generation, and claims two contributions. The first is a dedicated TDD benchmark which we use to evaluate 18 frontier LLMs. The second is the insights obtained via the evaluation. Specifically, instruction following and in-context learning are the key areas of improvement for LLMs and reasoning models to excel on more challenging TDD tasks.

There are two future directions. The first is to grow our benchmark to cover more application scenarios, meanwhile cross-examining learnings from this paper. The second is to explore practical hill-climbing ideas to address the vulnerability to long coded instructions.
\bibliography{icml2025}
\bibliographystyle{icml2025}
\newpage
\appendix
\onecolumn
\section{Benchmark Construction}\label{sec:construction}
The construction of WebApp1K follows the methodology of Self-Instruct\cite{selfinstruct}.  As the initial step, humans proposed 20 web application domains listed in Tab.~\ref{tab:scenarios}, referencing main applications of JavaScript and React\cite{react-book, mdn, fireship}.
\begin{table}[h!]
	\caption{Applications of WebApp1K}
	\centering
	\begin{tabular}{|c|p{12cm}|}
		\hline
		\textbf{Name} & \textbf{Overview} \\
		\hline
		blogging & A content management system for creating and managing blogs, with features like user registration, post creation, categorization, commenting, and SEO optimization. \\
		\hline
		customer support & A help desk application where users can submit support tickets, track their status, access a knowledge base, and chat with support agents. \\
		\hline
		e-commerce & A fully functional e-commerce site with features like product listings, shopping cart, user authentication, order processing, and payment integration. \\
		\hline
		event management & An app for organizing events, including event creation, ticket sales, attendee registration, and scheduling \\
		\hline
		fitness tracking & An application for tracking fitness activities, setting goals, monitoring progress, and integrating with wearable devices. \\
		\hline
		inventory management & A web application designed to help businesses track and manage their inventory. Features include product cataloging, stock level monitoring, automated reorder alerts, supplier management, sales and purchase order processing, and detailed reporting on inventory performance. \\
		\hline
		job board & A job listing site where employers can post job openings and job seekers can search and apply for jobs. \\
		\hline
		music streaming & A platform for streaming music, creating playlists, and discovering new artists. \\
		\hline
		news aggregator & A news platform that aggregates articles from various sources, categorizes them, and allows users to customize their news feed. \\
		\hline
		online learning & An LMS where users can enroll in courses, watch videos, complete quizzes, track progress, and receive certificates. \\
		\hline
		online marketplace & A platform for buying and selling goods, similar to eBay, with features like user ratings, bidding, and secure transactions. \\
		\hline
		personal finance & A tool for managing personal finances, including expense tracking, budget planning, report generation, and financial goal setting. \\
		\hline
		pet care & a web application designed to help pet owners maintain a detailed record of their pet's health, activities, and milestones. \\
		\hline
		photo gallery & An application for uploading, organizing, and sharing photos, with features like tagging, album creation, and social sharing. \\
		\hline
		real estate & A platform for listing and searching real estate properties, with features like property details, image galleries, map integration, and contact forms. \\
		\hline
		recipe sharing & A platform where users can share, search, and save recipes, with features like ingredient lists, cooking instructions, and user ratings. \\
		\hline
		social media & A social media platform where users can create profiles, post updates, follow others, like and comment on posts, and manage a feed of updates. \\
		\hline
		task management & An application for managing tasks and projects, with features like task creation, assignment, progress tracking, and notifications. \\
		\hline
		travel planning & An app for planning and booking travel, including flight and hotel searches, itinerary creation, and travel recommendations \\
		\hline
		weather & An app that provides real-time weather updates, forecasts, and severe weather alerts. \\
		\hline
	\end{tabular}
	\label{tab:scenarios}
\end{table}

Subsequently, five categories are proposed for each application domain, shown in Tab.~\ref{tab:categories}. Using these human-generated seeds, we further craft 10 scenarios for each category. Each scenario is described by a sentence. This results in a total of 1000 scenarios for the benchmark. As the final step, we prompt GPT-4o to generate a success test and failure test for each scenario, exemplified in Sec.~\ref{sec:benchmark.formulation}.
\begin{table}[h!]
\caption{Categories for each application of the benchmark}
\centering
\begin{tabular}{|l|p{12cm}|}
\hline
\textbf{Name} & \textbf{Categories} \\ \hline
blogging & Post Management, Categorization and Tag Management, Commenting System, SEO Optimization, Post Analytics \\ \hline
customersupport & Ticket Management, Agent and Collaboration, Knowledge Base, Notifications and Automation, Reporting and Analytics \\ \hline
ecommerce & Product Listings, Shopping Cart, Order Processing, Payment Integration, Product Reviews \\ \hline
eventmanagement & Event Creation, Ticket Sales, Attendee Registration, Scheduling, General Event Management \\ \hline
fitnesstracking & Activity Management, Goal Setting and Tracking, Progress Monitoring, Health and Nutrition, Device Integration and Data Management \\ \hline
inventorymanagement & Product Cataloging, Stock Level Monitoring, Supplier Management, Order Processing, Reporting \\ \hline
jobboard & Job Posting Management, Job Search and Viewing, Job Application Process, Employer Application Management, User and Profile Management \\ \hline
musicstreaming & Search and Discovery, Playback Control, Playlist Management, User Interaction, Advanced Features \\ \hline
newsaggregator & Article Management, User Preferences, Article Interactions, Content Customization, User Engagement \\ \hline
onlinelearning & Enrollment and Progress Tracking, Course Content and Interaction, Assessment and Certification, User Interaction and Communication, Course and Content Management \\ \hline
onlinemarketplace & Product Management, Checkout and Payment, Order Management, Search and Navigation, Bidding and Auctions \\ \hline
personalfinance & Expense Management, Income Management, Budget Planning, Report Generation, Financial Goal Setting \\ \hline
petcare & Pet Profiles, Daily Activities, Health Tracking, Reminders, Community \\ \hline
photogallery & Photo Upload and Management, Photo Tagging and Organization, Photo and Album Sharing, Photo Interaction and Social Features, Advanced Photo Features \\ \hline
realestate & Search and Filters, Sorting and Viewing, User Interaction, Property Management, Additional Features \\ \hline
recipesharing & Recipe Management, Search and Filtering, User Interactions, Recipe Viewing, User Profiles and Preferences \\ \hline
socialmedia & Profile Management, Post Management, User Interactions, Notifications, Feed Management \\ \hline
taskmanagement & Task Management, Project Management, User Management, Task Tracking, Advanced Features \\ \hline
travelplanning & Flight Search and Booking, Hotel Search and Booking, Itinerary Creation, Travel Recommendations, General Booking Logic \\ \hline
weather & Current Weather Data Retrieval, Weather Forecast Retrieval, Severe Weather Alerts, Location-based Services, User Preferences and Settings \\ \hline
\end{tabular}
\label{tab:categories}
\end{table}

\section{Experiment Setup}\label{sec:experiment}
The most straightforward way for us to access LLMs are public token-based APIs. For top close-sourced models, our only option is via the owners' APIs. The top open-sourced models are hosted by a few platforms, among which we choose Fireworks.

Although each API bears its minor difference, all APIs are heavily influenced by the design of OpenAI API. Tab.~\ref{tab:api_params} lists the tunable parameters exposed by each API. Since we do not know the default parameter value set by each API provider, we explicitly set the same parameter values to all LLMs under evaluation, whenever applicable. To limit the search space, we only tune $temperature$ and $top\_p$, the two most popular parameters available on all platforms. For other parameters, we assign fixed value to them across all LLMs.
\begin{table}[h!]
	\caption{Tunable parameters on different APIs}
	\centering
	\begin{tabular}{|l|c|c|c|c|c|}
		\hline
		& \textbf{temperature} & \textbf{top\_p} & \textbf{top\_k} & \textbf{presence\_penalty} & \textbf{frequency\_penalty} \\
		\hline
		\textbf{GPT4o} & Y & Y & N & Y & Y \\
		\hline
		\textbf{Claude} & Y & Y & Y & N & N \\
		\hline
		\textbf{Gemini} & Y & Y & Y & N & N \\
		\hline
		\textbf{Fireworks} & Y & Y & Y & Y & Y \\
		\hline
	\end{tabular}
	\label{tab:api_params}
\end{table}

We conducted a grid search to locate a sweet spot at which all LLMs deliver near-best results. We chose 100 random tasks from the benchmark, 5 out of each application domain. We then choose the large model out of the five leading model families, and measure their $pass@1$ ($n=1$) on the discrete 2D space of $temperature$ and $top\_n$, where $temperature = 0, 0.1, 0.2, ..., 1$, and $top\_p = 0, 0.1, 0.2, ..., 1$.
\begin{table}[h!]
	\caption{Parameter tuning results on pass@1}
	\centering
	\begin{tabular}{|l|c|c|c|}
		\hline
        		\textbf{Model} & \textbf{Lowest} & \makecell{\textbf{Chosen} \\ ($\text{temperature} = 0.2$, $\text{top\_p} = 0.8$)} & \textbf{Highest} \\
		\hline
		gpt-4o & 0.81 & \textbf{0.88} & 0.9 \\
		claude-3.5-sonnet & 0.82 & \textbf{0.85} & 0.86 \\
		deepsseek-v2 & 0.42 & \textbf{0.59} & 0.59 \\
		llama-v3-70b & 0.19 & \textbf{0.31} & 0.34 \\
		\hline
	\end{tabular}
	\label{tab:optimal_params}
\end{table}

Tab.~\ref{tab:optimal_params} presents the lowest and highest $pass@1$ value by each LLM in this grid search. Based on the results, we finalize our parameters as follows.
\begin{align*}
	temperature = 0.2 \\
	top\_p = 0.8 \\
	top\_k = 40 \\
	presence\_penalty = 0 \\
	frequency\_penalty = 0
\end{align*}

Results of our full-scale evaluations also align with this small-scale experiment, except for the deepseek-v2 model whose performance exceeds expectation. Also worth noting is that open-source models exhibit larger performance variation than closed-source models.
\section{Prompt Experiments}\label{sec:prompt}
We also study whether more sophisticated prompts can lift the model performance.

The first experiment is \textbf{system prompt}, which assigns an explicit role to the LLM and raises its awareness. Available in all APIs we run, it complements the user prompt (Equation (\ref{eq:prompt})) which gives detailed instructions to LLM. Equation (\ref{eq:system_prompt}) shows our system prompt.
\begin{equation}\label{eq:system_prompt}
	\text{You are a code generator.}
\end{equation}

The second experiment is \textbf{verbose comment}, which aims to help LLMs better understand the semantics of  tests it tries to pass. For each of the 1000 tasks, we feed its test code to GPT-4o and ask for English summary of the expectation in multiple sentences. The summary is then inserted into the test code. Tab.~\ref{tab:verbose} shows the verbose comment variant of the test code in Tab.~\ref{tab:success}.

\begin{table}[h!]
	\caption{Verbose cmment variant of the test case in Tab.~\ref{tab:success}}
	\centering
	\begin{tabular}{|l|}
		\hline
		\begin{minipage}{\dimexpr\textwidth-2\fboxsep-2\fboxrule}
			\vspace{2mm}
			\begingroup
			\renewcommand{\ttdefault}{pcr} 
			\scriptsize
			\begin{verbatim}
				test(
					"This test case verifies that a comment can be successfully added to a post by simulating
					a successful POST request to the '/api/comments' endpoint. The test ensures that the
					API call occurs exactly once and that a success message ('Comment added successfully')
					is displayed upon successful submission. This helps confirm the correct interaction
					between the frontend and backend components when adding comments.",
					async () => {

					// Lines identical to the original test case are ignored.

				}, 10000);
			\end{verbatim}
			\endgroup
		\end{minipage} \\
		\hline
	\end{tabular}
	\label{tab:verbose}
\end{table}
 
The third experiment is \textbf{error debugging}, simulating human behaviors to learn from test failures (Fig.~\ref{fig:tdd-human}). If the generated code fails the test, we add the failed code and the error log to the prompt, hoping the LLM will generate the correct code by learning from its own mistakes. Below is the prompt.
\begin{align*}
	&\{failed\_implementation\} \\
	&\text{The above code is the implementation of }\{file\_name\}\text{. It failed the tests below}\\
	&\{success\_test\_code\}\{failure\_test\_code\} \\
	&\text{Below is the test log}\\
	&\{error\_log\} \\
	&\text{Try to generate }\{file\_name\}\text{ again to pass the tests. RETURN CODE ONLY.}
\end{align*}

For all three prompt variants, we measure $pass@1$ ($n=1$) against all 1000 tasks of WebApp1K. Also in each experiment, we apply one prompt variant only, and compare it against the control test using the original prompt (Equation (\ref{eq:prompt})). Tab.~\ref{tab:prompt_experiment} summarizes the relative performance gains/loss of each variant.
\begin{table}[h]
	\caption{Prompt experiments: pass@1 gain/loss}
	\centering
	\begin{tabular}{|l|c|c|c|}
		\hline
		& \textbf{System Prompt} & \textbf{Verbose Comment} & \textbf{Error Debugging} \\
		\hline
		gpt-4o & -1.3\% & -4\% & -56\% \\
		claude-3.5-sonnet & 6.3\% & -1\% & 38\% \\
		deepsseek-v2 & -18.2\% & 7.5\% & -79\% \\
		llama-v3-70b & 8.5\% & -7.7\% & 111\% \\
		\hline
	\end{tabular}
	\label{tab:prompt_experiment}
\end{table}

We are unable to find a prompt variant delivering universally positive (or negative) impacts to all LLMs. Also we observe the huge swing in the error debugging column. The situation is unique here because this technique is not needed if the LLM output is correct on the first try. Strong LLMs with high $pass@1$ significantly shrink the sample size.

As such, we do not recommend any advanced prompting techniques for TDD tasks.
\section{Deep Dives to Reasoning Models}\label{sec:o1}
\subsection{Single-Feature Task}\label{sec:o1.customersupport}
We deep dive into \textbf{ticketSubmission} task under the \textbf{Customer Support} domain. o1 and DeepSeek R1\cite{deepseek-r1} solved this challenge, which all other LLMs failed. is the. Tab.~\ref{tab:customersupport}, lists the key steps of the test setup and expectations. We blacken the step which trapped non-reasoning models.
\begin{table}[h!]
	\caption{ticketSubmission problem}
	\centering
	\begin{tabular}{|l|}
		\hline
		\begin{minipage}{\dimexpr\textwidth-2\fboxsep-2\fboxrule}
			\vspace{2mm}
			\begingroup
			\renewcommand{\ttdefault}{pcr}
			\scriptsize
			\begin{alltt}
				test('shows error when submitting a ticket with missing fields', async () => {
					fetchMock.post('/api/tickets', { status: 400 });
					...
					fireEvent.click(screen.getByText('Submit'));
					...
					expect(fetchMock.calls('/api/tickets').length).toBe(1);
					expect(\textbf{screen.getByText('Title is required')}).toBeInTheDocument();
				}, 10000);
			\end{alltt}
			\endgroup
		\end{minipage} \\
		\hline
	\end{tabular}
	\label{tab:customersupport}
\end{table}

\begin{figure}[h!]
    \centering
    \begin{subfigure}[b]{0.49\textwidth}
        \centering
        \resizebox{\textwidth}{!}{ 
        \begin{tikzpicture}
            \draw[thick, rounded corners] (-1,1) rectangle (9,-5);

            \draw[thick] (0,0.5) -- (0,-4.5) node[below] {};
            \draw[thick] (4,0.5) -- (4,-4.5) node[below] {};
            \draw[thick] (8,0.5) -- (8,-4.5) node[below] {};

            \draw[->,thick] (0,-0.5) -- (4,-1) node[midway, above, sloped] {Form Submission};
            \draw[->,thick] (4,-1.5) -- (0,-2) node[midway, below, sloped] {Error: Title is required};

            \node at (6,-3.5) [anchor=center] {No Server Interaction};

            \node[align=center] at (0,0.8) {User};
            \node[align=center] at (4,0.8) {JS Client};
            \node[align=center] at (8,0.8) {Server};
        \end{tikzpicture}
        }
        \caption{Frontend Validation}
    \end{subfigure}
    \begin{subfigure}[b]{0.49\textwidth}
        \centering
        \resizebox{\textwidth}{!}{ 
        \begin{tikzpicture}
            \draw[thick, rounded corners] (-1,1) rectangle (9,-5);

            \draw[thick] (0,0.5) -- (0,-4.5) node[below] {};
            \draw[thick] (4,0.5) -- (4,-4.5) node[below] {};
            \draw[thick] (8,0.5) -- (8,-4.5) node[below] {};

            \draw[->,thick] (0,-0.5) -- (4,-1) node[midway, above, sloped] {Form Submission};
            \draw[->,thick] (4,-1.5) -- (8,-2) node[midway, above, sloped] {API Request};
            \draw[->,thick] (8,-2.5) -- (4,-3) node[midway, below, sloped] {400 Error: Title is required};
            \draw[->,thick] (4,-3.5) -- (0,-4) node[midway, below, sloped] {Display Error to User};

            \node[align=center] at (0,0.8) {User};
            \node[align=center] at (4,0.8) {JS Client};
            \node[align=center] at (8,0.8) {Server};
        \end{tikzpicture}
        }
        \caption{Backend Validation}
    \end{subfigure}

    \caption{Comparison of frontend and backend validation}
    \label{fig:validation}
\end{figure}
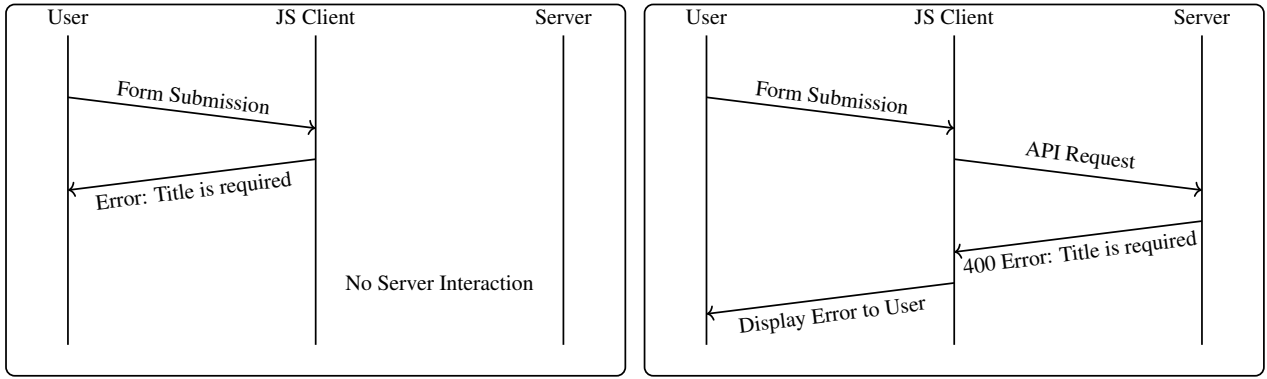

Similar to all test cases, the mocked API is first setup, followed by simulated user action, then expectations on API access and error message. Non-reasoning models understand the semantics, write functioning code, but fail expectations. The root cause here is the string \textbf{Title is required}, which is akin to a technique not requiring API access, aka frontend validation.  As a best practice (hence prevelance in pretraining dataset), frontend valiation is lightweight and fast, therefore preferred over backend validation, as shown in Fig.~\ref{fig:validation}. As such, all non-reasoning models are misled to implement frontend validation instead of expected behaviors which is backend validation.

On the other hand, o1 models discover the unpopular yet correct implementation: unconditionally visit the API, and output the \textbf{Title is required} error message upon a 400 response. Below is the ChatGPT reasoning chain, in which steps reasoning the 400 response is blackened.
\begin{flushleft}
\hspace{1.5cm}Mapping out the component $\longrightarrow$ Setting up event handlers $\longrightarrow$ \\
\hspace{1.5cm}Setting up the form $\longrightarrow$ Writing test cases $\longrightarrow$ \\
\hspace{1.5cm}\textbf{Refining the approach} $\longrightarrow$ \textbf{Refining error handling} $\longrightarrow$ \\
\hspace{1.5cm}\textbf{Adjusting error handling} $\longrightarrow$ Adjusting code logic $\longrightarrow$ \\
\hspace{1.5cm}Updating JavaScript code
\end{flushleft}

The most crucial step here is \textbf{Refining the approach}. Below is its detailed wording.
\begin{flushleft} 
\hspace{1.5cm}I’m updating the code to ensure a fetch request is \textbf{always} sent, even without a title.\\
\hspace{1.5cm}The server will respond with a 400 status if the title is absent.
\end{flushleft}

Evidently, the step before it (Writing test cases) conducted certain verification, which leads the model to pivot to the right path.

The reasoning process of DeepSeek R1 is not explicitly divided into steps, but a mixture of natural language and code. After long struggling, the model comes to the right path. Below are excerpts from its verbose self reflection.
\begin{flushleft}
\hspace{1.5cm}So maybe the frontend does client-side validation. $\longrightarrow$ \\
\hspace{1.5cm}But the second test's fetchMock returns 400, which might mean the backend validation. $\longrightarrow$ \\
\hspace{1.5cm}But in the test, the fetchMock is called once, which would mean the request is sent. $\longrightarrow$ \\
\hspace{1.5cm}So that's a contradiction. Hmm, this is confusing. $\longrightarrow$ \\
\hspace{1.5cm}\textbf{That's not a good practice}, but perhaps that's \textbf{what the tests require}. $\longrightarrow$ \\
\hspace{1.5cm}Maybe the test is written incorrectly, but I have to \textbf{write the component to pass the tests as given}.
\end{flushleft}

Unfortunatelly the reasoning models can also fall for the same trap. Below is a ChatGPT reasoning chain leading o1-preview to the faulty implementation like previous models. 
\begin{flushleft}
\hspace{1.5cm}Mapping out test strategy $\longrightarrow$ Setting up the test $\longrightarrow$ \\
\hspace{1.5cm}\textbf{Customer service improvement} $\longrightarrow$ Setting up for data $\longrightarrow$ \\
\hspace{1.5cm}Setting up the form $\longrightarrow$ \textbf{Verifying form submission} $\longrightarrow$ \\
\hspace{1.5cm}SHOWING ERRORS $\longrightarrow$ Refining the form handling 
\end{flushleft}

On a closer look, step \textbf{Customer service improvement} derails the model from backend validation to frontend validation.
\begin{flushleft} 
\hspace{1.5cm}I’m thinking about creating a TicketSubmission component with\\
\hspace{1.5cm}a 'Title' input and 'Submit' button. Submitting the form will trigger\\
\hspace{1.5cm}a POST request to '/api/tickets', validating the 'Title' field \textbf{before} submission.
\end{flushleft}

More interestingly, the step \textbf{Verifying form submission} does not correct the wrong direction, but solidify it.
\begin{flushleft} 
\hspace{1.5cm}I’m thinking about how the form ensures 'Title' must be filled.\\
\hspace{1.5cm}It sends a POST request \textbf{if} 'Title' is entered, showing success\\
\hspace{1.5cm}or 'Title is required' based on the response status.
\end{flushleft}

With these superficial clues, we speculate that the derailing is due to preemption of original expectations by model's inherent knowledge. The subsequent verification step is derived from neighboring steps already derailed, instead of orginal expectations only accessible from the input tokens. 
\subsection{Duo-Feature Task}
The duo-feature task was composed in two ways. The first way is shown in Tab.~\ref{tab:webapp1kduo} (a), in which the original export name of the single-feature benchmark is preserved as is. The second way is shown in Tab.~\ref{tab:webapp1kduo} (b), where the export names are normalized to a unified name \textbf{App}.
\begin{table}[h!]
    \caption{Two formats of the duo-feature benchmark}
    \centering
    \begin{minipage}{0.48\textwidth}
        \centering
        \begin{tabular}{|l|}
            \hline
            \begin{minipage}{\dimexpr\textwidth-2\fboxsep-2\fboxrule}
                \vspace{2mm}
                \begingroup
                \renewcommand{\ttdefault}{pcr}
                \scriptsize
                \begin{alltt}
...
import \textbf{TaskA} from './TaskA_B';
import \textbf{TaskB} from './TaskA_B';

test("Success at task A", async () => {
  ...
  render(
    <MemoryRouter><\textbf{TaskA} /></MemoryRouter>
  );
  ...
}, 10000);

test("Failure at task A", async () => {
  ...
  render(
    <MemoryRouter><\textbf{TaskA} /></MemoryRouter>
  );
  ...
}, 10000);

test("Success at task B", async () => {
  ...
  render(
    <MemoryRouter><\textbf{TaskB} /></MemoryRouter>
  );
  ...
}, 10000);

test("Failure at task B", async () => {
  ...
  render(
    <MemoryRouter><\textbf{TaskB} /></MemoryRouter>
  );
  ...
}, 10000);
                \end{alltt}
                \endgroup
            \end{minipage} \\
            \hline
        \end{tabular}
        \subcaption{Raw format}
    \end{minipage}
    \hspace{0.02\textwidth}
    \begin{minipage}{0.48\textwidth}
        \centering
        \begin{tabular}{|l|}
            \hline
            \begin{minipage}{\dimexpr\textwidth-2\fboxsep-2\fboxrule}
                \vspace{2mm}
                \begingroup
                \renewcommand{\ttdefault}{pcr}
                \scriptsize
                \begin{alltt}
...
...
import \textbf{App} from './TaskA_B';

test("Success at task A", async () => {
  ...
  render(
    <MemoryRouter><\textbf{App} /></MemoryRouter>
  );
  ...
}, 10000);

test("Failure at task A", async () => {
  ...
  render(
    <MemoryRouter><\textbf{App} /></MemoryRouter>
  );
  ...
}, 10000);

test("Success at task B", async () => {
  ...
  render(
    <MemoryRouter><\textbf{App} /></MemoryRouter>
  );
  ...
}, 10000);

test("Failure at task B", async () => {
  ...
  render(
    <MemoryRouter><\textbf{App} /></MemoryRouter>
  );
  ...
}, 10000);
                \end{alltt}
                \endgroup
            \end{minipage} \\
            \hline
        \end{tabular}
        \subcaption{Normalized format}
    \end{minipage}
    \label{tab:webapp1kduo}
\end{table}

Tab.~\ref{tab:leaderboard_duo} shows results from the normalized format. Under the raw format, all models struggle. Most strikingly, o1 models fail all problems (Tab.~\ref{tab:webapp1kduo_raw}).
\begin{table}[h!]
\caption{Duo-feature benchmark raw format: pass@1 results for selected models}
\centering
\begin{tabular}{|l|c|}
\hline
\textbf{Model} & \textbf{pass@1} \\ \hline
claude-3.5-sonnet & 0.32 \\ \hline
gpt-4o-2024-08-06 & 0.026 \\ \hline
deepseek-v2.5 & 0.02 \\ \hline
mistral-large-2 & 0.02 \\ \hline
o1-mini & 0 \\ \hline
o1-preview & 0 \\ \hline
\end{tabular}
\label{tab:webapp1kduo_raw}
\end{table}

To find the root cause, we find the raw format (Tab.~\ref{tab:webapp1kduo} (a)) has two imports of different names, i.e. \textbf{TaskA} and \textbf{TaskB}. But they are actually default imports (without curly braces) which are name-agnostic. Also since only one default export is allowed per module, this format is in fact semantically equivalent to the normalized format in Tab.~\ref{tab:webapp1kduo} (b). Both formats demand the models to build a single module implementing all expectations, with a single default export. To help readers understand related concepts, we explain JavaScript export rules in Tab.~\ref{tab:exports}.
\begin{table}[h!]
\caption{Illustration of JavaScript default export in comparison to named imports}
\centering
\begin{tabular}{|l|l|l|}
\hline
                            & \textbf{Named Exports}                          & \textbf{Default Export}                       \\ \hline
\textbf{Purpose}            & Export multiple items from a module            & Export a single item from a module       \\ \hline
\textbf{Syntax}             & \texttt{export const x = ...;}                 & \texttt{export default ...;}                  \\ 
                            & \texttt{export function y() \{...\}}           &                                               \\ \hline
\textbf{Import Syntax}      & \texttt{import \{ x, y \} from}    & \texttt{import anyName from}      \\
                            &  \texttt{'./module';}                &   \texttt{'./module';}                                             \\ \hline
\textbf{Curly Braces}       & Required during import                         & Not required during import                    \\ \hline
\textbf{Import Naming}      & Must use the exact exported names              & Can be imported with any name                 \\ 
                            & (can use \texttt{as} to rename)                &                                               \\ \hline
\textbf{Multiplicity}  & Multiple named exports per module              & Only one default export per module            \\ \hline
\textbf{Use Case}    & Utility functions, constants, classes          & Main functionality of a module                \\ \hline
\textbf{Export Location}    & Anywhere in the module           & Bottom or after the main logic \\ \hline
\end{tabular}
\label{tab:exports}
\end{table}

Tab.~\ref{tab:solutions_raw} collects different ways models cope with this challenge. Tab.~\ref{tab:solutions_raw} (d) is the only right answer, but also the least straightforward, challenging the intuition trap that two exports from two separate modules are needed. Both non-reasoning and reasoning models fall for the trap and attempt to split the implementation into two modules, (Tab.~\ref{tab:solutions_raw} (a), (b), (c)), resulting in very high failure rates.
\begin{table}[h!]
    \caption{Patterns to address the duo-feature benchmark raw format (Tab.~\ref{tab:webapp1kduo} (a))}
    \centering
    \begin{minipage}{0.48\textwidth}
        \centering
        \begin{tabular}{|l|}
            \hline
            \begin{minipage}{\dimexpr\textwidth-2\fboxsep-2\fboxrule}
                \vspace{2mm}
                \begingroup
                \renewcommand{\ttdefault}{pcr}
                \scriptsize
                \begin{verbatim}
function TaskA() {
  // Implementation of TaskA
}

function TaskB() {
  // Implementation of TaskB
}
export default TaskA;
export { TaskB };
                \end{verbatim}
                \endgroup
            \end{minipage} \\
            \hline
        \end{tabular}
        \subcaption{One default export and one named export}
    \end{minipage}
    \hspace{0.02\textwidth}
    \begin{minipage}{0.48\textwidth}
        \centering
        \begin{tabular}{|l|}
            \hline
            \begin{minipage}{\dimexpr\textwidth-2\fboxsep-2\fboxrule}
                \vspace{2mm}
                \begingroup
                \renewcommand{\ttdefault}{pcr}
                \scriptsize
                \begin{verbatim}
function TaskA() {
  // Implementation of TaskA
}

function TaskB() {
  // Implementation of TaskB
}

export { TaskA, TaskB };
                \end{verbatim}
                \endgroup
            \end{minipage} \\
            \hline
        \end{tabular}
        \subcaption{Two named exports}
    \end{minipage}

    \vspace{0.4cm}

    \begin{minipage}{0.48\textwidth}
        \centering
        \begin{tabular}{|l|}
            \hline
            \begin{minipage}{\dimexpr\textwidth-2\fboxsep-2\fboxrule}
                \vspace{2mm}
                \begingroup
                \renewcommand{\ttdefault}{pcr}
                \scriptsize
                \begin{verbatim}
function TaskA_or_B() {
  // Implementation of TaskA or TaskB
}

export default TaskA_or_B;
                \end{verbatim}
                \endgroup
            \end{minipage} \\
            \hline
        \end{tabular}
        \subcaption{Only one task is implemented and exported}
    \end{minipage}
    \hspace{0.02\textwidth}
    \begin{minipage}{0.48\textwidth}
        \centering
        \begin{tabular}{|l|}
            \hline
            \begin{minipage}{\dimexpr\textwidth-2\fboxsep-2\fboxrule}
                \vspace{2mm}
                \begingroup
                \renewcommand{\ttdefault}{pcr}
                \scriptsize
                \begin{verbatim}
function TaskA_or_B() {
  // Implementation of both TaskA and TaskB
}

export default TaskA_or_B;
                \end{verbatim}
                \endgroup
            \end{minipage} \\
            \hline
        \end{tabular}
        \subcaption{Two tasks jointly implemented and exported}
    \end{minipage}
    \label{tab:solutions_raw}
\end{table}

Next, we try to understand why non-reasoning models occasionally succeed by following the pattern of Tab.~\ref{tab:solutions_raw} (d), but non-reasoning models never do so. We suspect that the normalized format (Tab.~\ref{tab:webapp1kduo} (b)) definitely dominates the pretraining/posttraining dataset, but does not exclude the raw format (Tab.~\ref{tab:webapp1kduo} (a)), as well as the matching solutions. This makes the success possible.

On the other hand, from the first reasoning step which often plays the role of planning, reasoning models commit to the wrong judgment, and do not get a chance to correct the course in subsequent steps. Below is the detailed wording of the first reasoning step from a ChatGPT reeactment.
\begin{flushleft} 
\hspace{1.5cm}To progress, the key task is creating components TaskA and TaskB in TaskA\_B.js \\
\hspace{1.5cm}to ensure all tests are successfully passed.
\end{flushleft}

Comparing to the mistakes made in Sec.~\ref{sec:o1.customersupport}, the mistake in the above step covers a larger scope. It is reasonable to argue that mistakes made in large-scoped steps are more fatal and harder to correct.
\section{Line-of-Code (LOC) Analysis}\label{sec:loc}
Since top LLMs with SOTAs are proprietary, mechanistic studies are impossible. Therefore, we can only seek insights from model outputs. Thanks to the modularized design of the React framework, the solutions output by all models universally follow the template outlined in Tab.~\ref{tab:template}, with no need for any explicit prompting. As such, we use LOC (line-of-code) as the proxy signal. Results in this appendix are from the single-feature benchmark.
\subsection{LOC Distribution by Models}
\begin{table}[h!]
    \caption{Models ranked by median LOC with pass@1}
    \centering
    \begin{tabular}{|l|l|l|}
        \hline
        \textbf{Model} & \textbf{Median LOC} & \textbf{pass@1} \\ \hline
        mixtral-8x7b          & 35 & 0.1269 \\ \hline
        llama-v3-8b           & 39 & 0.0679 \\ \hline
        llama-v3p1-405b       & 40 & 0.3020 \\ \hline
        gpt-4o-2024-08-06              & 40 & 0.8850 \\ \hline
        deepseek-v2     & 40 & 0.7002 \\ \hline
        gpt-4o-mini                    & 40 & 0.8271 \\ \hline
        mistral-large-2                & 41 & 0.7804 \\ \hline
        gemini-2.0-flash               & 41 & 0.8220 \\ \hline
        llama-v3p1-8b         & 42 & 0.2512 \\ \hline
        mixtral-8x22b         & 43 & 0.3074 \\ \hline
        claude-3.5-sonnet              & 43 & 0.8808 \\ \hline
        llama-v3-70b          & 43 & 0.3323 \\ \hline
        gemini-2.0-thinking                 & 45 & 0.8590 \\ \hline
        llama-v3p1-70b        & 46 & 0.1027 \\ \hline
    \end{tabular}
    \label{tab:loc_by_models}
\end{table}

In Tab.~\ref{tab:loc_by_models}, we rank models by their median LOC alongside their respective $pass@1$ scores. Picking one $pass@k$ is sufficient because all scores produced basically the same model rankings as shown in Tab.~\ref{tab:leaderboard}.

We observe that the median LOCs across all models stay close, ranging from 35 to 46. We believe this narrow range is largely enforced by the conciseness and expressiveness of the React framework itself. Also there is no strong correlation between the conciseness (median LOC) and correctness ($pass@1$). For example, mixtral-8x7b, which has the shortest median LOC, ranks quite low on $pass@1$ (0.1269). Conversely, stronger models like claude-3.5-sonnet and gpt-4o-2024-08-06, generate longer code. Other models, e.g. deepseek-v2, strike a balance between median.

\begin{figure}[h!]
	\centering
	\begin{subfigure}{0.32\textwidth}
		\includegraphics[width=\textwidth]{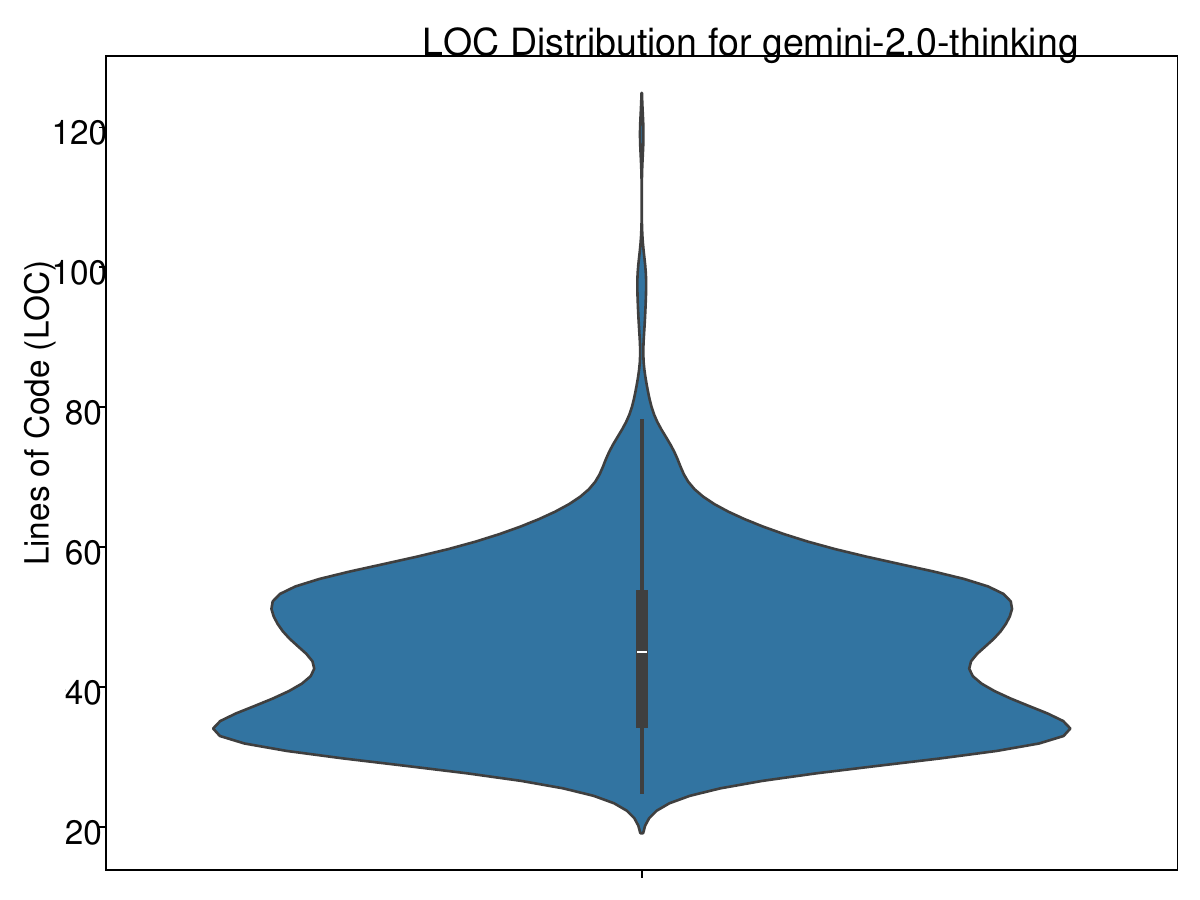}
		\caption{gemini-2.0-thinking}
	\end{subfigure}
	\begin{subfigure}{0.32\textwidth}
		\includegraphics[width=\textwidth]{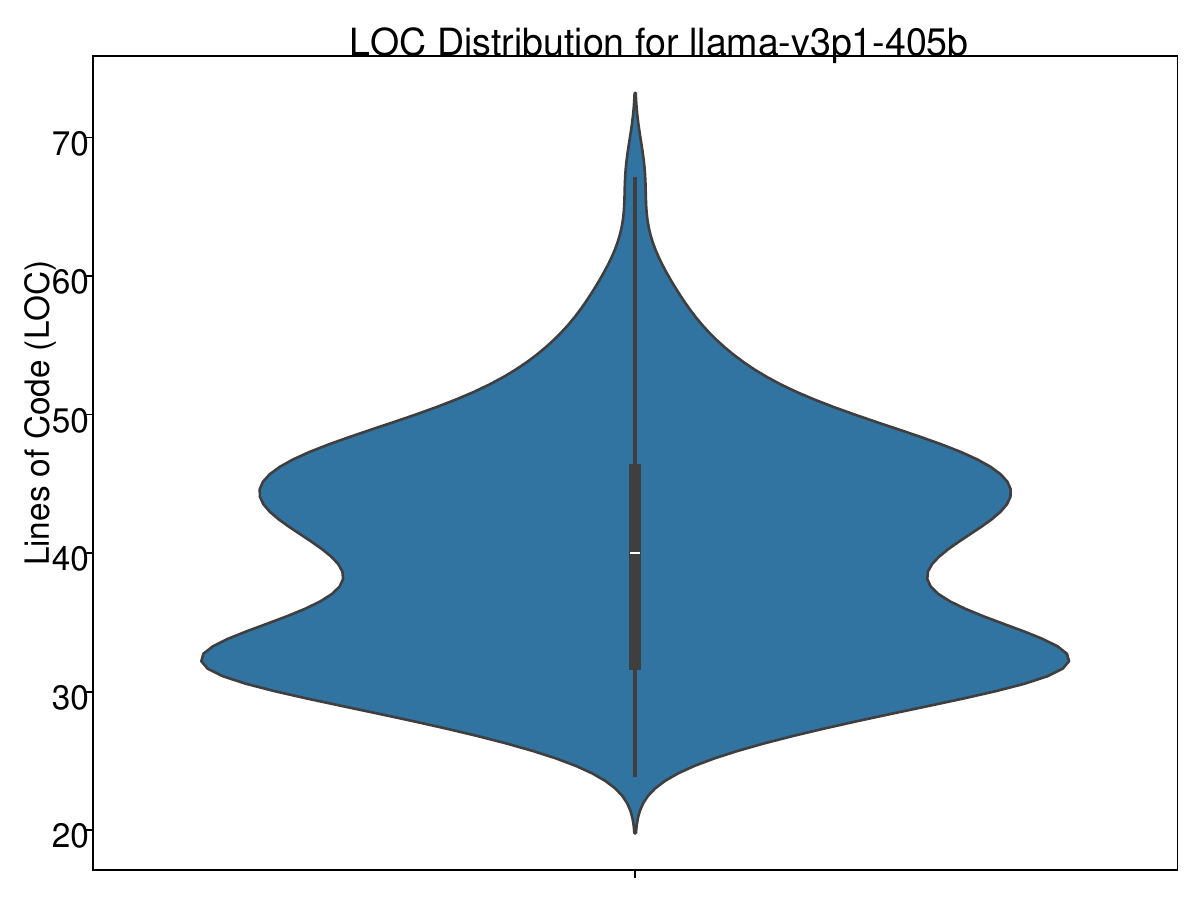}
		\caption{llama-v3p1-405b}
	\end{subfigure}
	\begin{subfigure}{0.32\textwidth}
		\includegraphics[width=\textwidth]{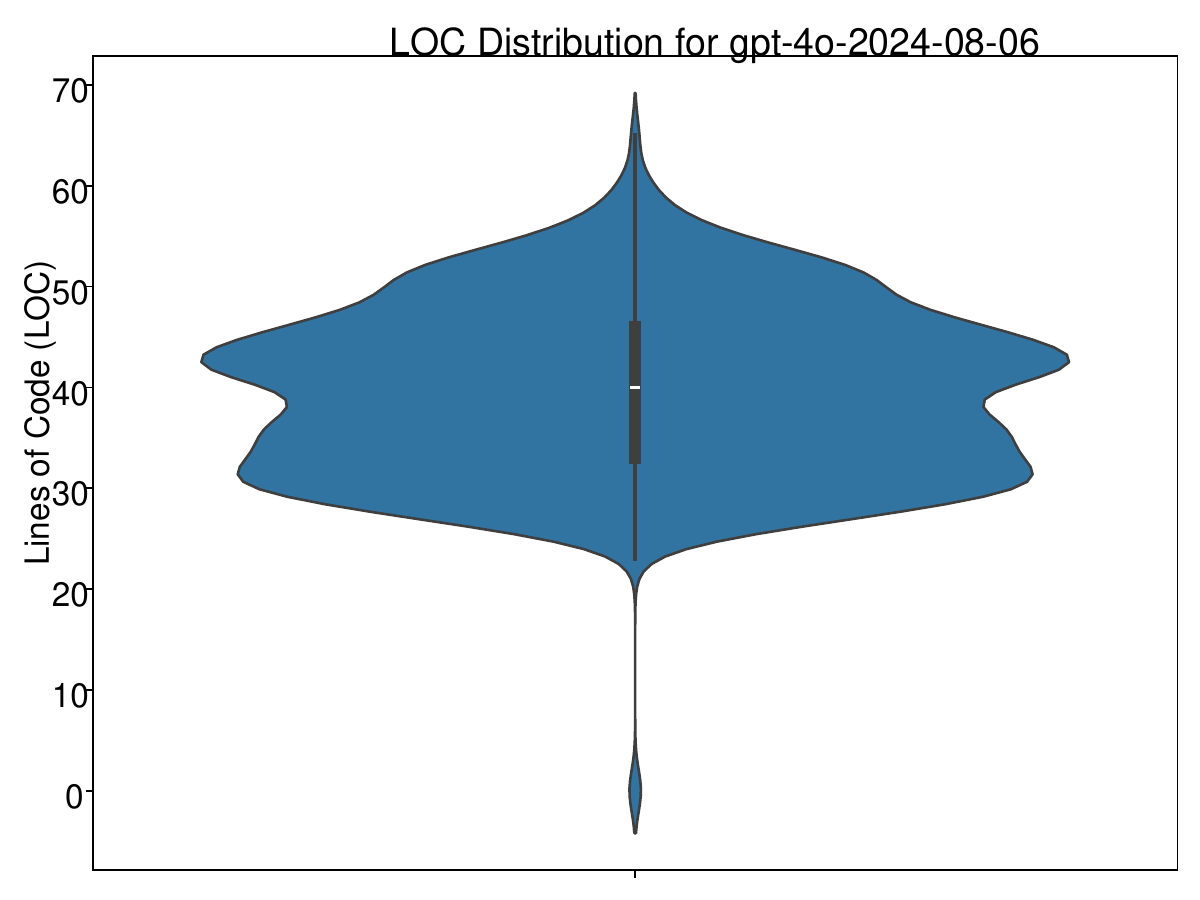}
		\caption{gpt-4o-2024-08-06}
	\end{subfigure}
	
	\begin{subfigure}{0.32\textwidth}
		\includegraphics[width=\textwidth]{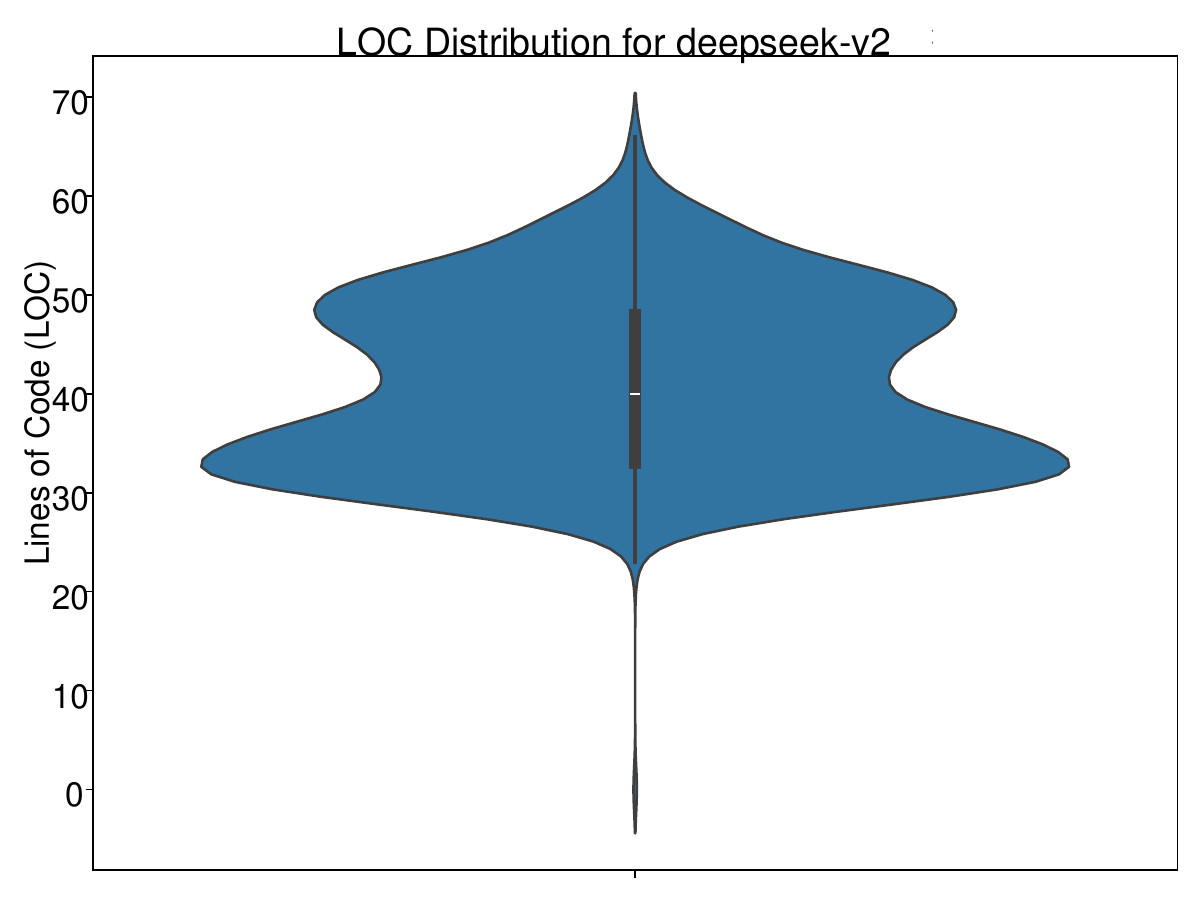}
		\caption{deepseek-coder-v2}
	\end{subfigure}
	\begin{subfigure}{0.32\textwidth}
		\includegraphics[width=\textwidth]{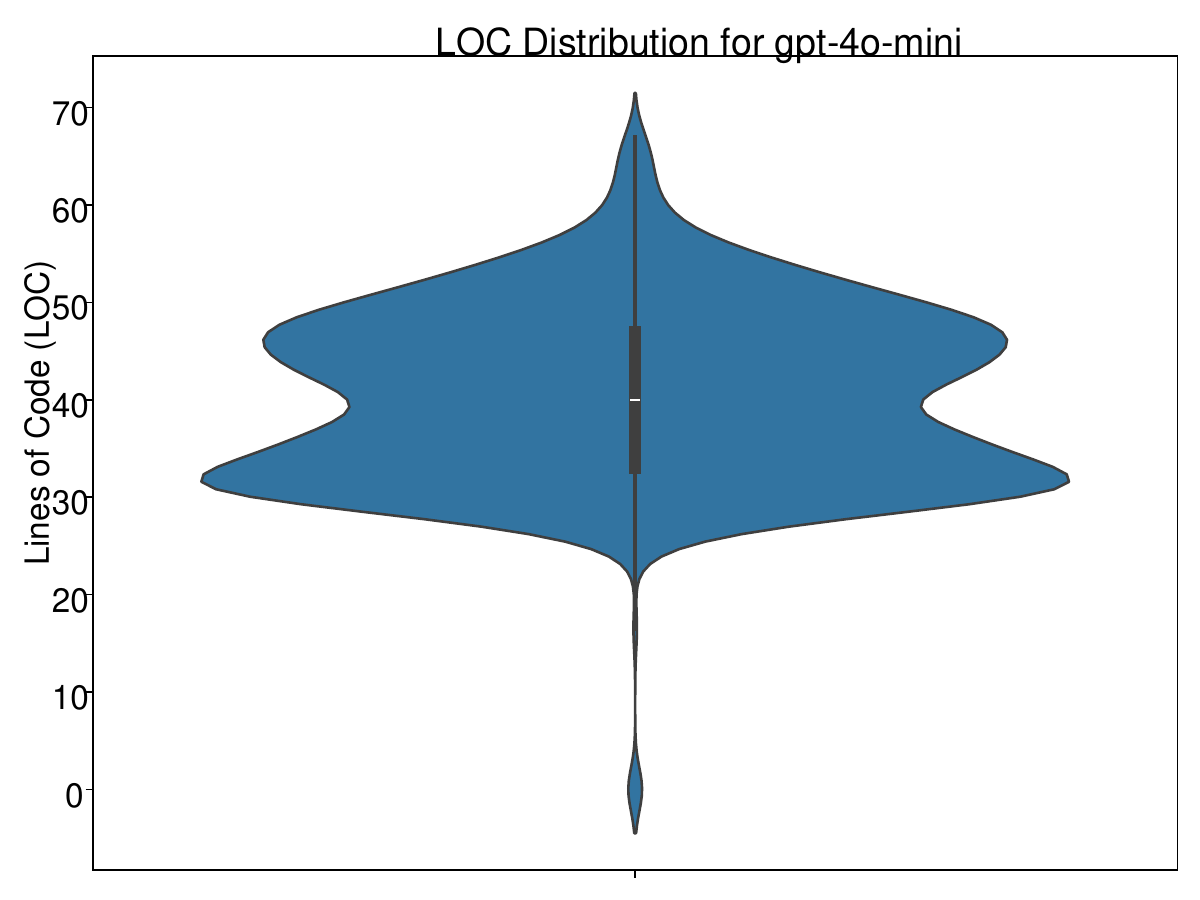}
		\caption{gpt-4o-mini}
	\end{subfigure}
	\begin{subfigure}{0.32\textwidth}
		\includegraphics[width=\textwidth]{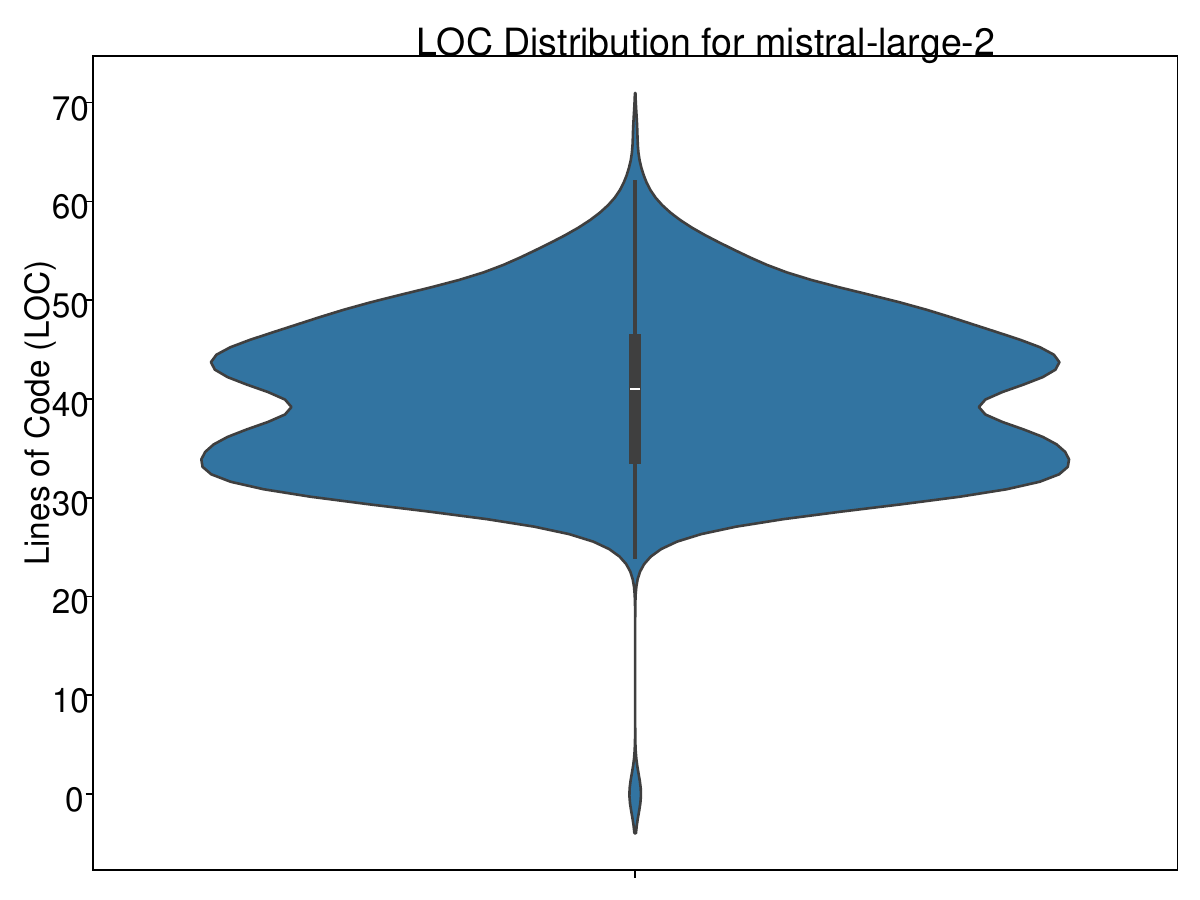}
		\caption{mistral-large-2}
	\end{subfigure}
	
	\begin{subfigure}{0.32\textwidth}
		\includegraphics[width=\textwidth]{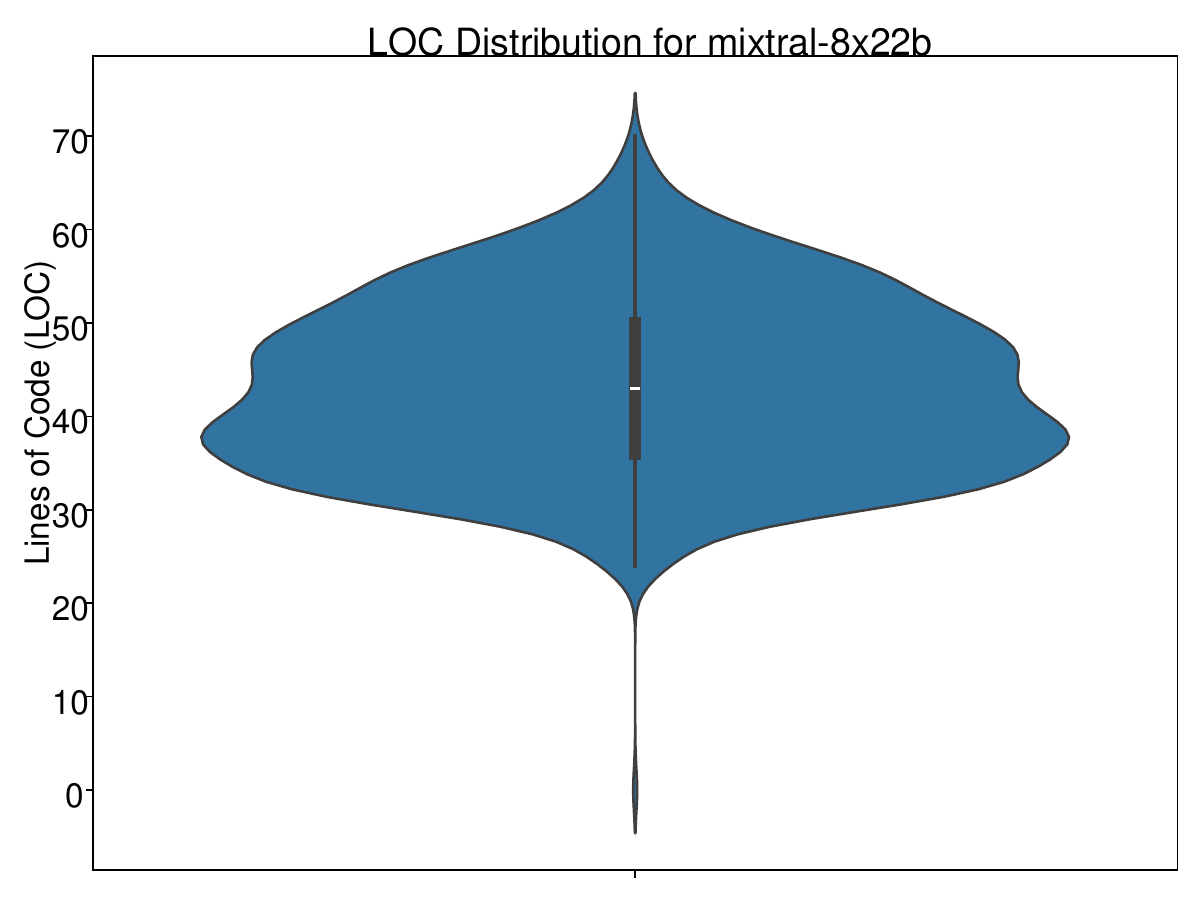}
		\caption{mixtral-8x22b}
	\end{subfigure}
	\begin{subfigure}{0.32\textwidth}
		\includegraphics[width=\textwidth]{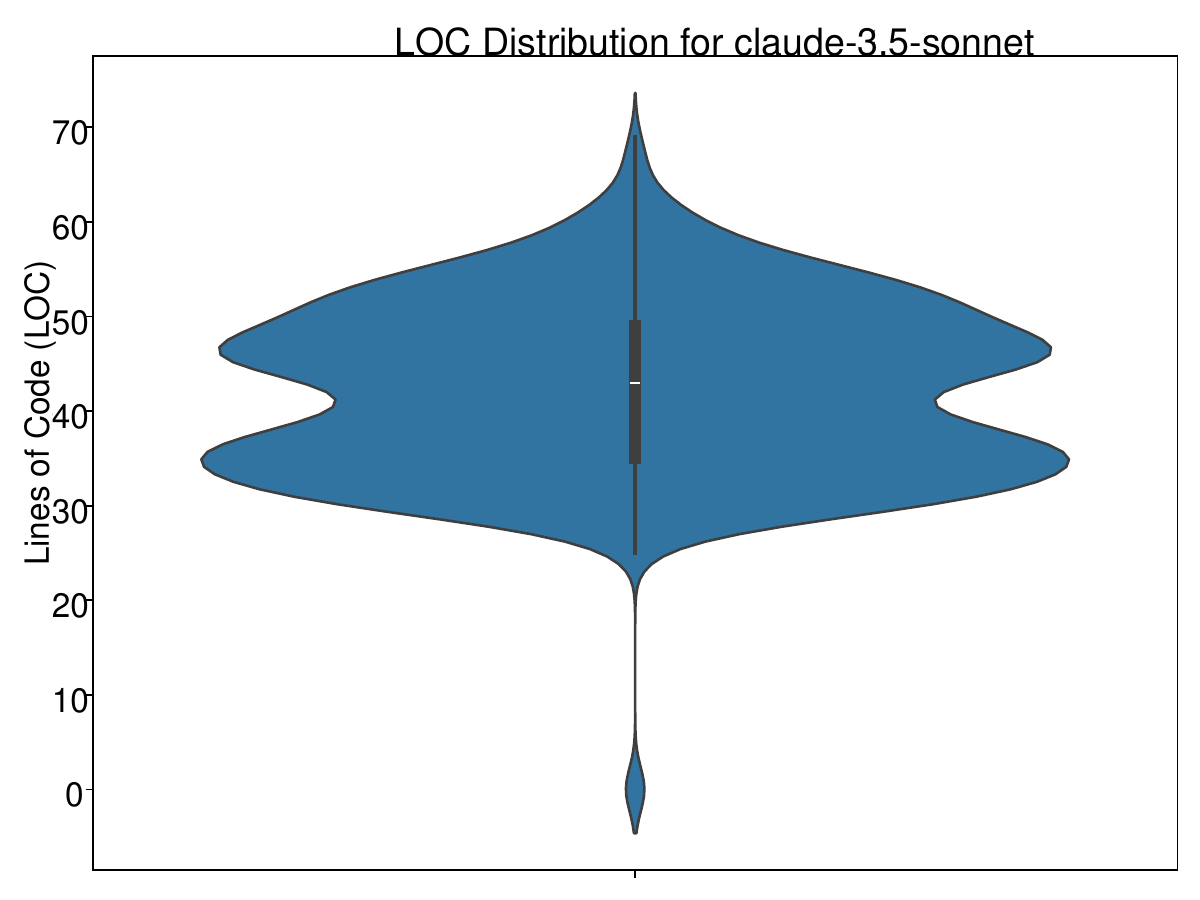}
		\caption{claude-3.5-sonnet}
	\end{subfigure}
		
	\caption{LOC distribution by model (bimodal)}
	\label{fig:loc_distribution_models_bimodal}
\end{figure}

\begin{figure}[h!]
	\centering
	\begin{subfigure}{0.32\textwidth}
		\includegraphics[width=\textwidth]{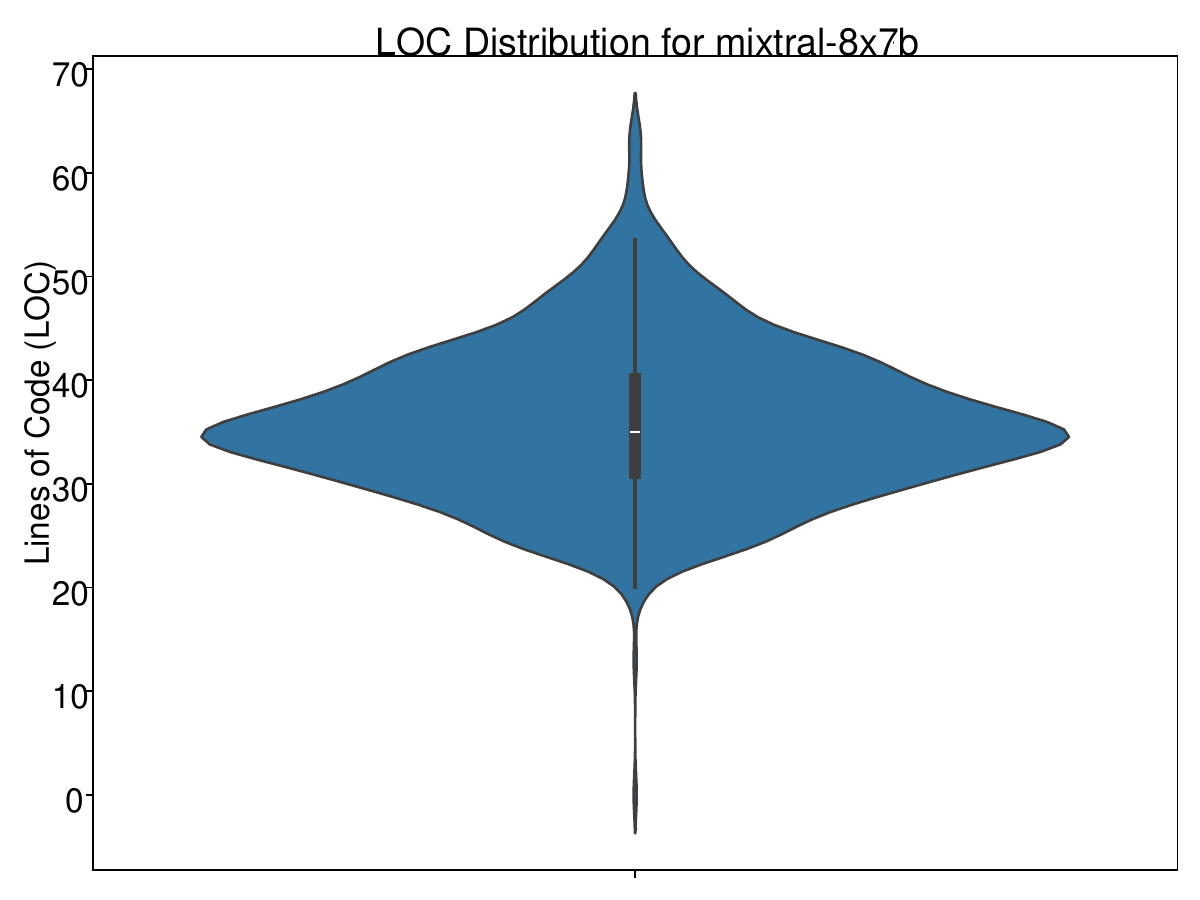}
		\caption{mixtral-8x7b}
	\end{subfigure}
	\begin{subfigure}{0.32\textwidth}
		\includegraphics[width=\textwidth]{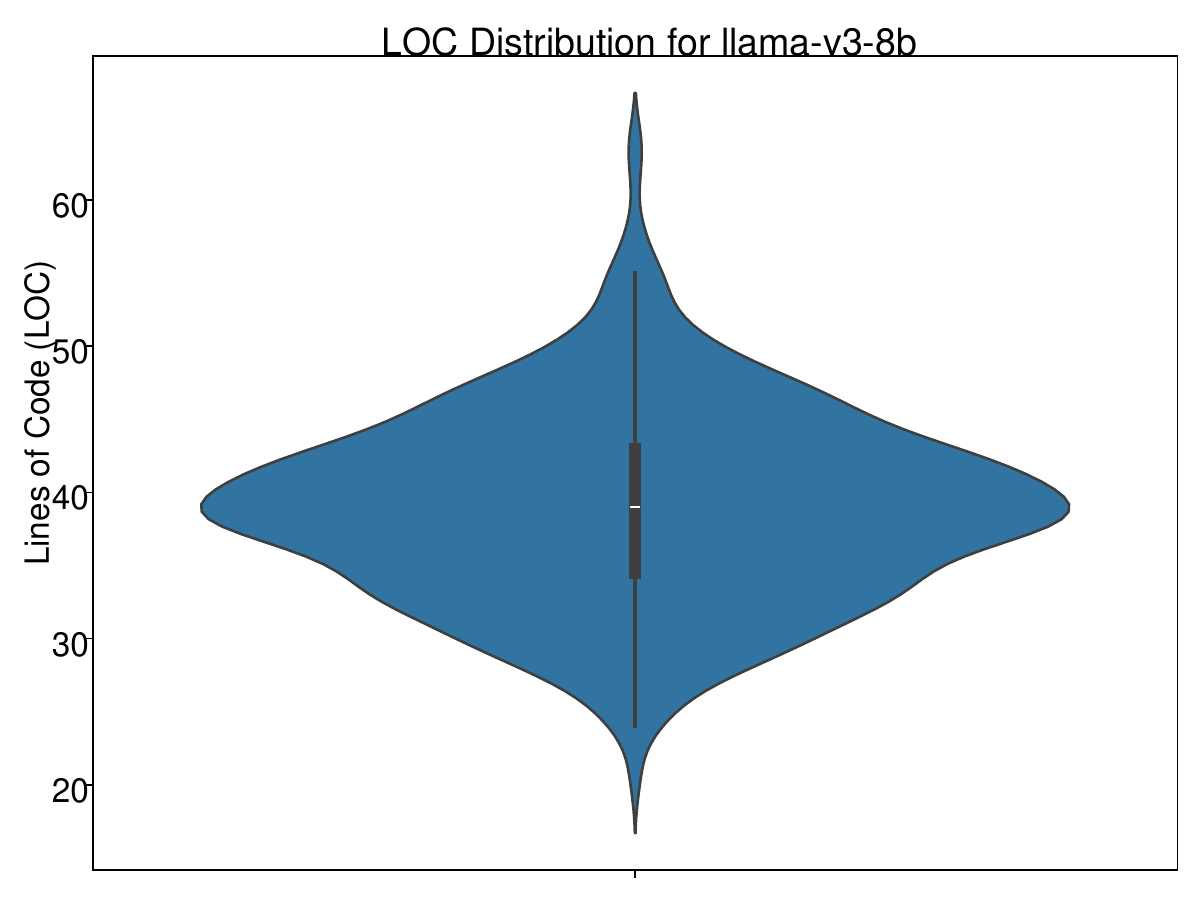}
		\caption{llama-v3-8b}
	\end{subfigure}
	\begin{subfigure}{0.32\textwidth}
		\includegraphics[width=\textwidth]{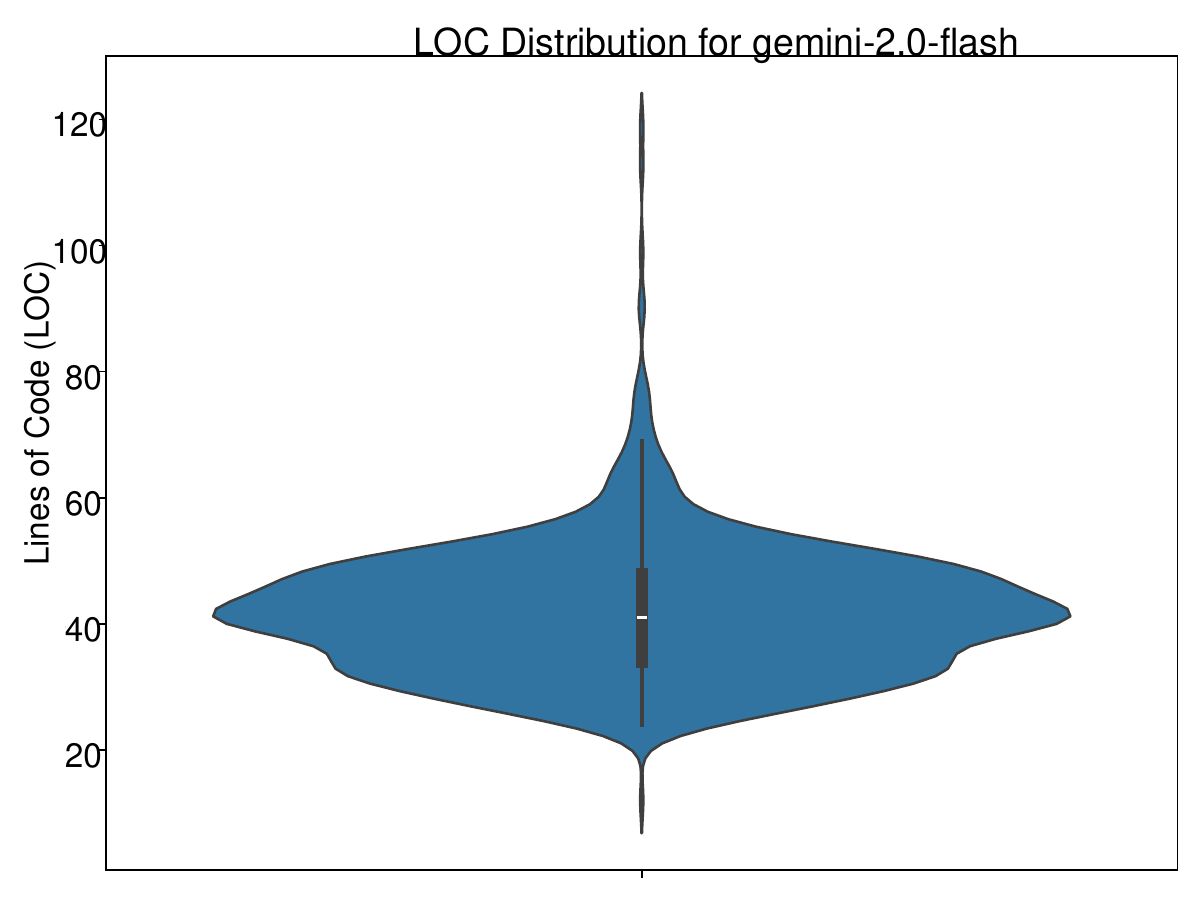}
		\caption{gemini-2.0-flash}
	\end{subfigure}
	
	\begin{subfigure}{0.32\textwidth}
		\includegraphics[width=\textwidth]{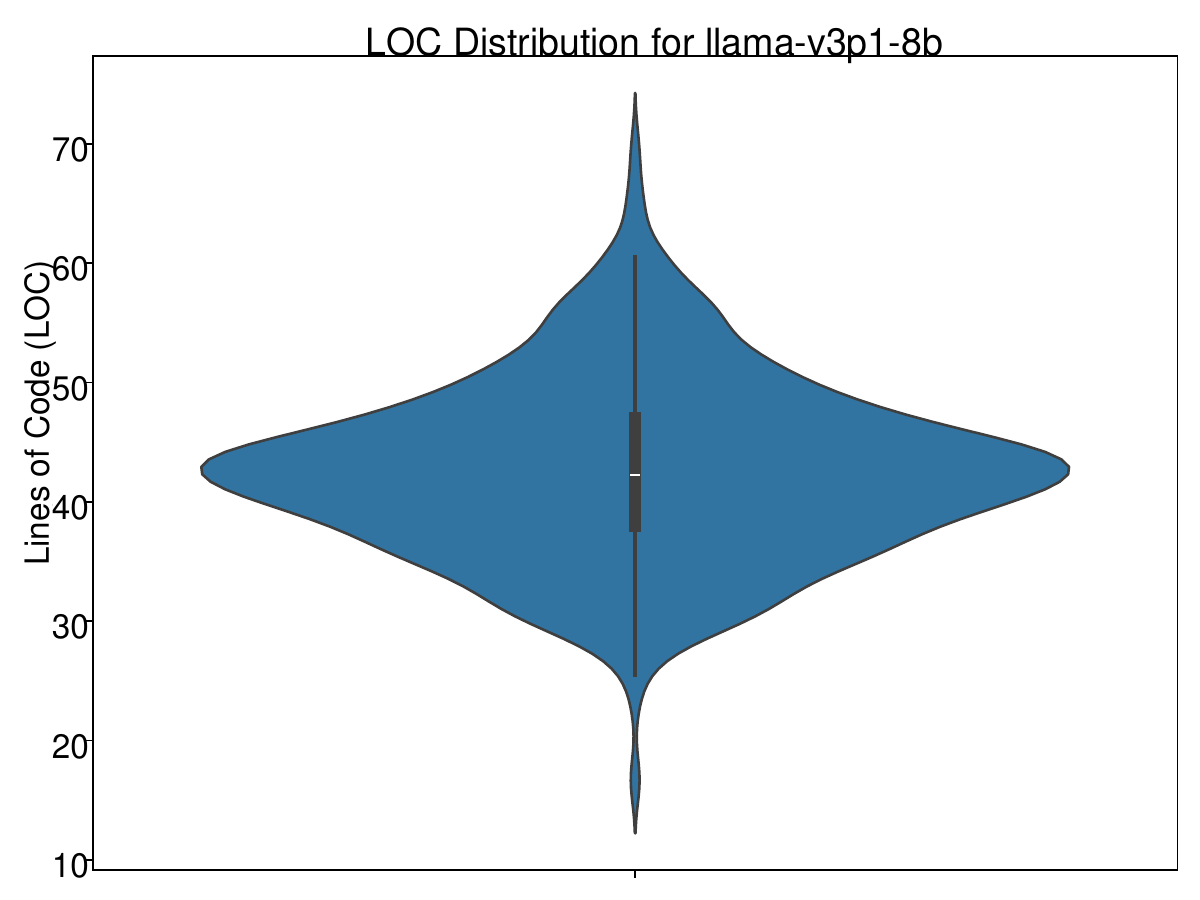}
		\caption{llama-v3p1-8b}
	\end{subfigure}
	\begin{subfigure}{0.32\textwidth}
		\includegraphics[width=\textwidth]{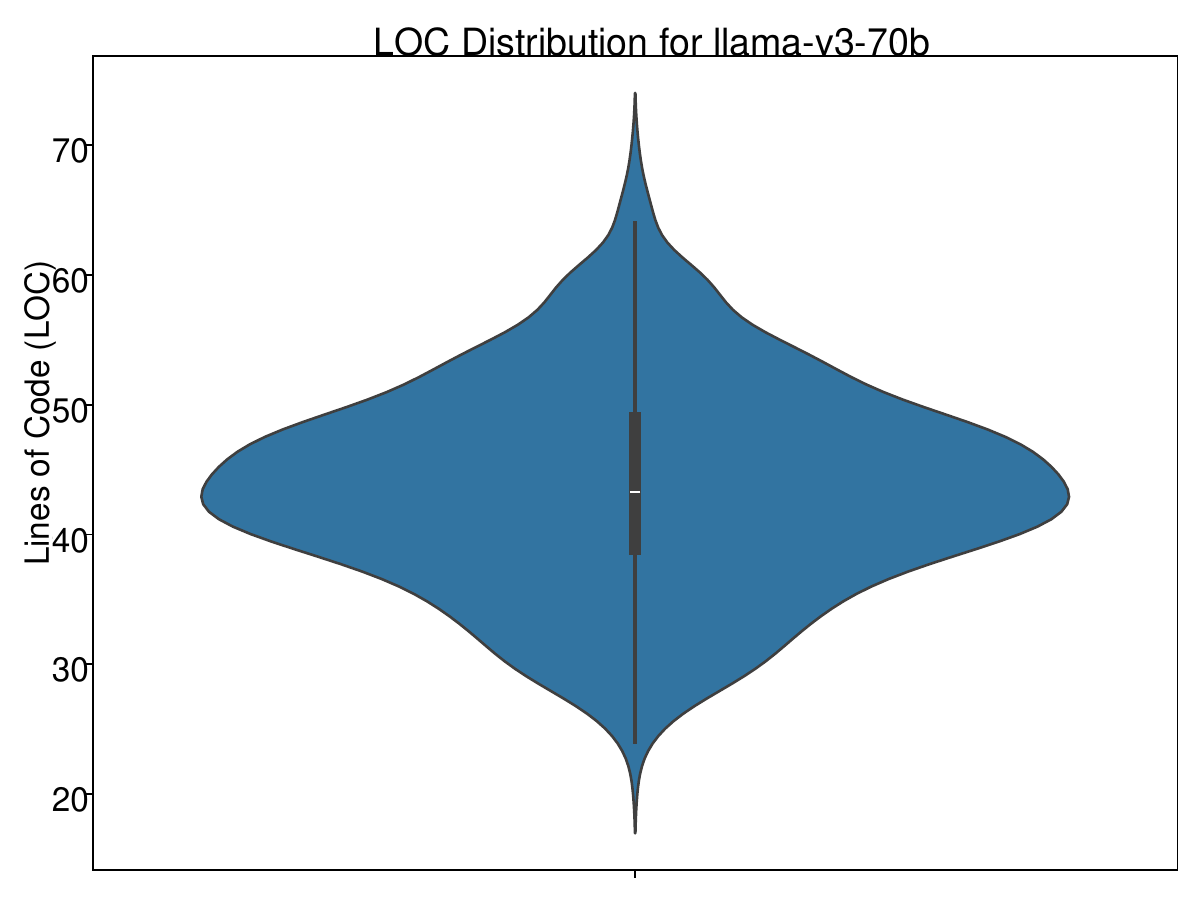}
		\caption{llama-v3-70b}
	\end{subfigure}
	\begin{subfigure}{0.32\textwidth}
		\includegraphics[width=\textwidth]{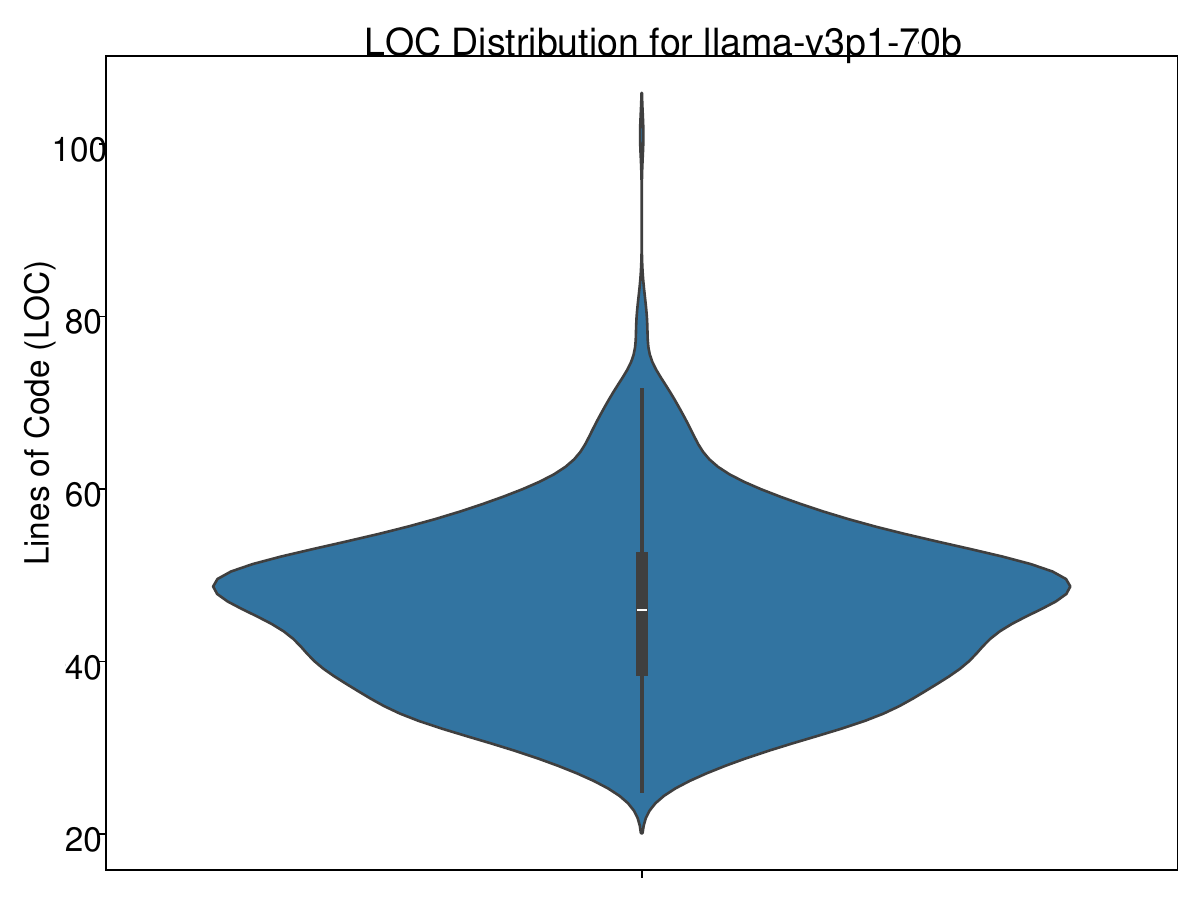}
		\caption{llama-v3p1-70b}
	\end{subfigure}
	
	\caption{LOC distribution by model (unimodal)}
	\label{fig:loc_distribution_models_unimodal}
\end{figure}

Next, we use violin charts to visualize LOC distribution of each model. The distributions are either bimodal or unimodal, and they are collected in Fig.~\ref{fig:loc_distribution_models_bimodal} and Fig.~\ref{fig:loc_distribution_models_unimodal} respectively.

Notably, all high-performing models with high $pass@1$ scores are located in Fig.~\ref{fig:loc_distribution_models_bimodal}. These models, such as the gpt-4o variants and deepseek-coder series, demonstrate higher variability in their LOC distributions, i.e. bimodal. The two distinct peaks in these models' distributions suggests that they generate both shorter and longer code lengths, depending on the task. Importantly, the median LOC values for these bimodal models consistently fall between the two peaks, highlighting a balance in their code generation. Also the higher of the two peaks often corresponds to smaller LOC. This suggests that while these models can produce longer code when necessary, they tend to generate shorter, more optimized code in most cases. 

In contrast, Fig.~\ref{fig:loc_distribution_models_unimodal} contains smaller models. Some exhibit near-perfect normal distributions, e.g. mixtral-8x7b and llama-v3-8b. These models generate LOC distributions that are tightly centered around their medians, indicating more consistent and predictable behavior. The lack of bimodal characteristics in these distributions reflects a more stable output across tasks, but with lower complexity compared to the larger models in Fig.~\ref{fig:loc_distribution_models_bimodal}.
\subsection{Impact of Success/Failure}\label{sec:loc_successfail}
\begin{figure}[h!]
    \centering
    \begin{subfigure}{0.49\textwidth}
        \centering
        \includegraphics[width=\linewidth]{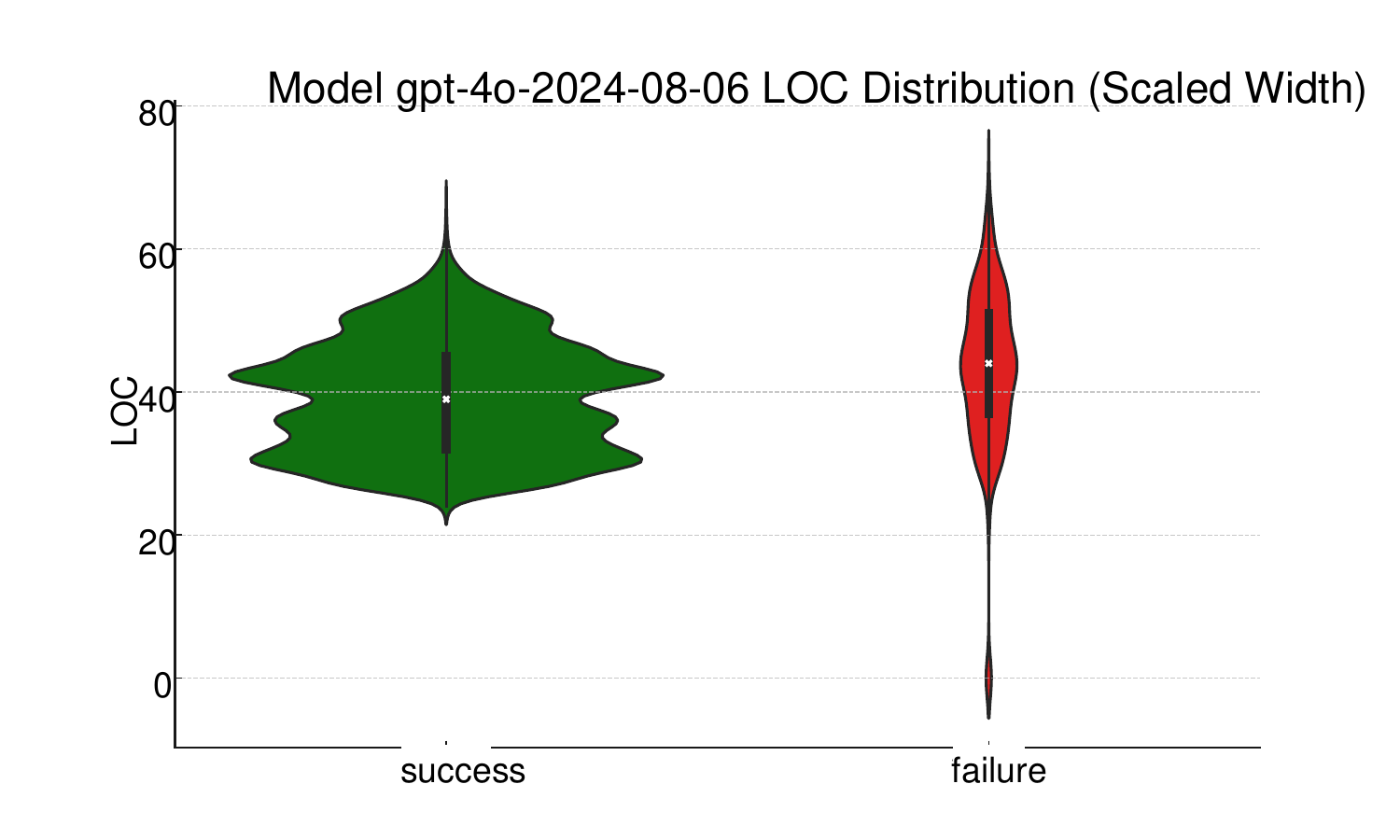}
        \caption{gpt-4o-2024-08-06 ($pass@1$ = 0.885)}
    \end{subfigure}
    \begin{subfigure}{0.49\textwidth}
        \centering
        \includegraphics[width=\linewidth]{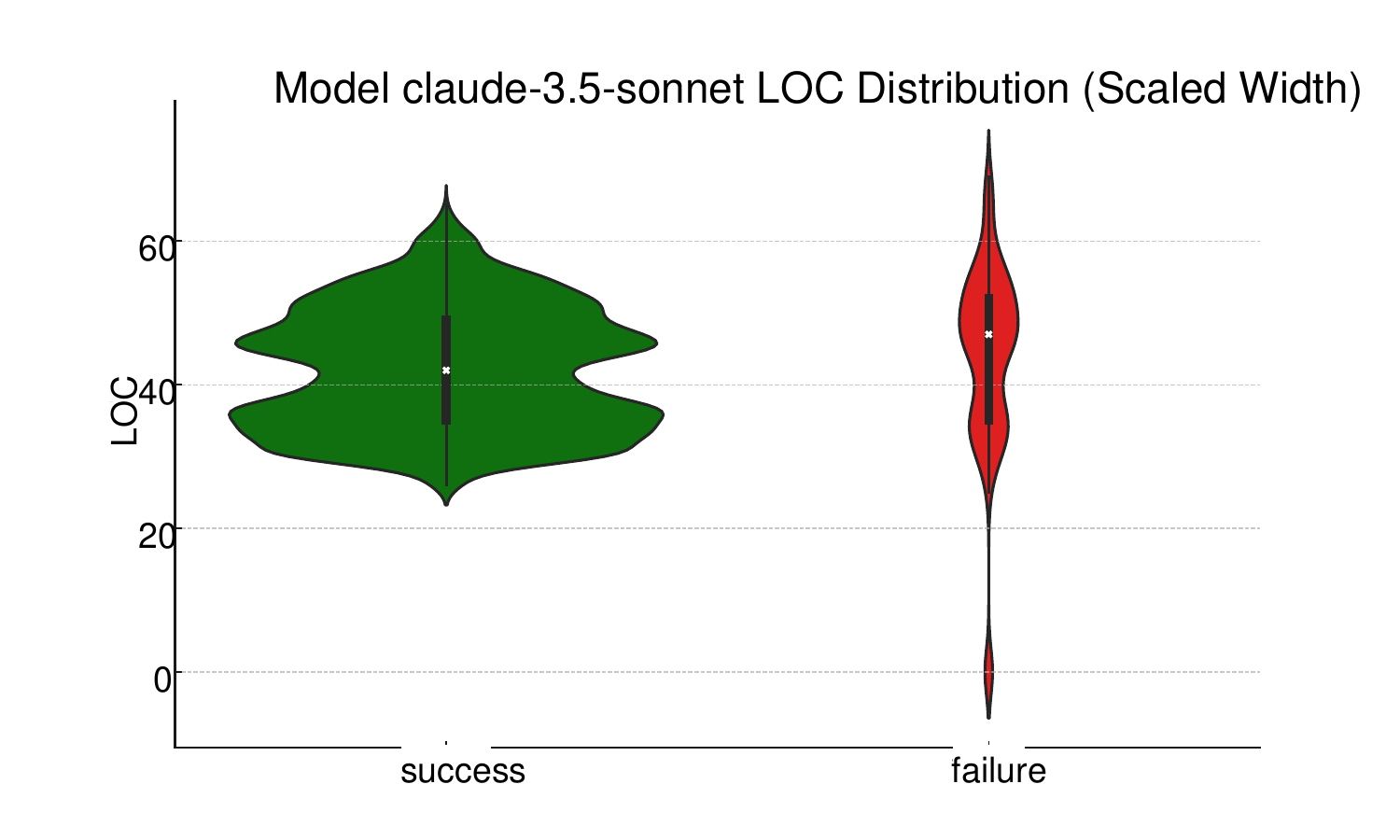}
        \caption{claude-3.5-sonnet ($pass@1$ = 0.8808)}
    \end{subfigure}

    \begin{subfigure}{0.49\textwidth}
        \centering
        \includegraphics[width=\linewidth]{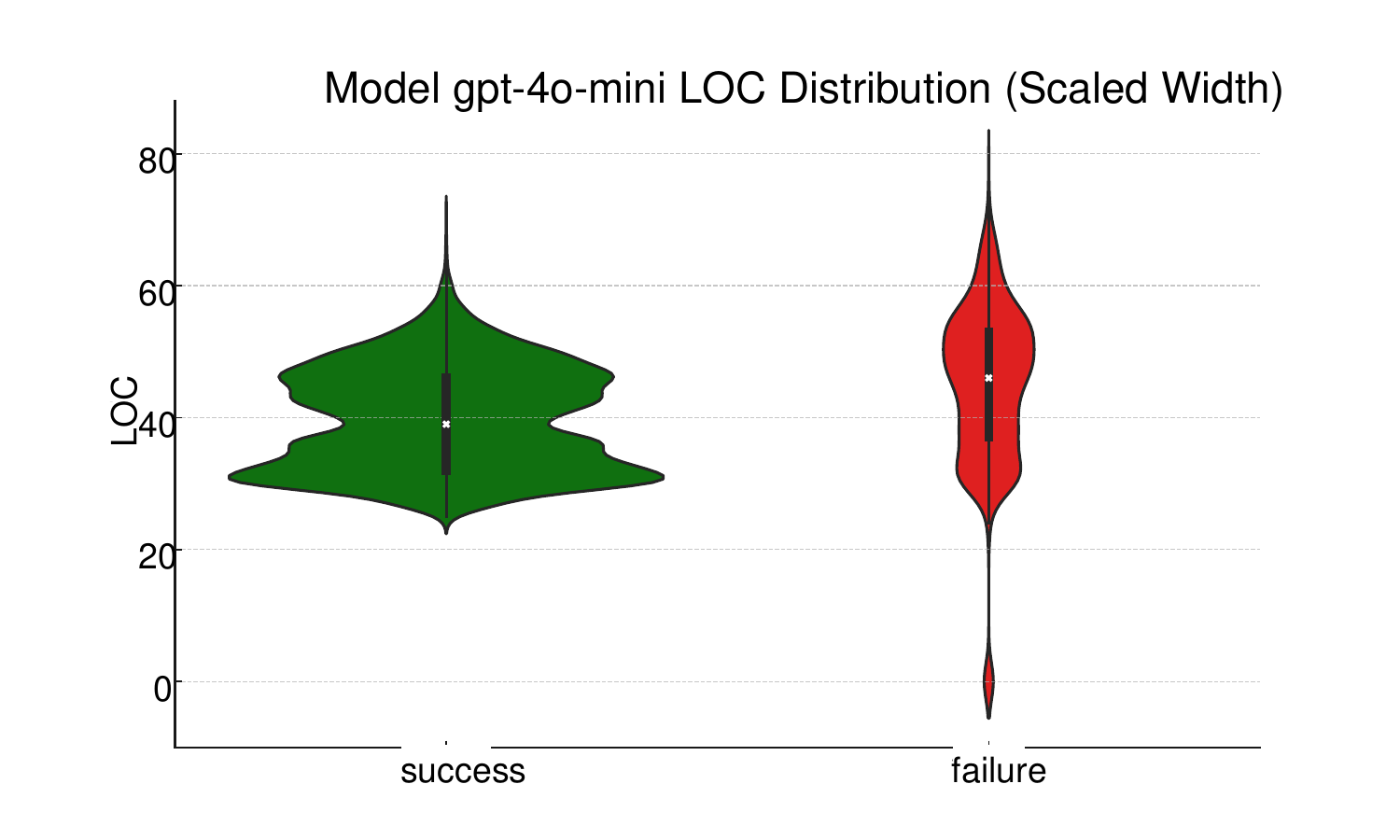}
        \caption{gpt-4o-mini ($pass@1$ = 0.8271)}
    \end{subfigure}
    \caption{LOC distribution by model of high pass@1: success vs failure}
    \label{fig:loc_success_distribution_models}
\end{figure}

\begin{figure}[h!]
    \centering
    \begin{subfigure}{0.49\textwidth}
        \centering
        \includegraphics[width=\linewidth]{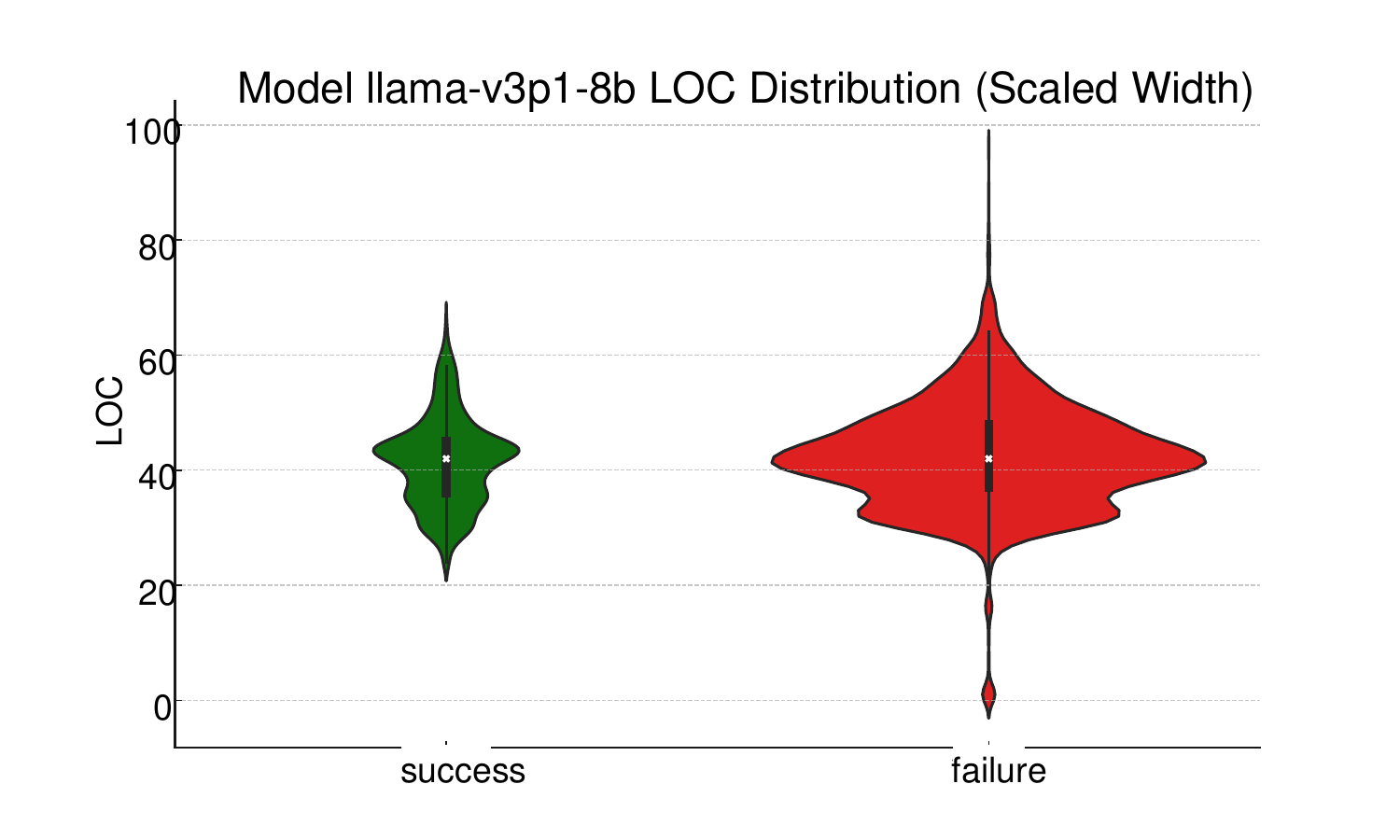}
        \caption{llama-v3p1-8b (pass@1 = 0.2512)}
    \end{subfigure}
    \begin{subfigure}{0.49\textwidth}
        \centering
        \includegraphics[width=\linewidth]{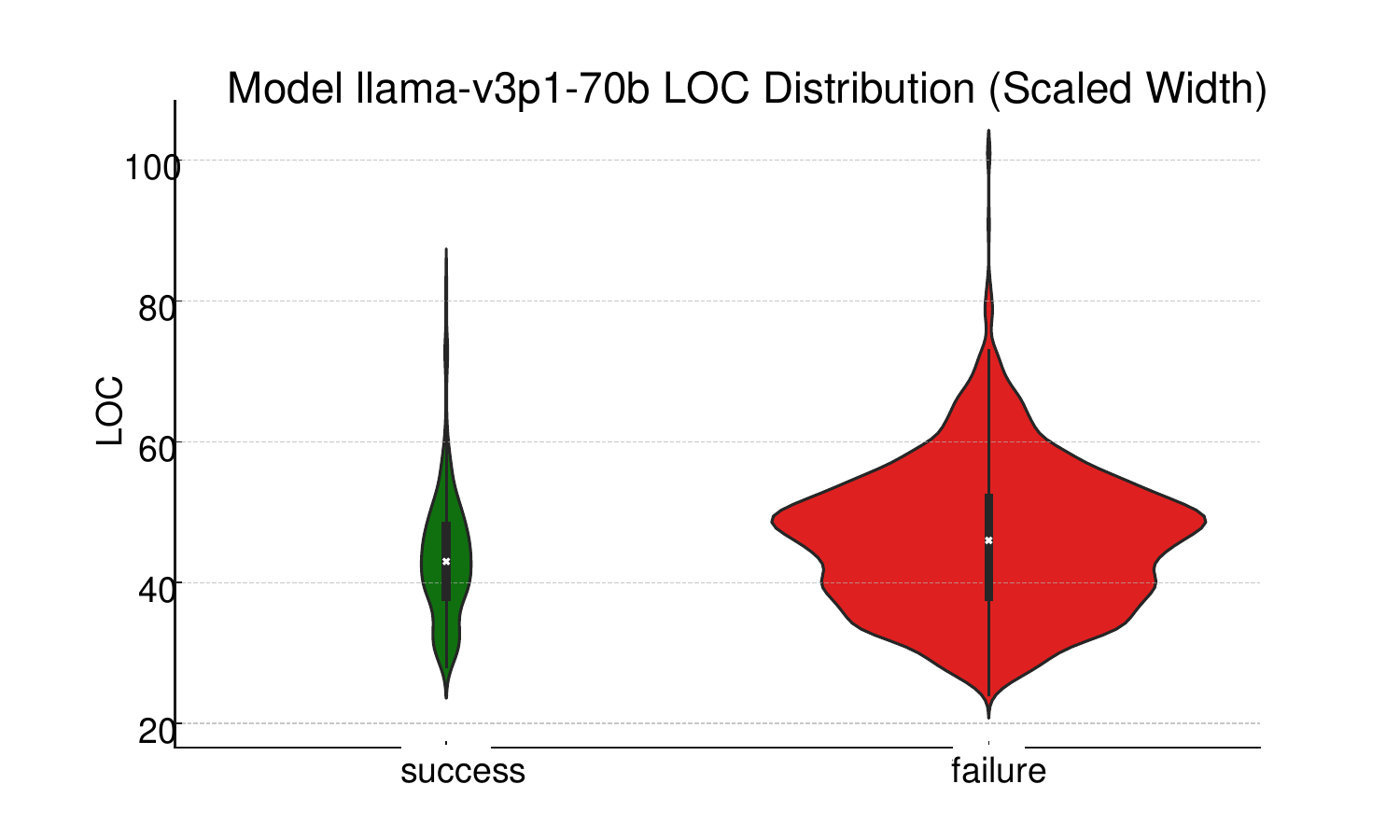}
        \caption{llama-v3p1-70b (pass@1 = 0.1027)}
    \end{subfigure}

    \begin{subfigure}{0.49\textwidth}
        \centering
        \includegraphics[width=\linewidth]{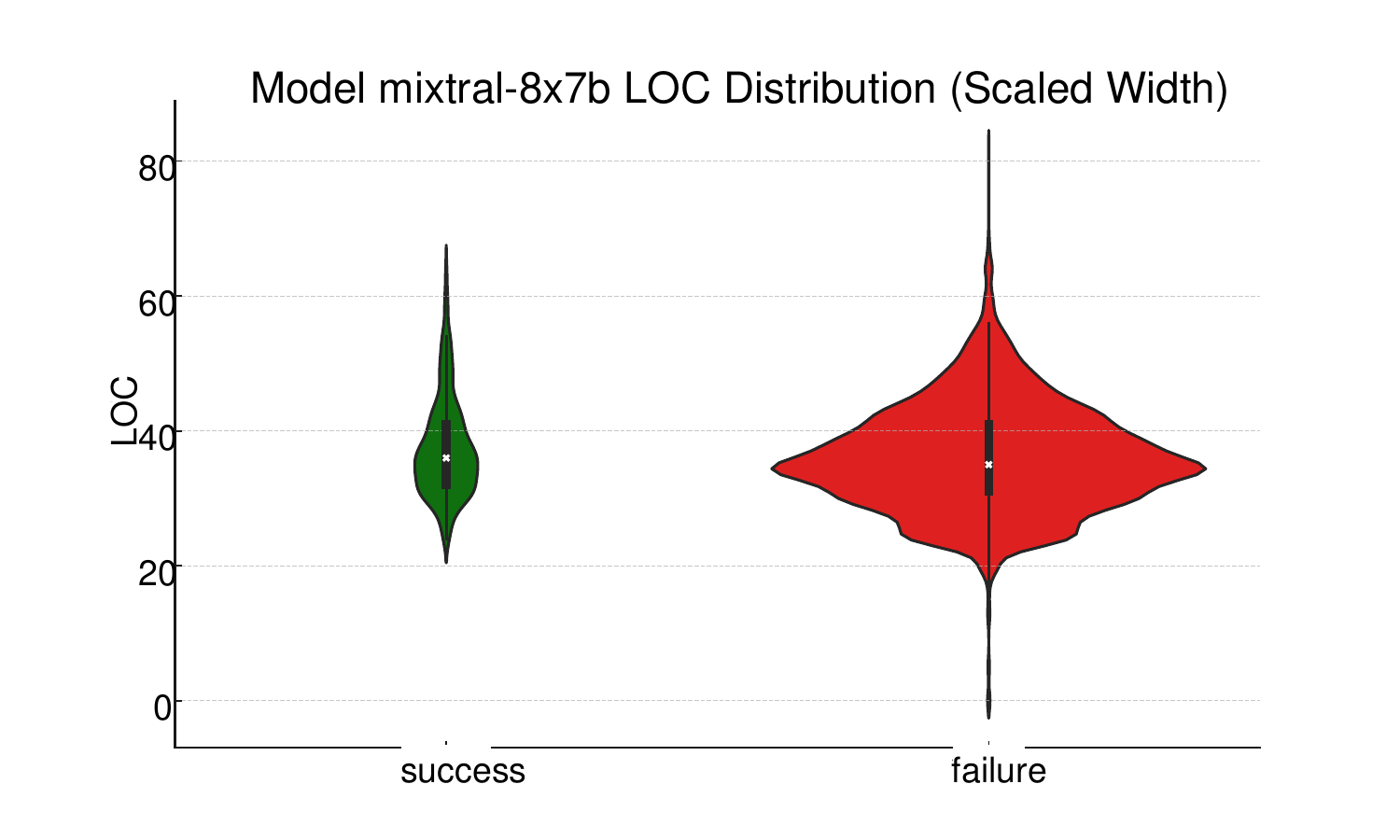}
        \caption{mixtral-8x7b (pass@1 = 0.1269)}
    \end{subfigure}
    \begin{subfigure}{0.49\textwidth}
        \centering
        \includegraphics[width=\linewidth]{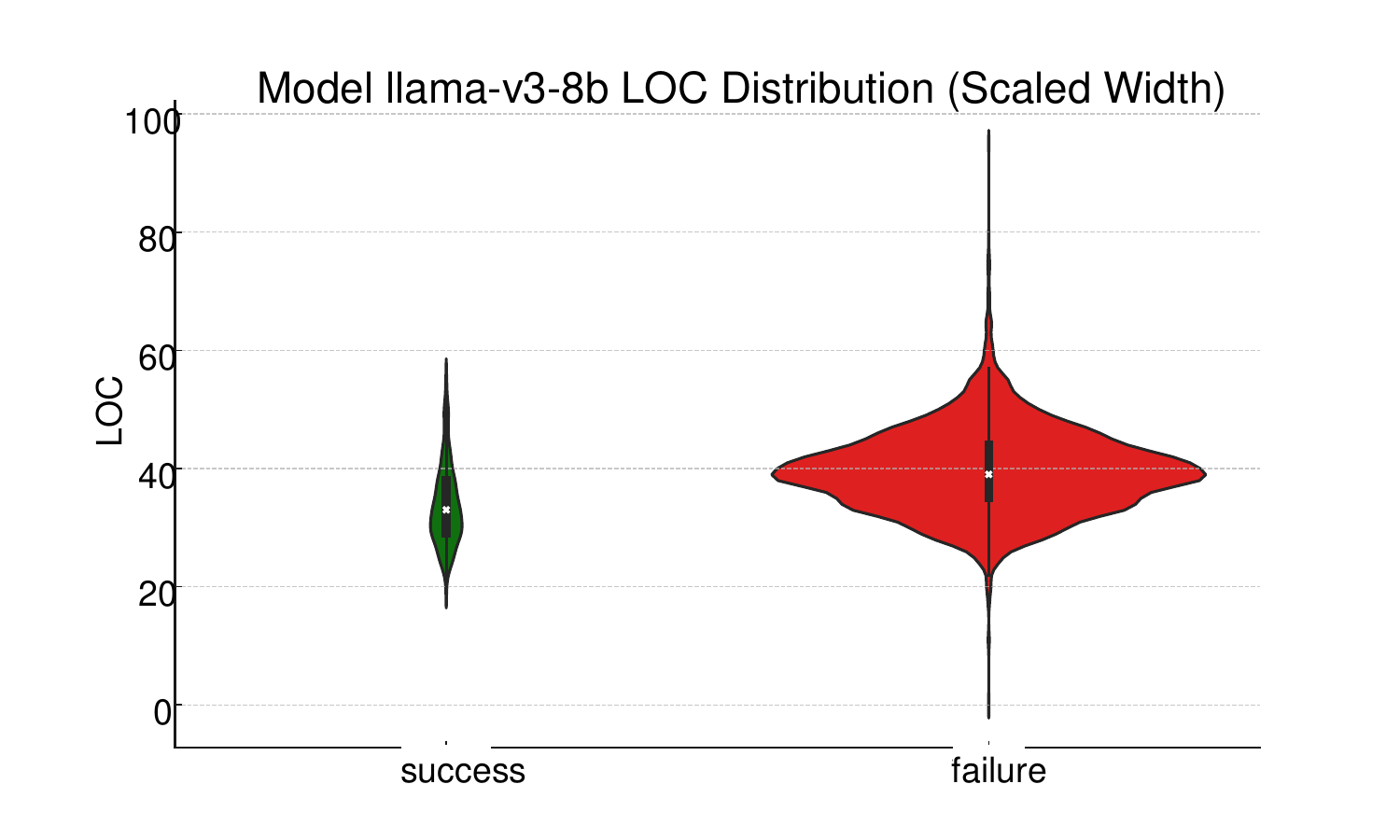}
        \caption{llama-v3-8b (pass@1 = 0.0679)}
    \end{subfigure}
    \caption{LOC distribution by model of low pass@1: success vs failure}
    \label{fig:loc_fail_distribution_models}
\end{figure}

To get more insights, we search for statistical distinction between successful model outputs and failed outputs. In Fig.~\ref{fig:loc_success_distribution_models} and \ref{fig:loc_fail_distribution_models}, we visualize the LOC distribution separately for succssful outputs and failed ones, for each model. The graphs are ranked by $pass@1$, where higher $pass@1$ means bigger success sample set and smaller failure sample set. We normalize the width of each violin chart by its sample set size, hence resulting in the thinnest failure graph for the model with the highest $pass@1$. The graph gradually grows wider as the model performance degrades. The opposite pattern is observed for the success violin chart.

An important finding here is that the success distribution is always more complex than its failure counterpart, with more peaks and deviations. Fig.~\ref{fig:loc_fail_distribution_models} groups lower performing models whose failure sample set dominates the success sample set. The failure LOC distributions are unimodal, in contrast with the multimodal distributions of top models in Fig.~\ref{fig:loc_success_distribution_models}. This implies the inherent complexity involved in writing correct code even when the mean LOC is less than 50.

\begin{figure}[h!]
    \centering
    \begin{subfigure}{0.49\textwidth}
        \centering
        \includegraphics[width=\linewidth]{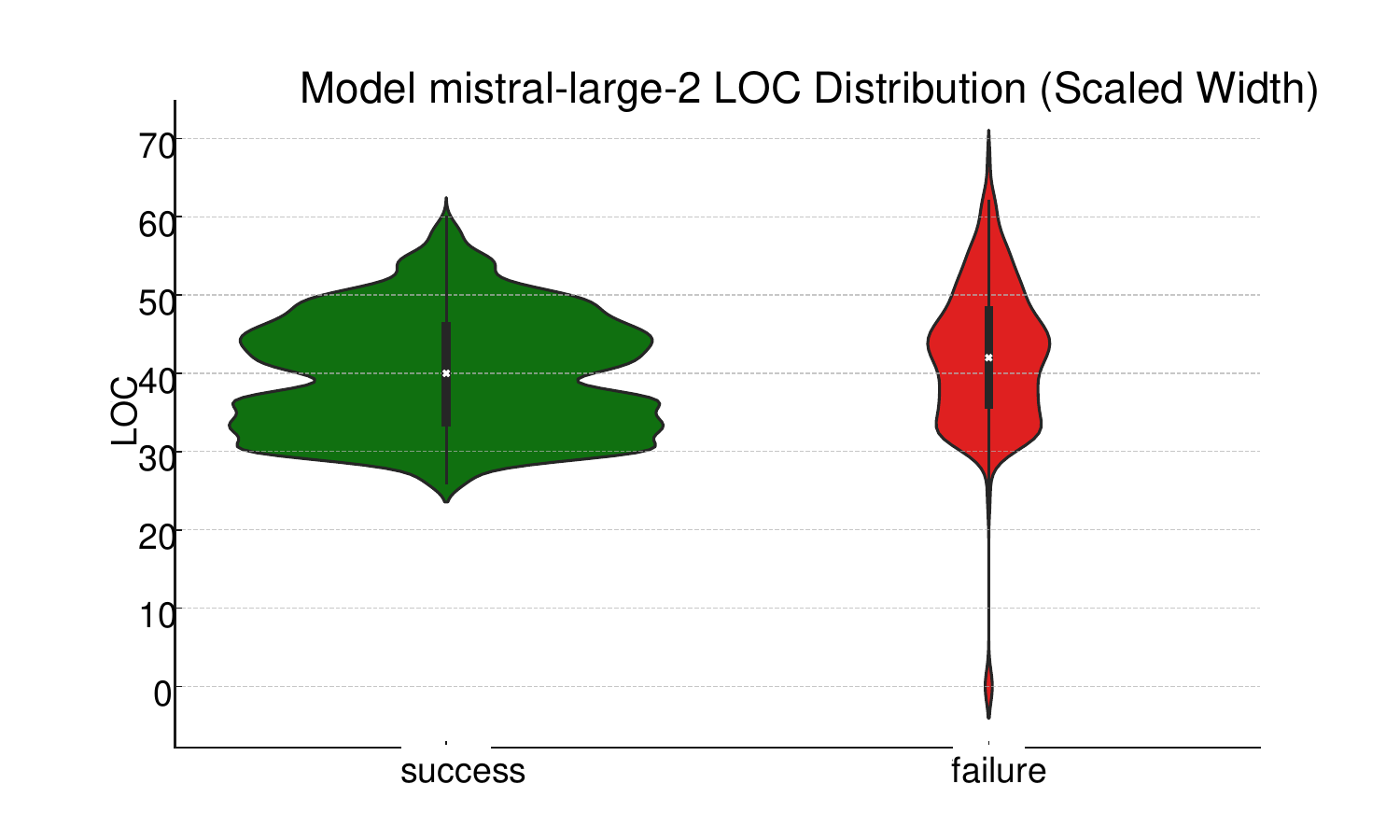}
        \caption{mistral-large-2 ($pass@1$ = 0.7804)}
    \end{subfigure}
    \begin{subfigure}{0.49\textwidth}
        \centering
        \includegraphics[width=\linewidth]{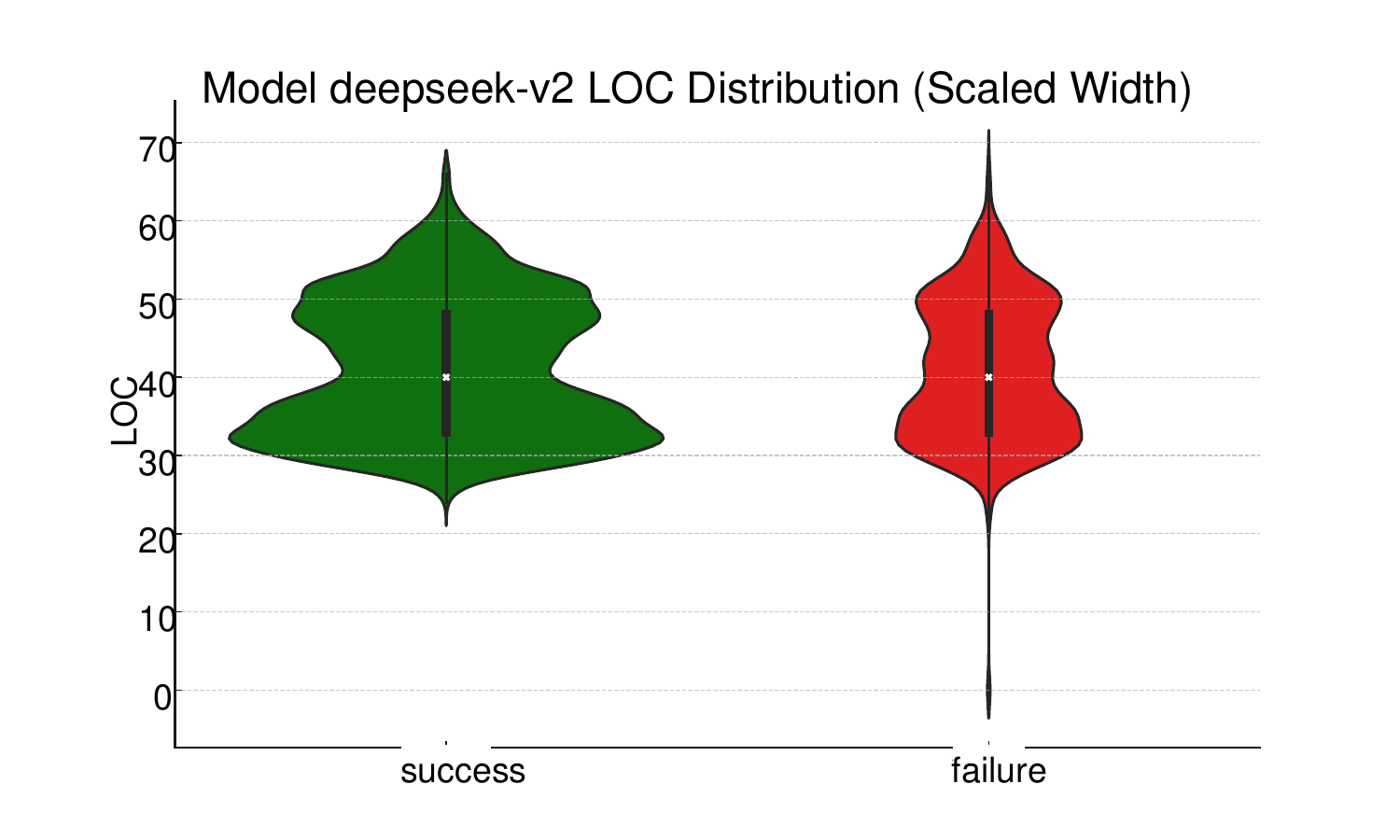}
        \caption{deepseek-coder-v2 ($pass@1$ = 0.7002)}
    \end{subfigure}

    \begin{subfigure}{0.49\textwidth}
        \centering
        \includegraphics[width=\linewidth]{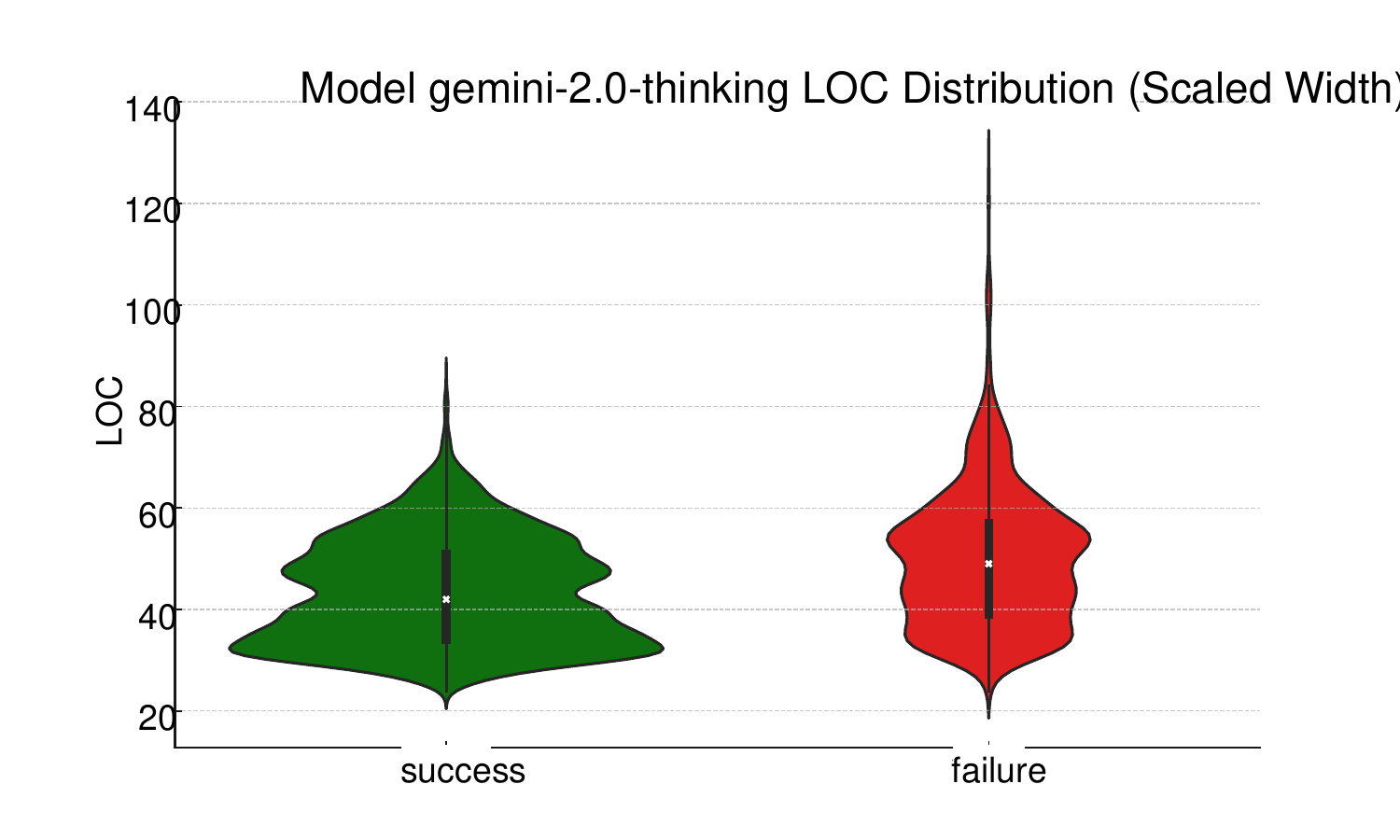}
        \caption{gemini-2.0-thinking ($pass@1$ = 0.6813)}
    \end{subfigure}
    \begin{subfigure}{0.49\textwidth}
        \centering
        \includegraphics[width=\linewidth]{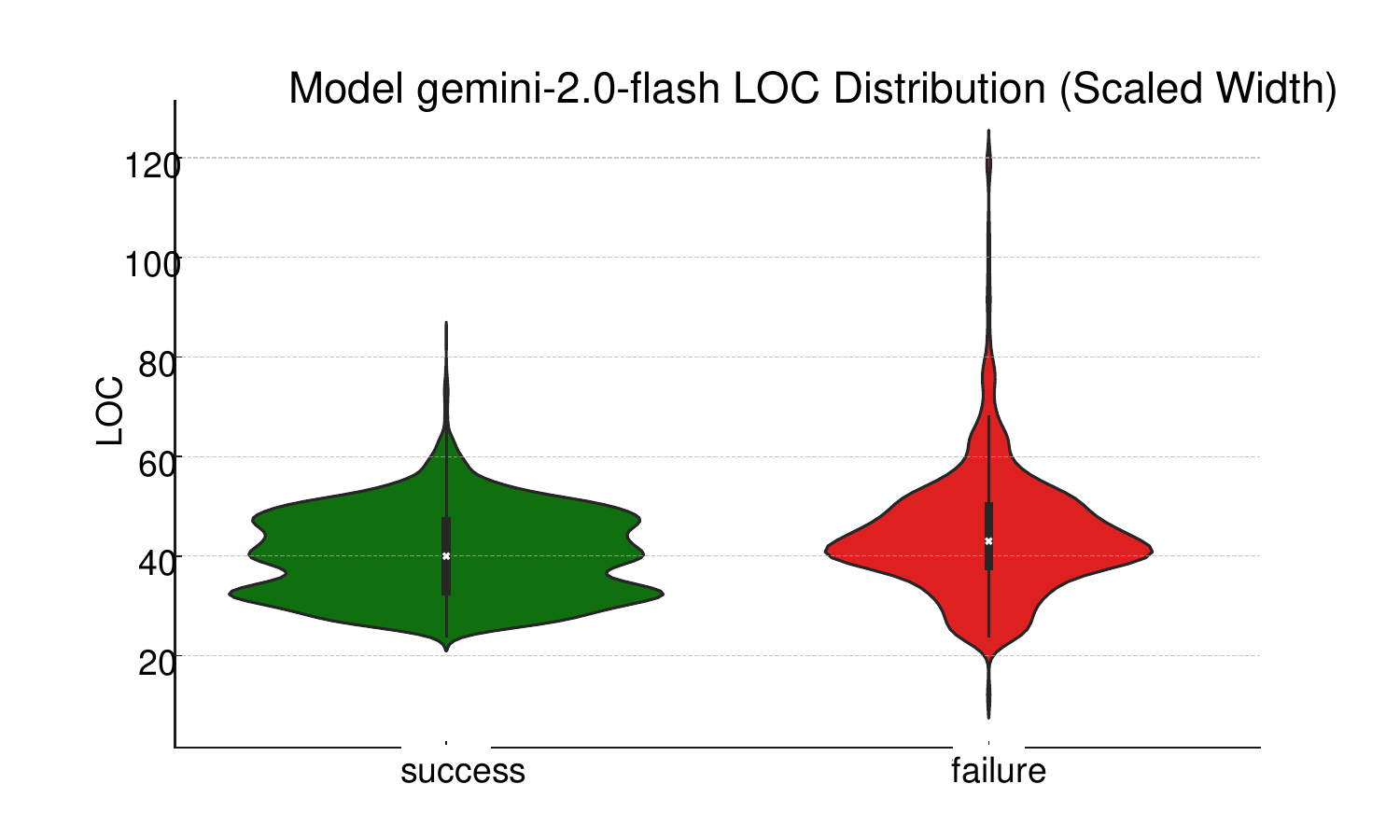}
        \caption{gemini-2.0-flash ($pass@1$ = 0.57)}
    \end{subfigure}

    \begin{subfigure}{0.49\textwidth}
        \centering
        \includegraphics[width=\linewidth]{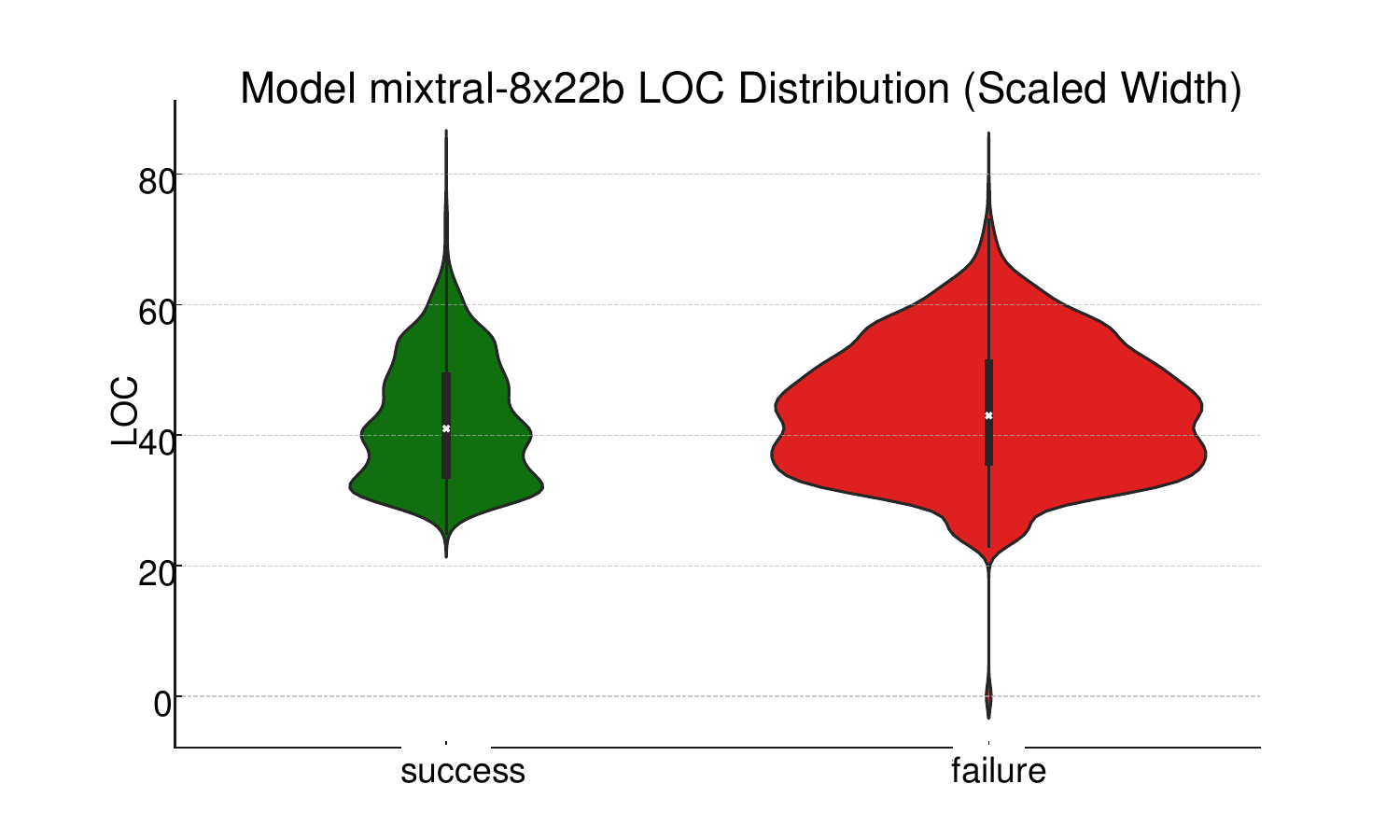}
        \caption{mixtral-8x22b ($pass@1$ = 0.3074)}
    \end{subfigure}
    \begin{subfigure}{0.49\textwidth}
        \centering
        \includegraphics[width=\linewidth]{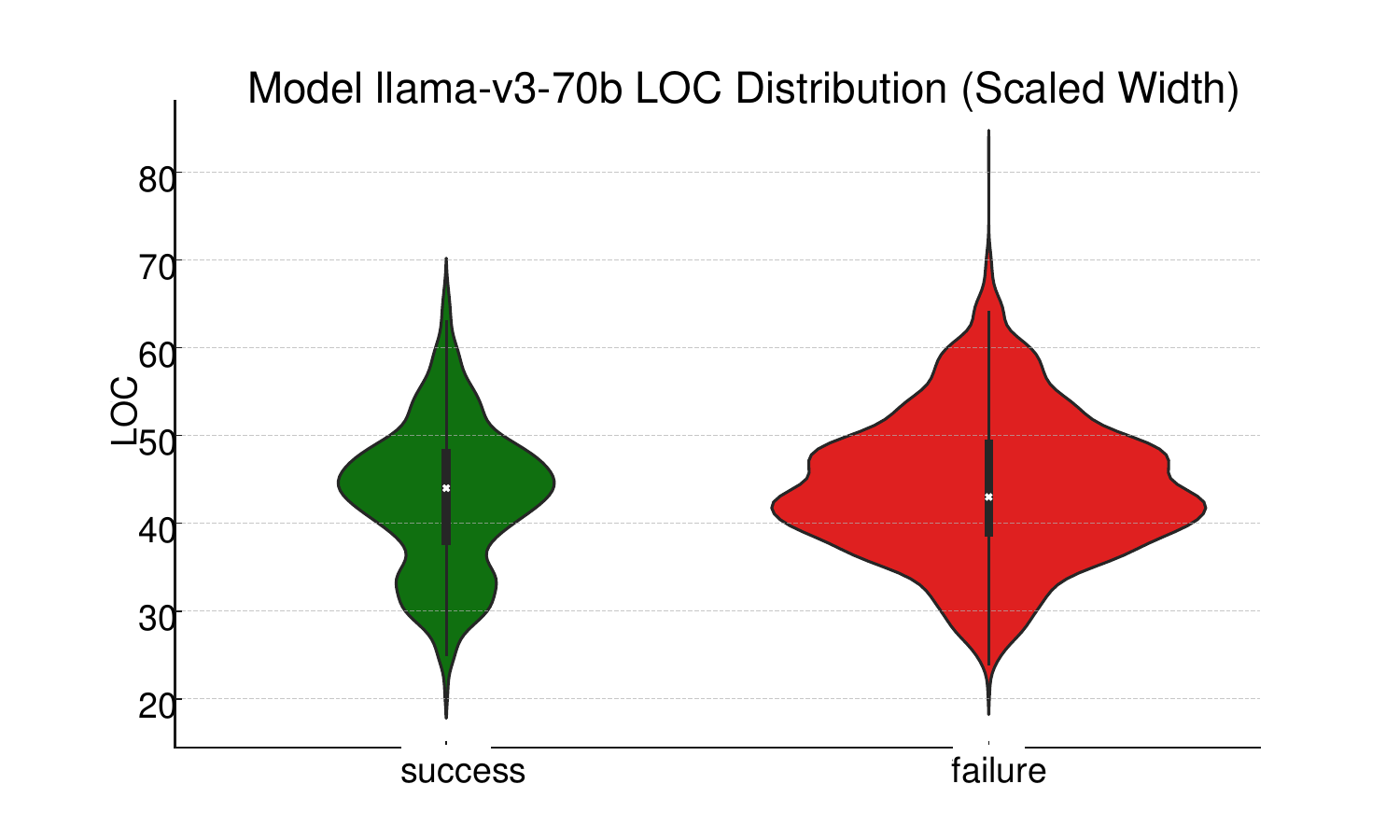}
        \caption{llama-v3-70b (pass@1 = 0.3323)}
    \end{subfigure}

    \begin{subfigure}{0.49\textwidth}
        \centering
        \includegraphics[width=\linewidth]{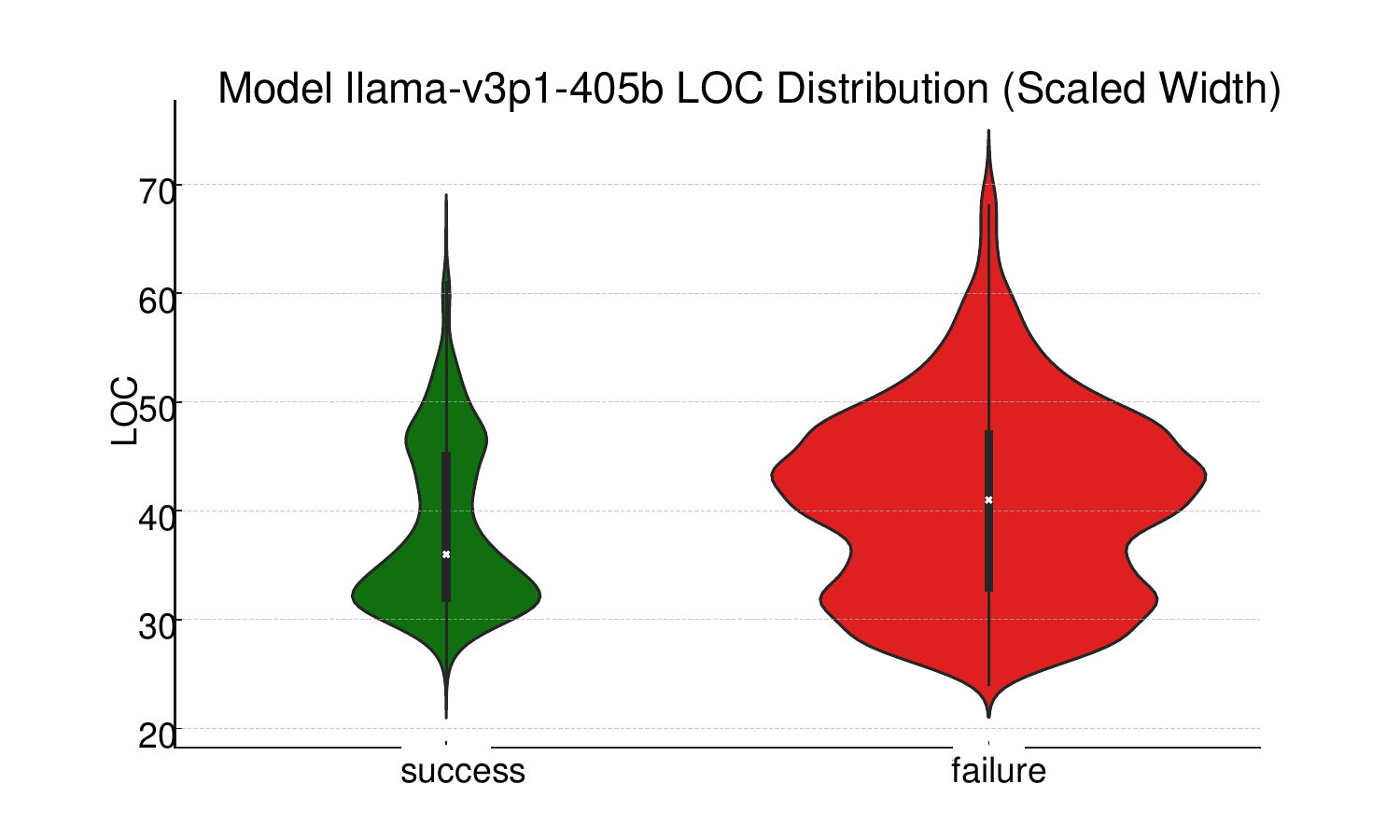}
        \caption{llama-v3p1-405b (pass@1 = 0.302)}
    \end{subfigure}
    \caption{LOC distribution by model: success and failure}
    \label{fig:loc_successfail_distribution_models}
\end{figure}

The success/fail LOC distribution of remaining 8 models are shown in Fig.~\ref{fig:loc_successfail_distribution_models}.
\subsection{LOC Distribution by Applications}
In Tab.~\ref{tab:loc_by_apps}, we rank median LOC for each application. Consistent with the case for model ranking (Tab.~\ref{tab:loc_by_models}), the median values stay within a narrow range (37 to 46). This suggests that all models consistently produce solutions of similar length, irrespective of the task complexity or domain.

\begin{table}[h!]
\caption{Applications ranked by mean LOC}
\centering
\begin{tabular}{|l|l|l|}
\hline
\textbf{Application}          & \textbf{Mean LOC} \\ \hline
News Aggregator       & 37    \\ \hline
Music Streaming       & 37    \\ \hline
Online Marketplace    & 37    \\ \hline
E-commerce            & 37    \\ \hline
Recipe Sharing        & 38    \\ \hline
Fitness Tracking      & 38    \\ \hline
Online Learning       & 38    \\ \hline
Blogging              & 39    \\ \hline
Weather               & 40    \\ \hline
Real Estate           & 42    \\ \hline
Social Media          & 42    \\ \hline
Job Board             & 42    \\ \hline
Inventory Management  & 42    \\ \hline
Pet Care              & 42    \\ \hline
Travel Planning       & 42    \\ \hline
Personal Finance      & 43    \\ \hline
Customer Support      & 44    \\ \hline
Photo Gallery         & 44    \\ \hline
Event Management      & 45    \\ \hline
Task Management       & 46    \\ \hline
\end{tabular}
\label{tab:loc_by_apps}
\end{table}

\begin{figure}[h]
    \centering
    \begin{subfigure}{0.3\textwidth}
        \centering
        \includegraphics[width=\textwidth]{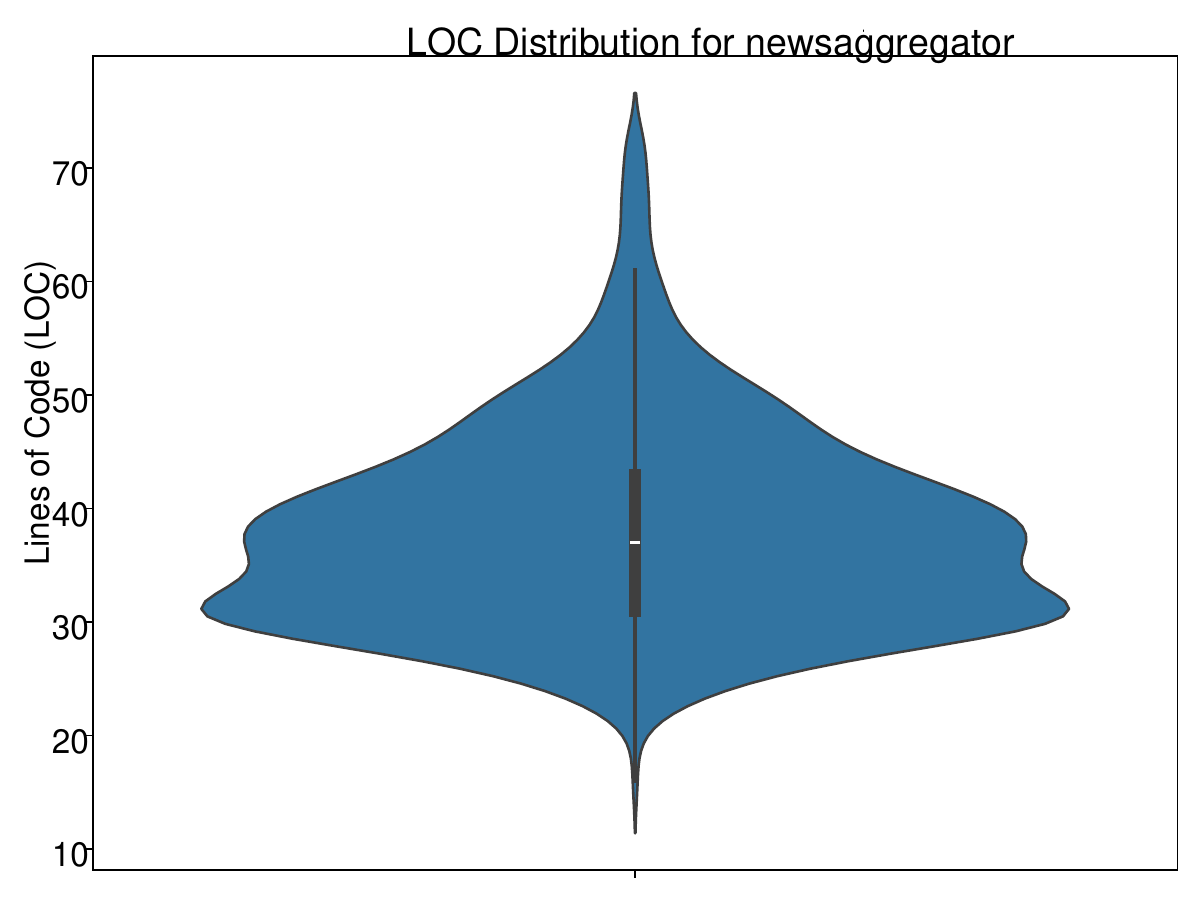}
        \caption{News aggregator}
    \end{subfigure}
    \begin{subfigure}{0.3\textwidth}
        \centering
        \includegraphics[width=\textwidth]{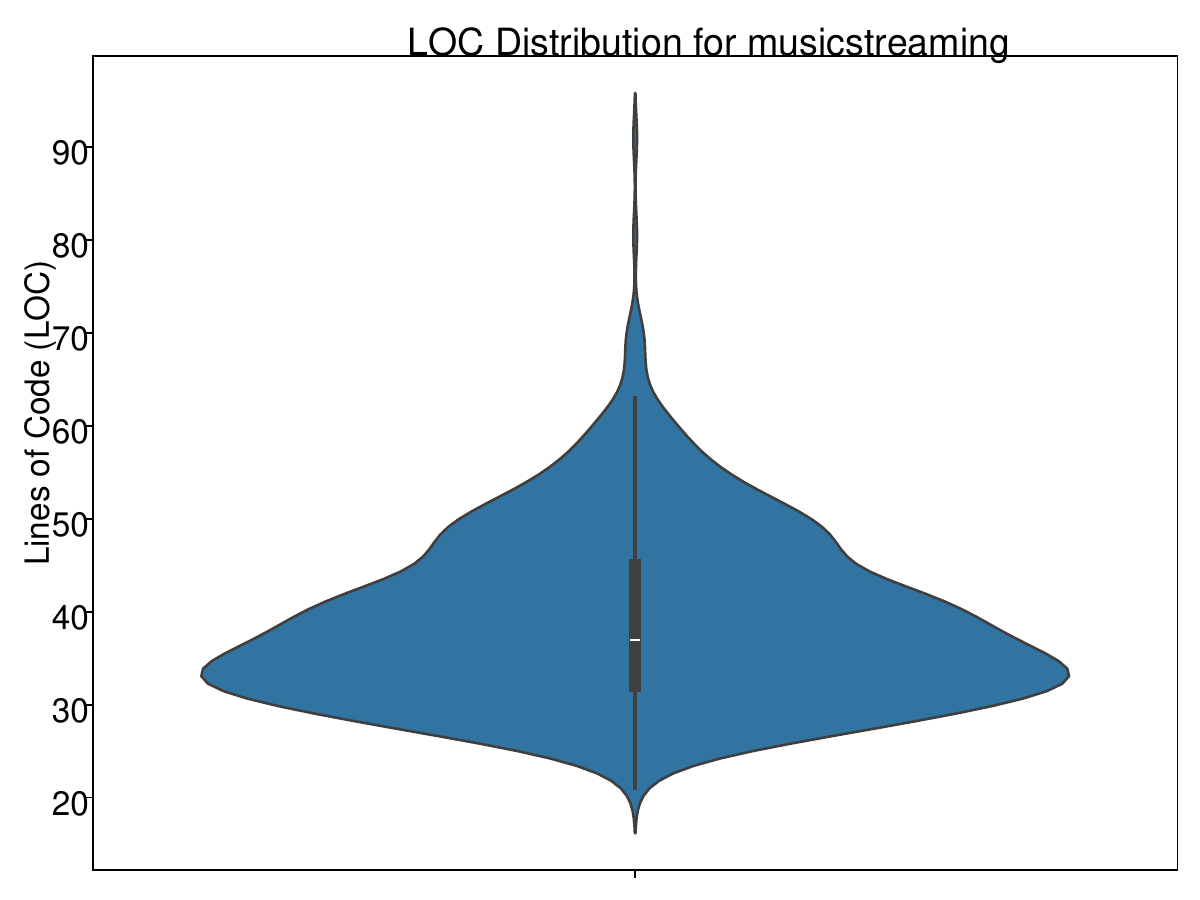}
        \caption{Music streaming}
    \end{subfigure}
    \begin{subfigure}{0.3\textwidth}
        \centering
        \includegraphics[width=\textwidth]{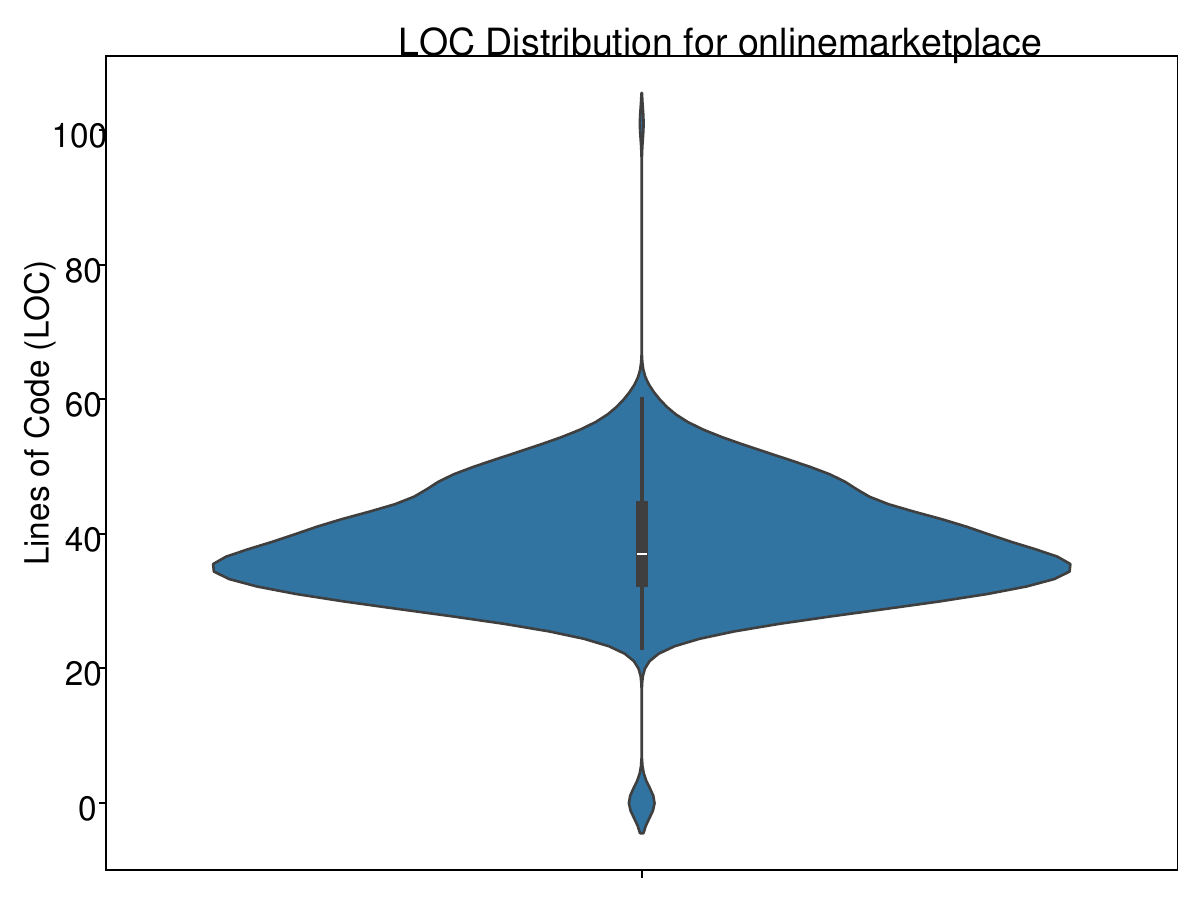}
        \caption{Online marketplace}
    \end{subfigure}
    
    \begin{subfigure}{0.3\textwidth}
        \centering
        \includegraphics[width=\textwidth]{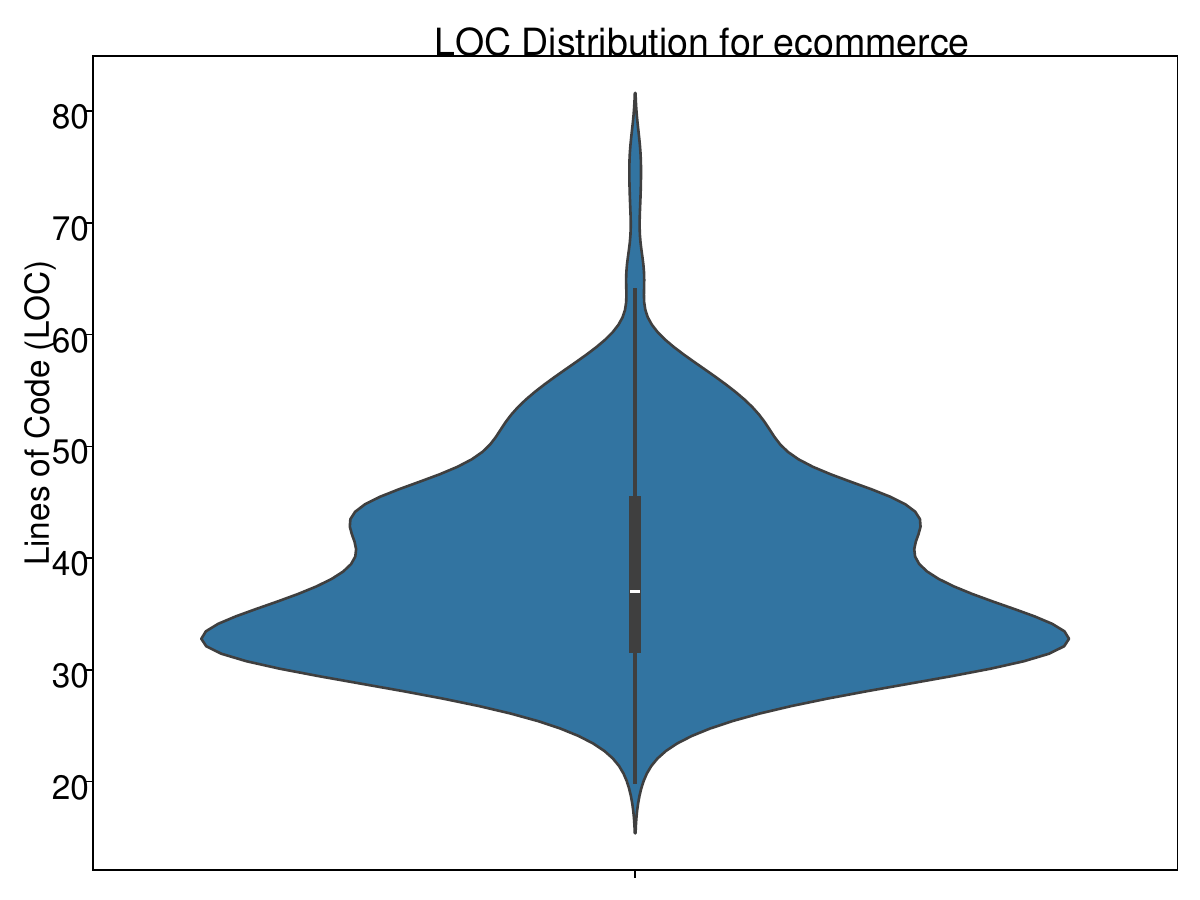}
        \caption{E-commerce}
    \end{subfigure}
    \begin{subfigure}{0.3\textwidth}
        \centering
        \includegraphics[width=\textwidth]{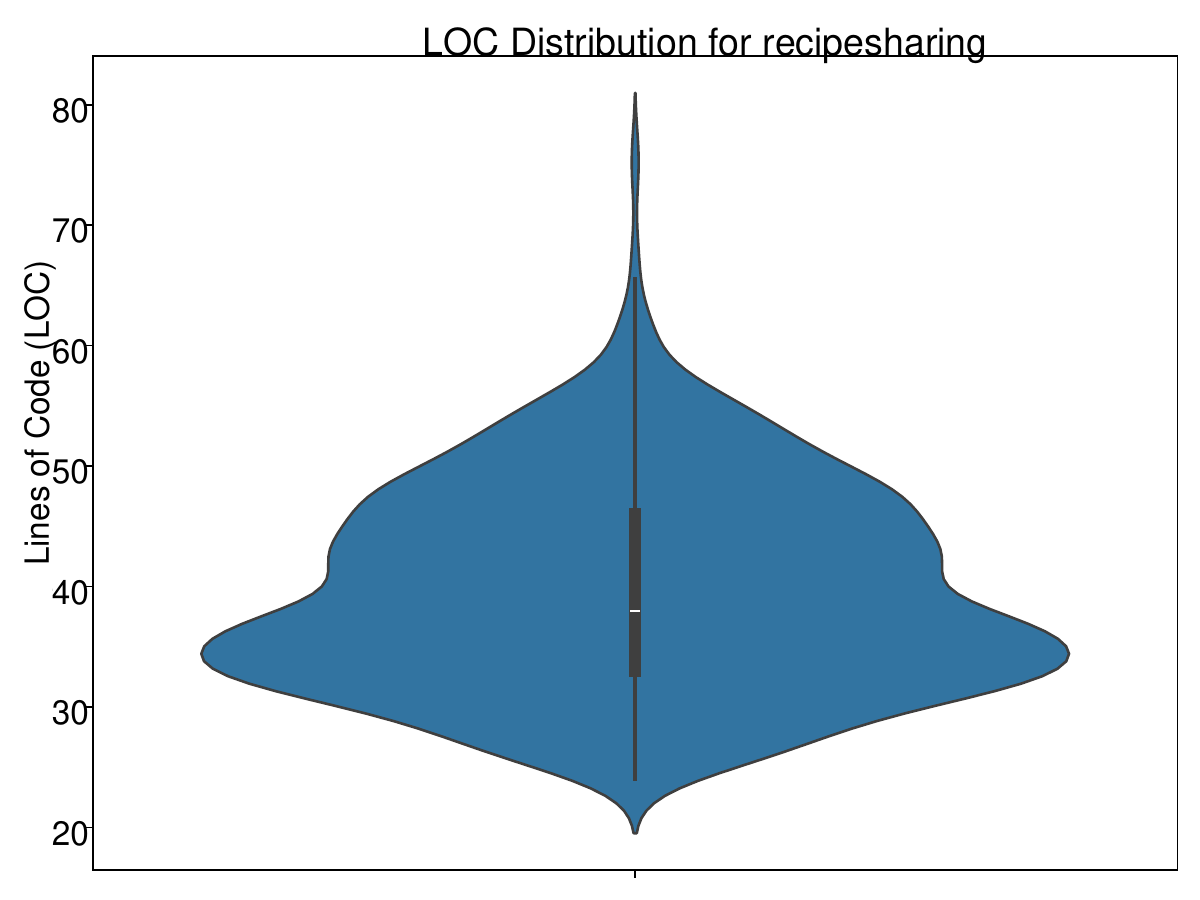}
        \caption{Recipe sharing}
    \end{subfigure}
    \begin{subfigure}{0.3\textwidth}
        \centering
        \includegraphics[width=\textwidth]{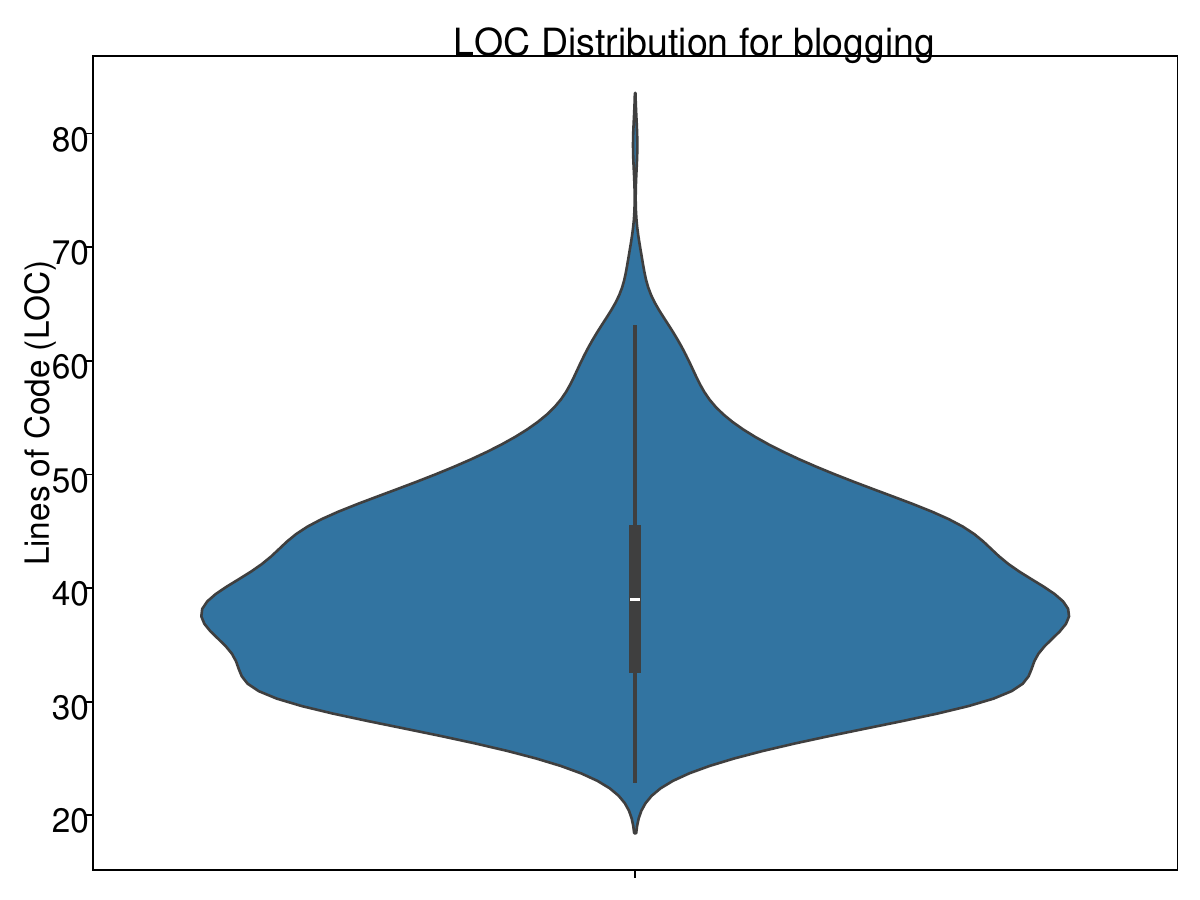}
        \caption{Blogging}
    \end{subfigure}
    
    \begin{subfigure}{0.3\textwidth}
        \centering
        \includegraphics[width=\textwidth]{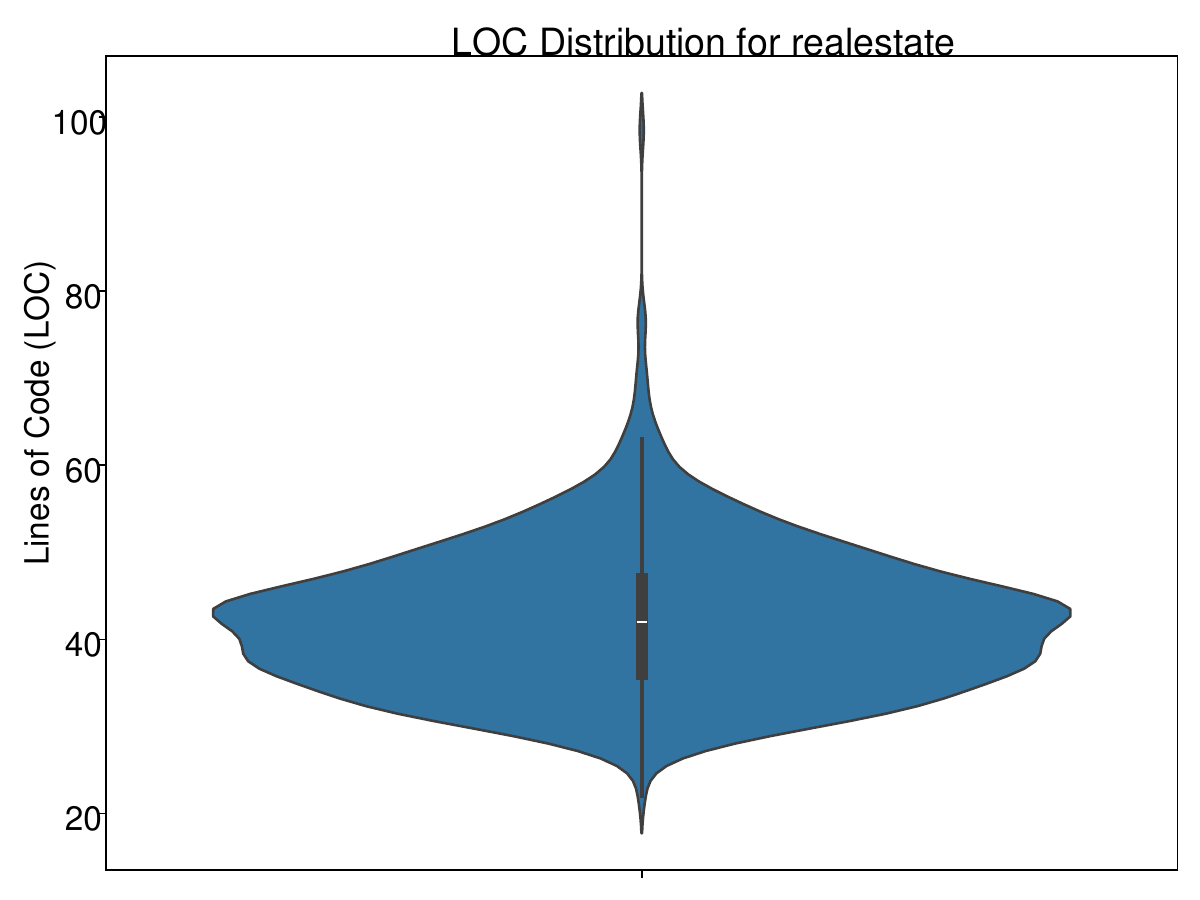}
        \caption{Real estate}
    \end{subfigure}
    \begin{subfigure}{0.3\textwidth}
        \centering
        \includegraphics[width=\textwidth]{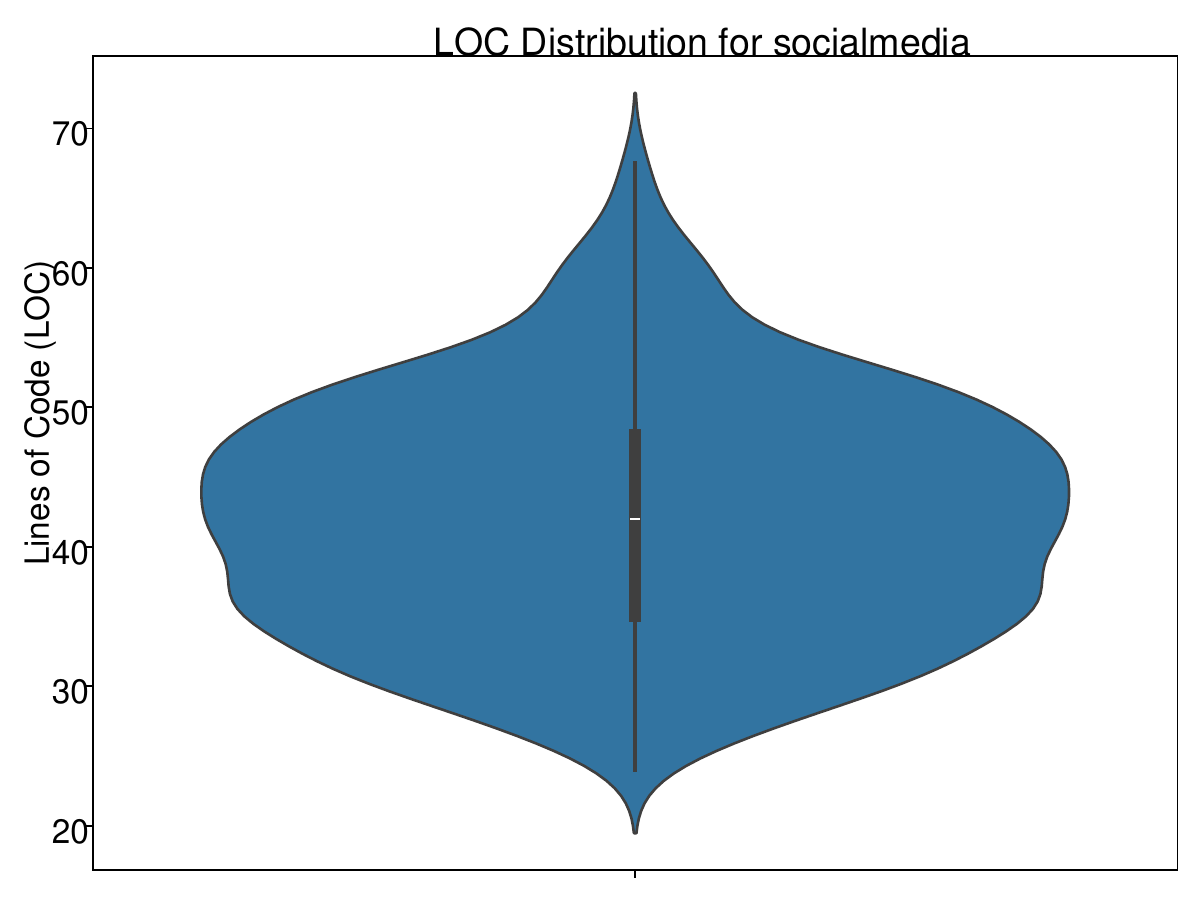}
        \caption{Social media}
    \end{subfigure}
    \begin{subfigure}{0.3\textwidth}
        \centering
        \includegraphics[width=\textwidth]{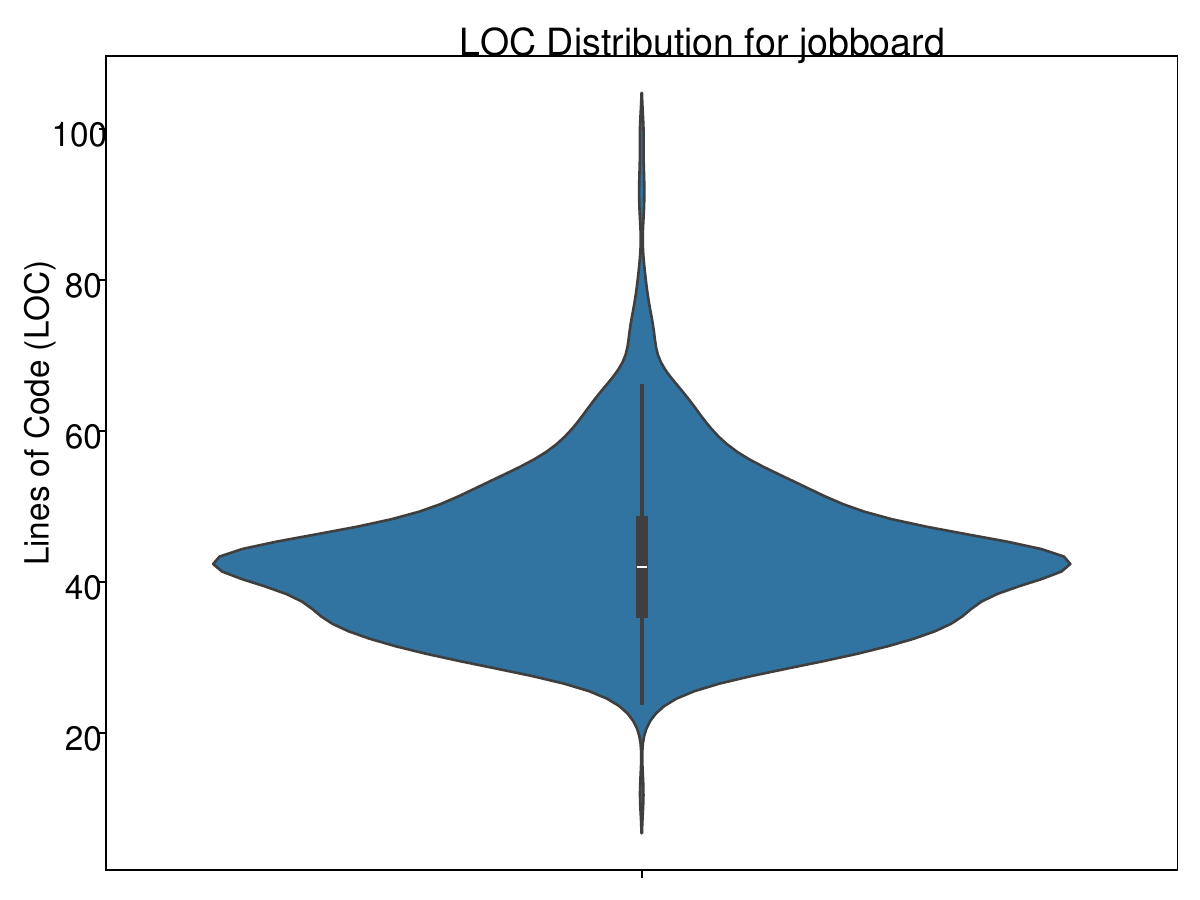}
        \caption{Job board}
    \end{subfigure}
 
    \begin{subfigure}{0.3\textwidth}
        \centering
        \includegraphics[width=\textwidth]{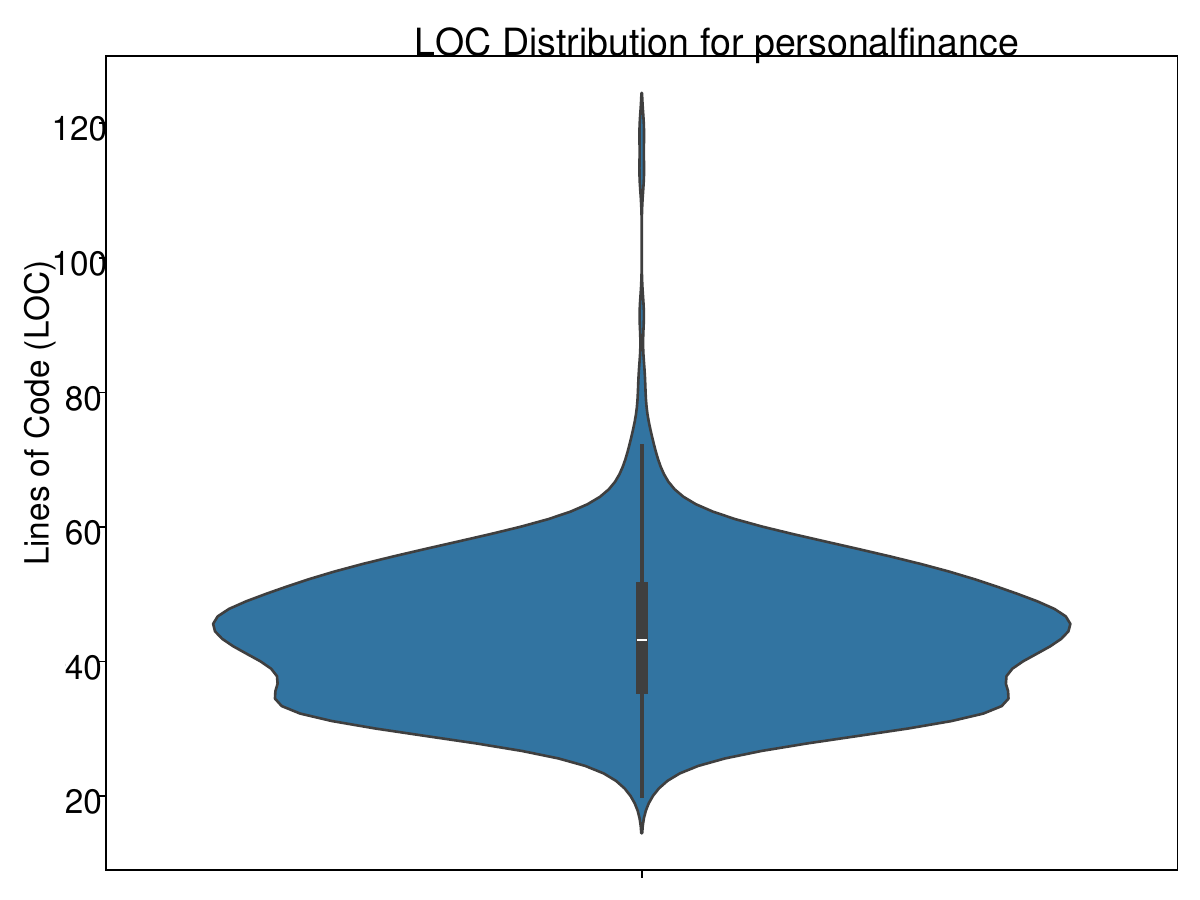}
        \caption{Personal finance}
    \end{subfigure}
    \begin{subfigure}{0.3\textwidth}
        \centering
        \includegraphics[width=\textwidth]{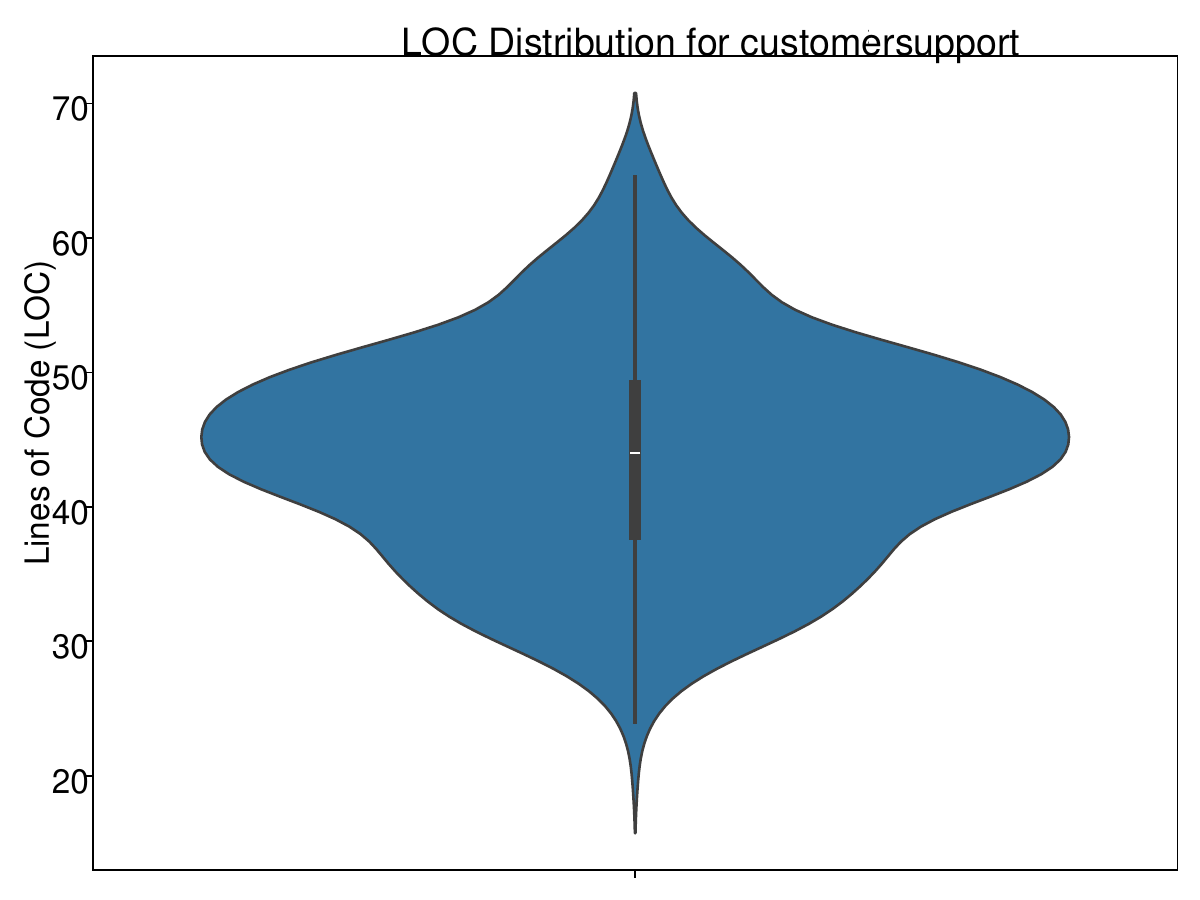}
        \caption{Customer support}
    \end{subfigure}
        \begin{subfigure}{0.3\textwidth}
        \centering
        \includegraphics[width=\textwidth]{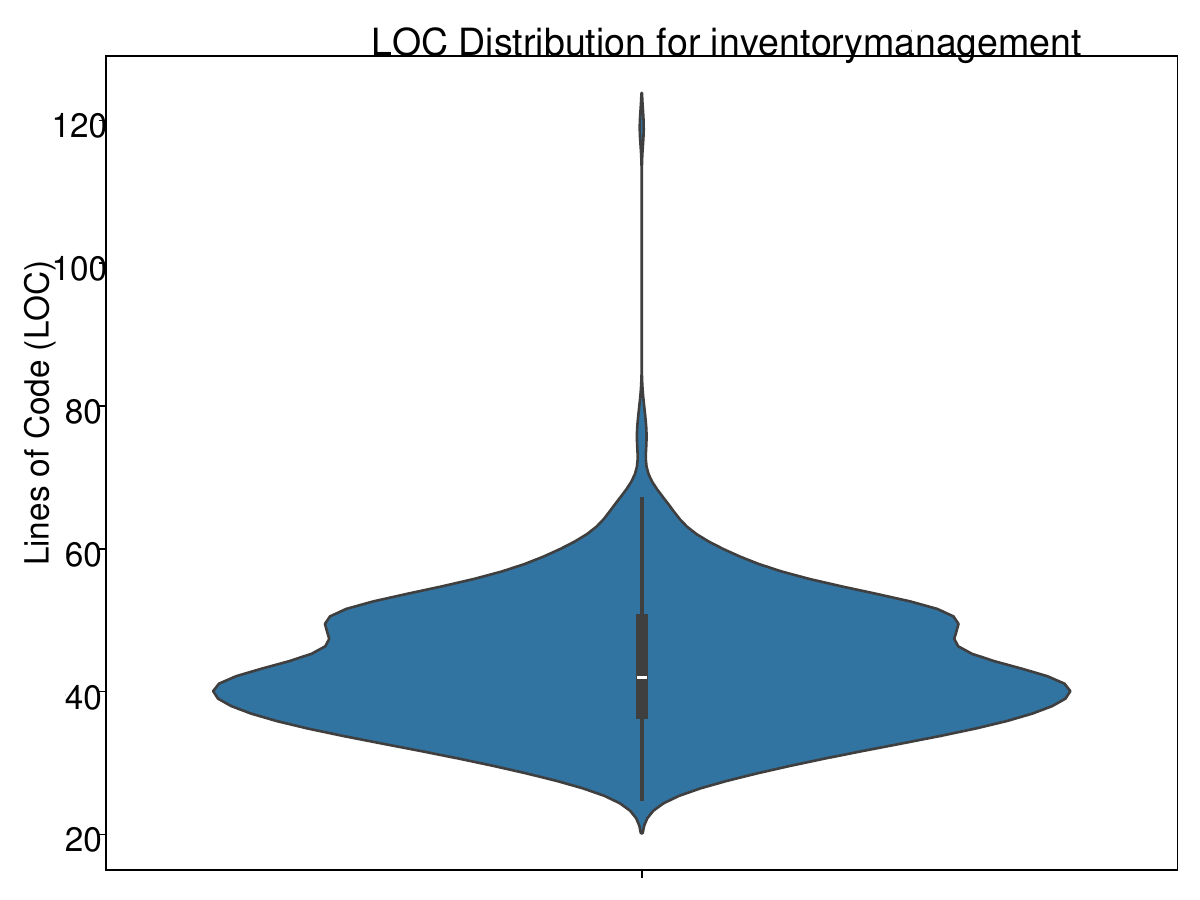}
        \caption{Inventory management}
    \end{subfigure}

    \begin{subfigure}{0.3\textwidth}
        \centering
        \includegraphics[width=\textwidth]{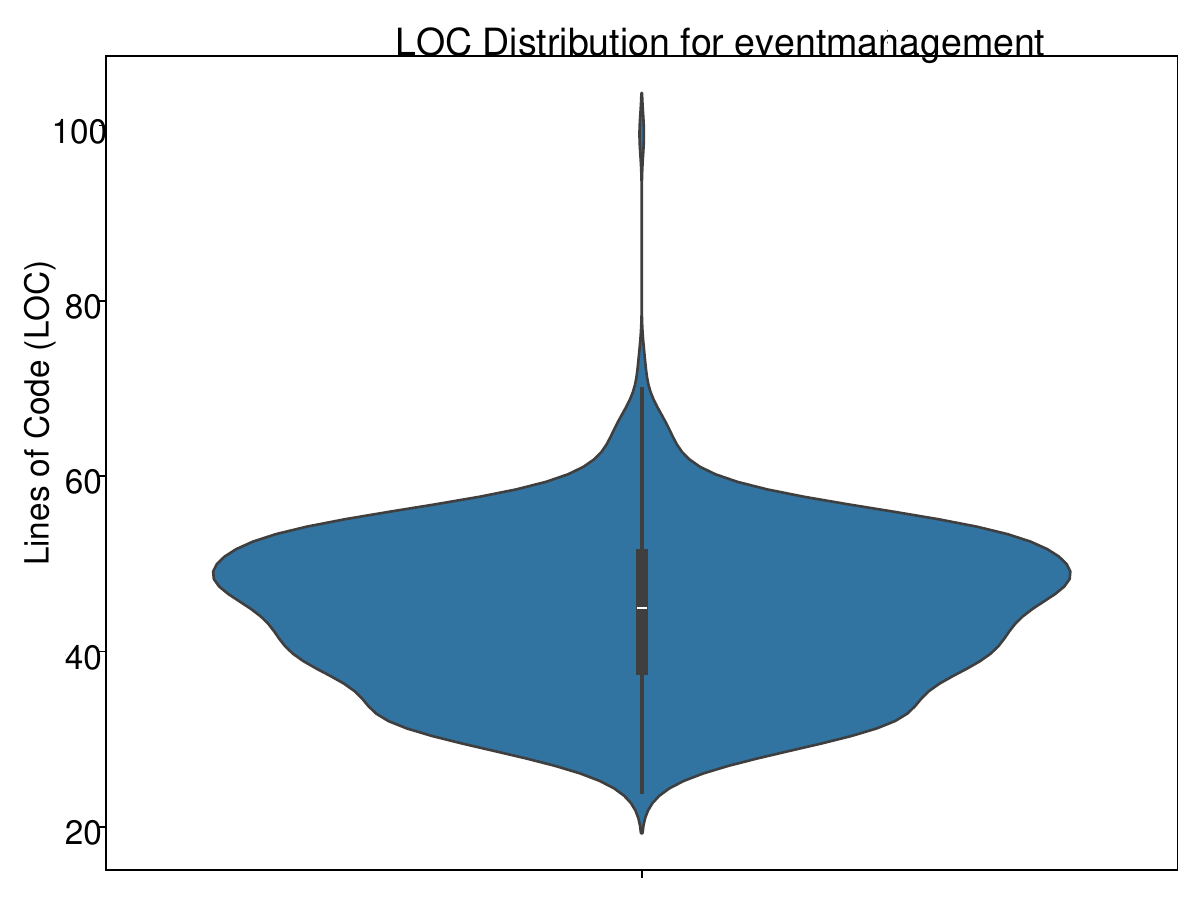}
        \caption{Event management}
    \end{subfigure}
    \begin{subfigure}{0.3\textwidth}
        \centering
        \includegraphics[width=\textwidth]{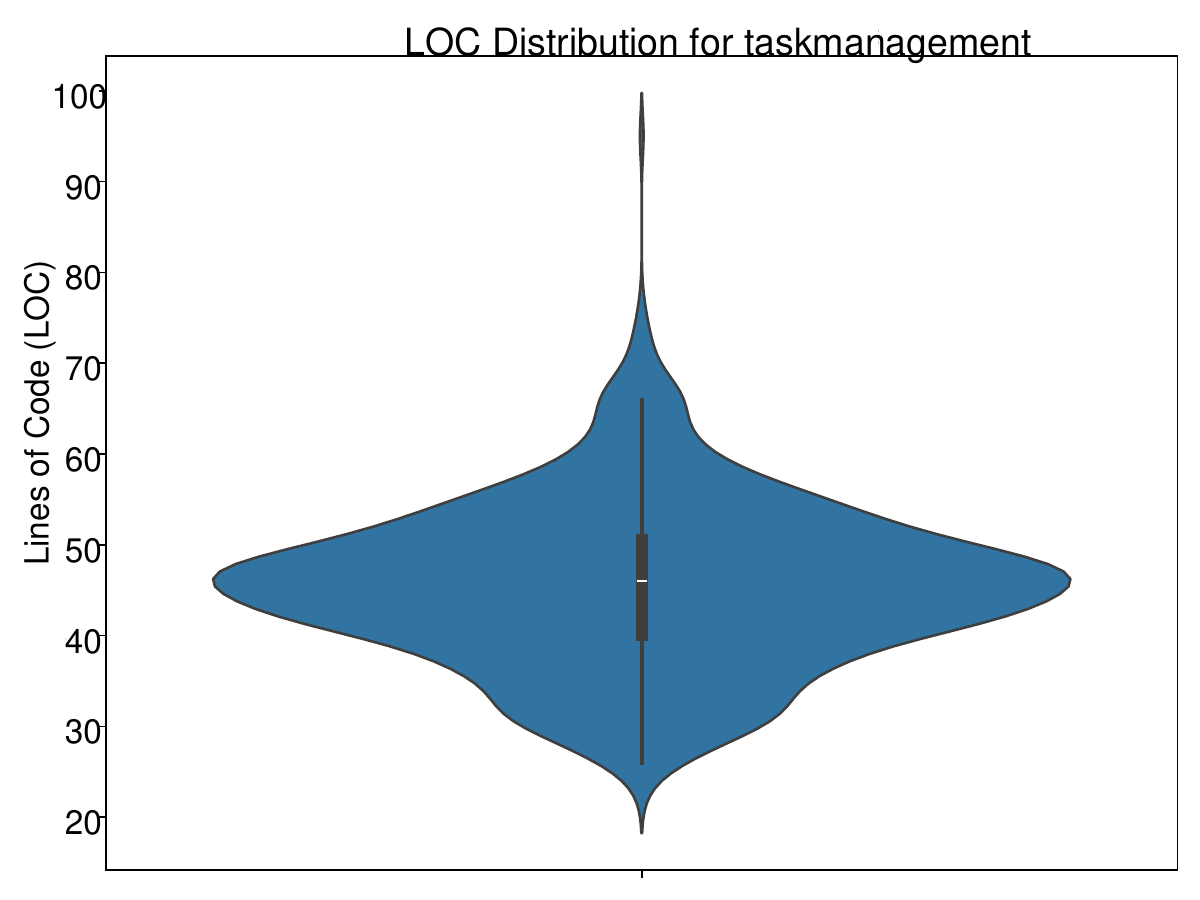}
        \caption{Task management}
    \end{subfigure}
    \caption{LOC distribution by applications: unimodal}
    \label{fig:loc_distribution_apps_unimodal}
\end{figure}

\begin{figure}[h]
    \centering
    \begin{subfigure}{0.3\textwidth}
        \centering
        \includegraphics[width=\textwidth]{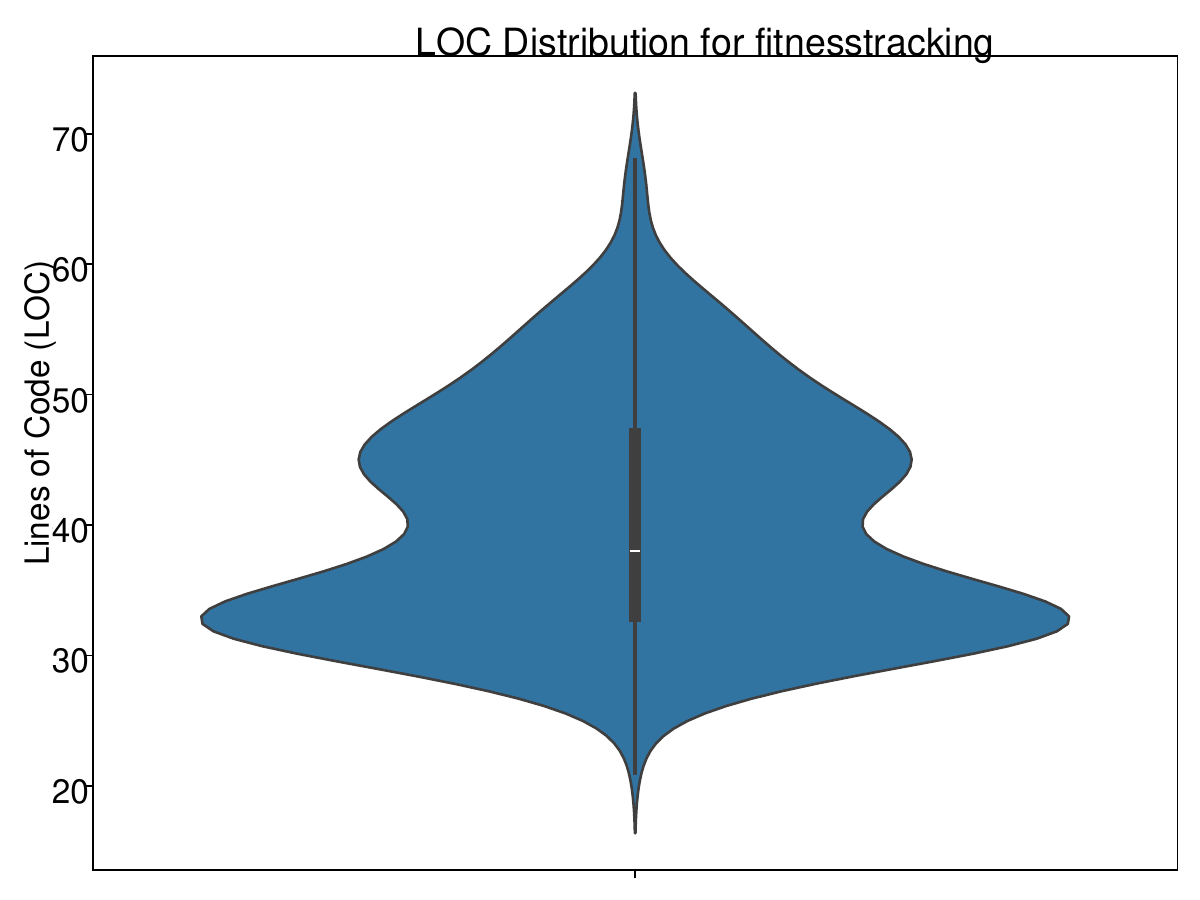}
        \caption{Fitness tracking}
    \end{subfigure}
    \begin{subfigure}{0.3\textwidth}
        \centering
        \includegraphics[width=\textwidth]{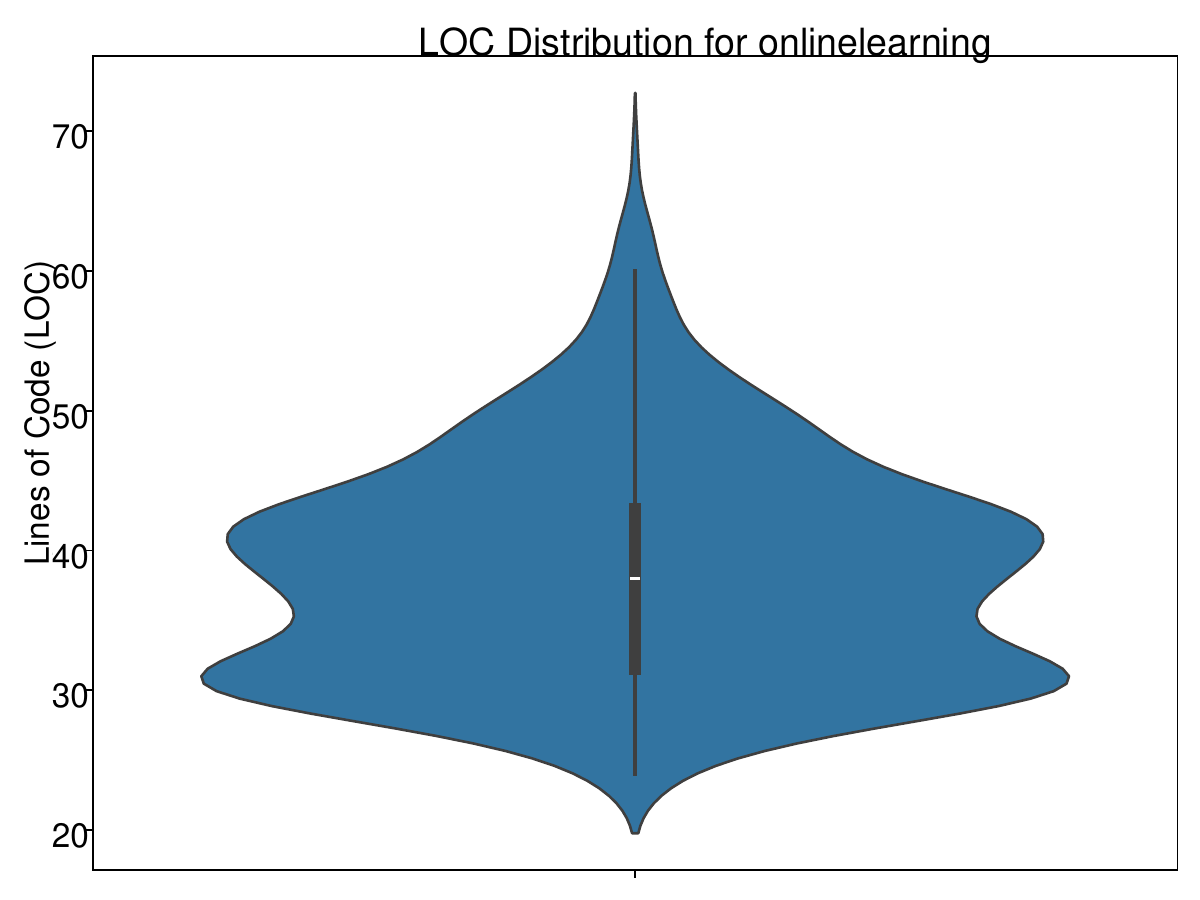}
        \caption{Online learning}
    \end{subfigure}
    \begin{subfigure}{0.3\textwidth}
        \centering
        \includegraphics[width=\textwidth]{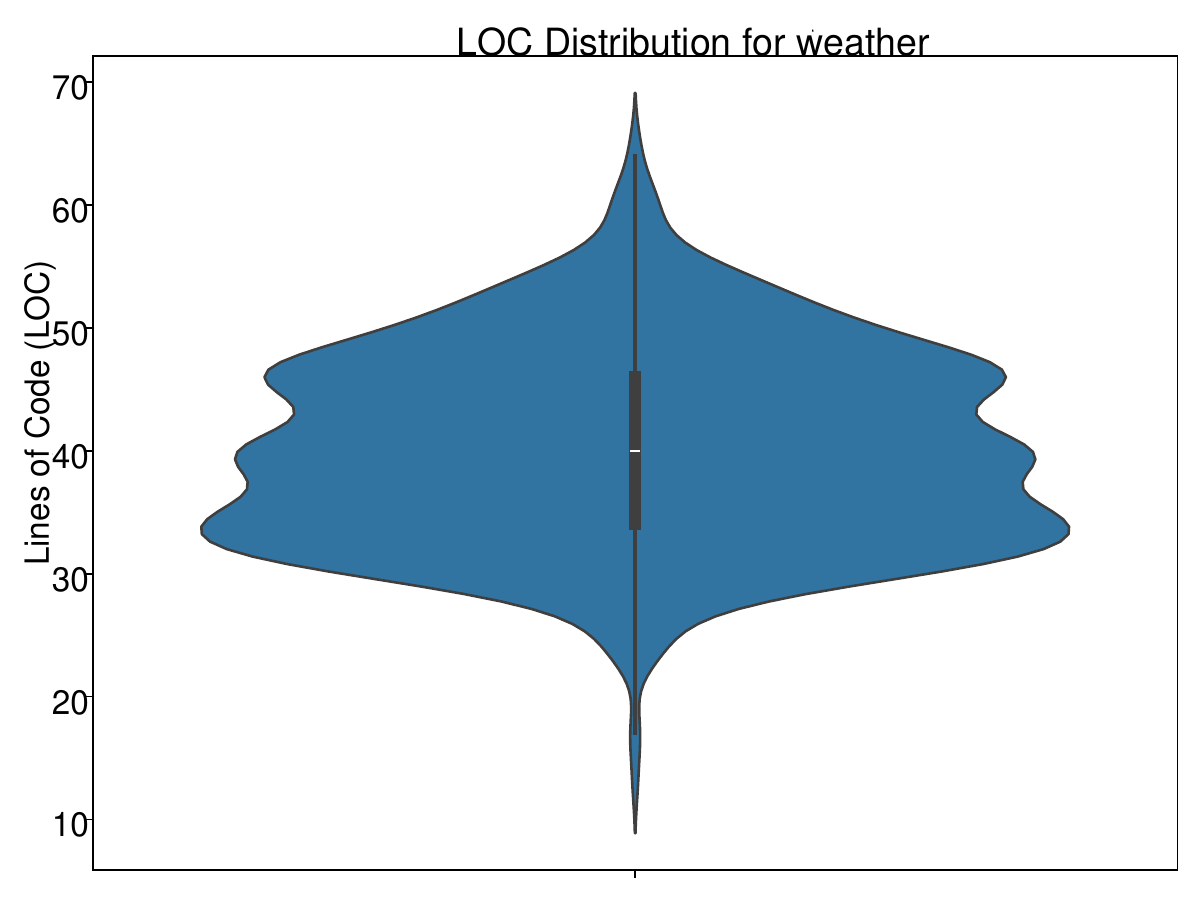}
        \caption{Weather}
    \end{subfigure}
    
    \begin{subfigure}{0.3\textwidth}
        \centering
        \includegraphics[width=\textwidth]{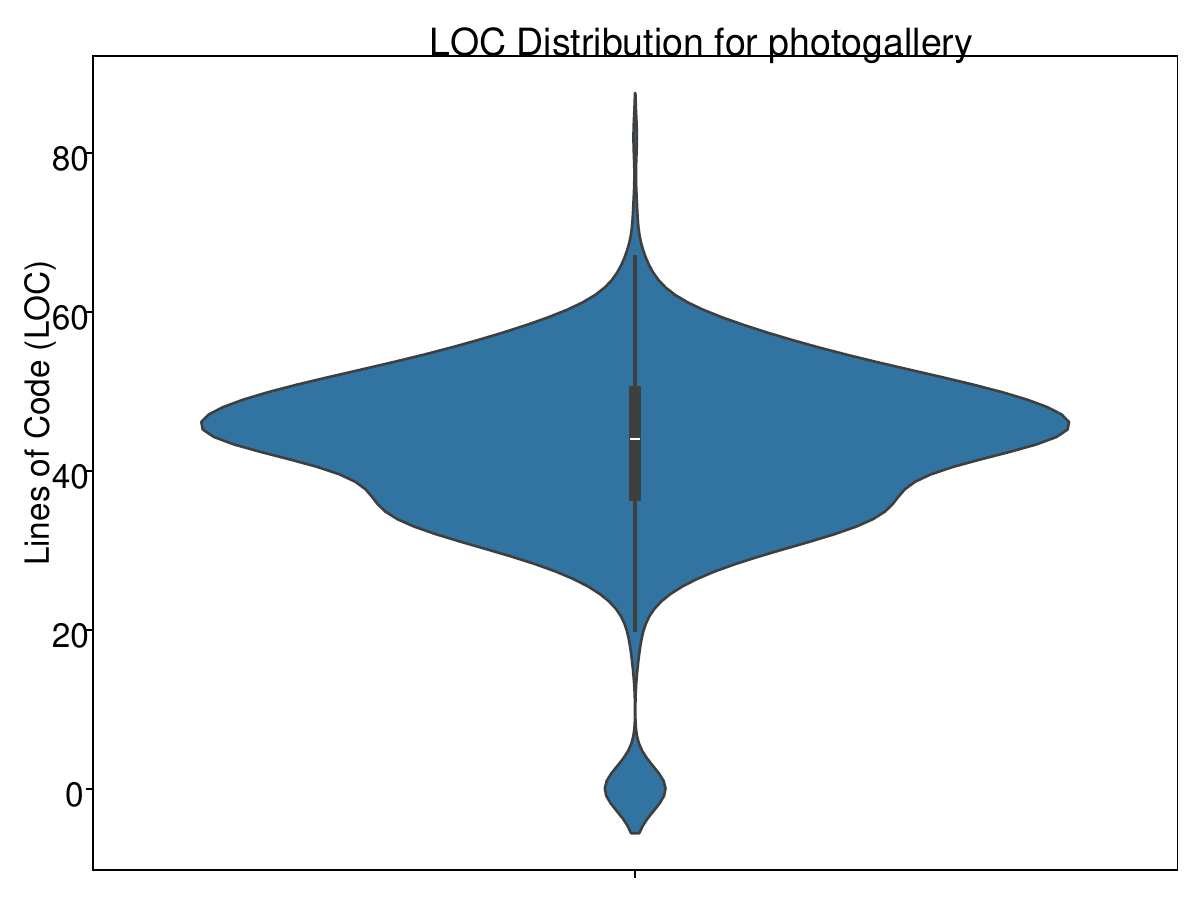}
        \caption{Photo gallery}
    \end{subfigure}
    \begin{subfigure}{0.3\textwidth}
        \centering
        \includegraphics[width=\textwidth]{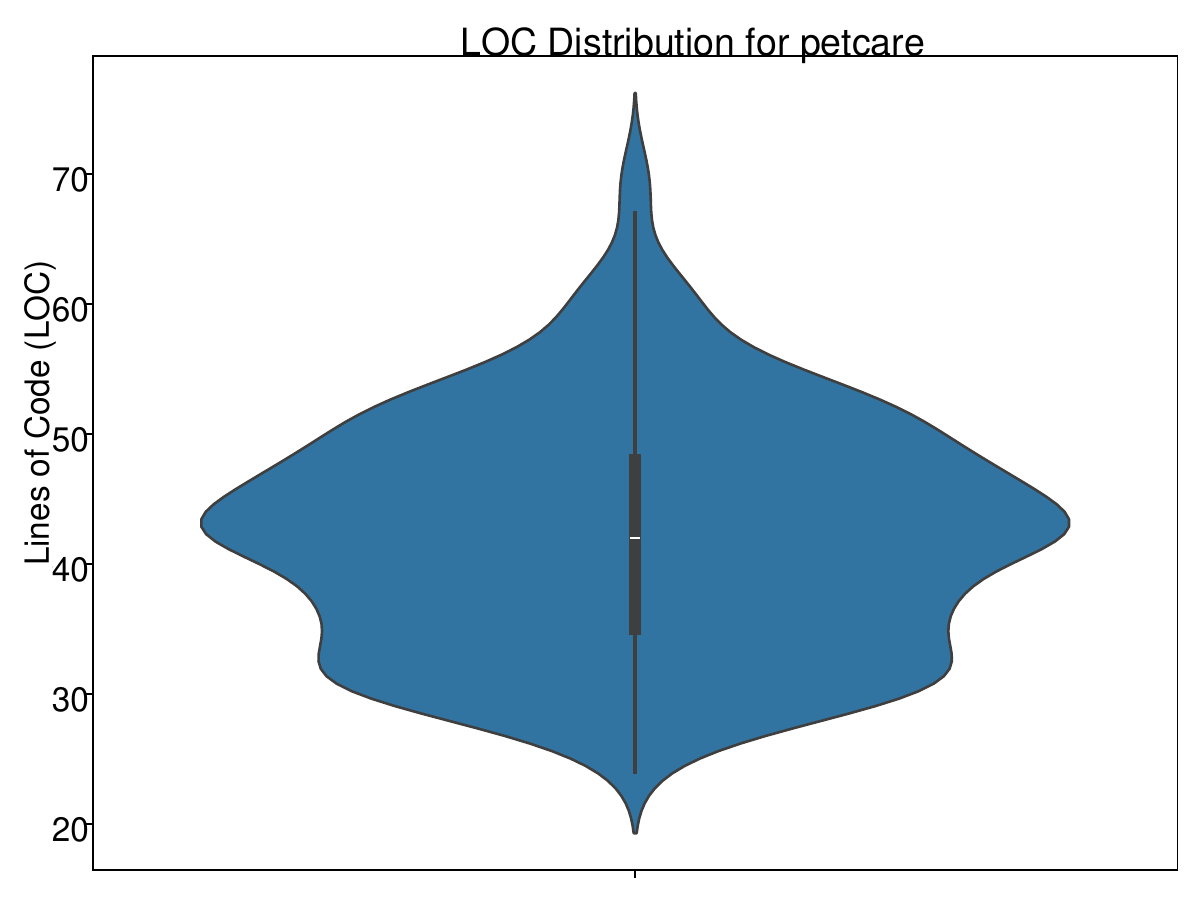}
        \caption{Pet care}
    \end{subfigure}
    \begin{subfigure}{0.3\textwidth}
        \centering
        \includegraphics[width=\textwidth]{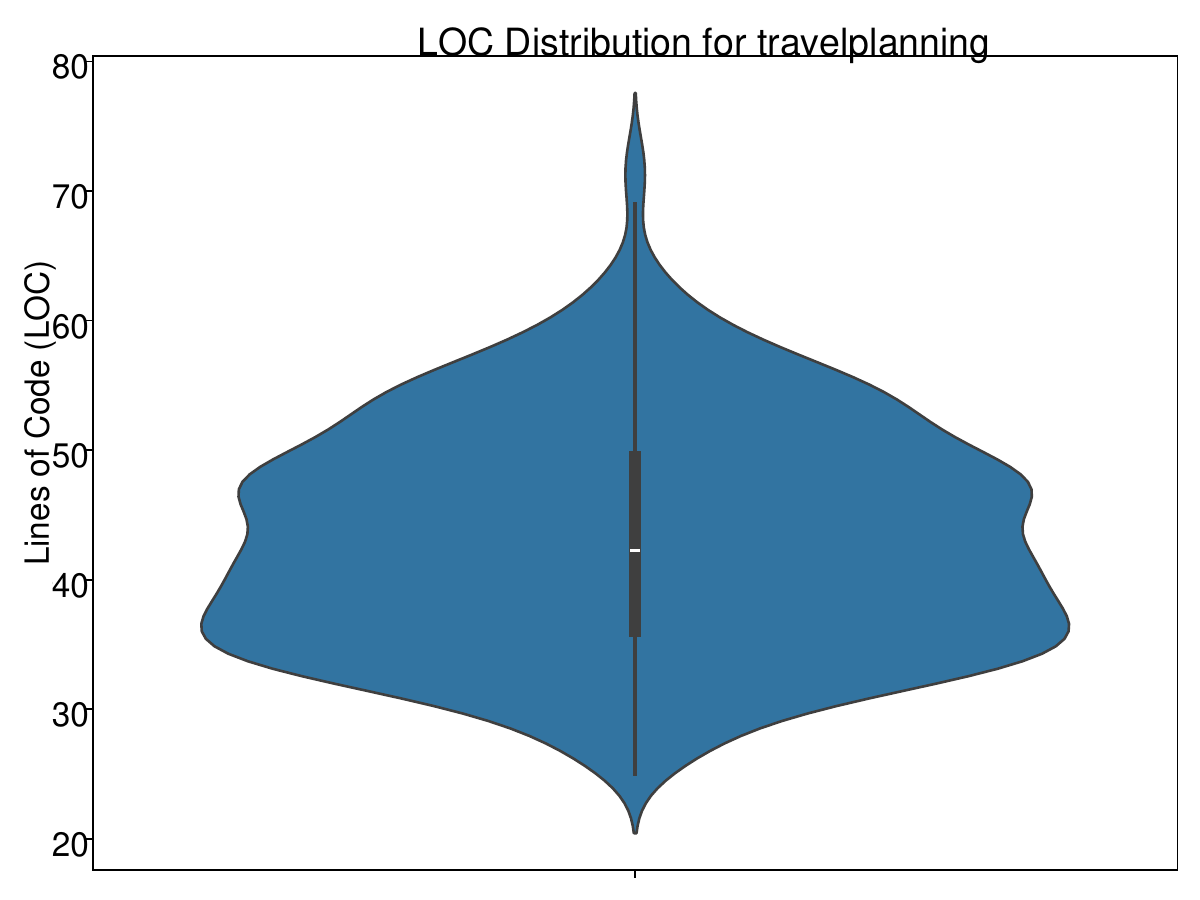}
        \caption{Travel planning}
    \end{subfigure}
    \caption{LOC distribution by applications: multimodal}
    \label{fig:loc_distribution_apps_multimodal}
\end{figure}

Fig.~\ref{fig:loc_distribution_apps_unimodal} collects violin charts of 14 applications following unimodal distribution, where the model outputs are centered around a common length, with less variation between extremes. The remaining 6 applications are in Fig.~\ref{fig:loc_distribution_apps_multimodal}, following multimodal distribution. In both cases, the median LOC is always positioned centrally in each distribution, which suggests that the code generation is stable across applications. Applications in Fig.~\ref{fig:loc_distribution_apps_multimodal} exhibit more complex patterns, but the distributions remain balanced with the median value positioned at the center of the distribution.

\subsection{LOC Distribution by Applications: Success vs Failure}
We conduct the same study described in Sec.~\ref{sec:loc_successfail}, except we shard the LOC distribution across applications instead of models. The results are collected in Fig.~\ref{fig:loc_successfail_distribution_apps}.
\begin{figure}[h!]
    \centering
    \begin{subfigure}{0.4\textwidth}
        \centering
        \includegraphics[width=\linewidth]{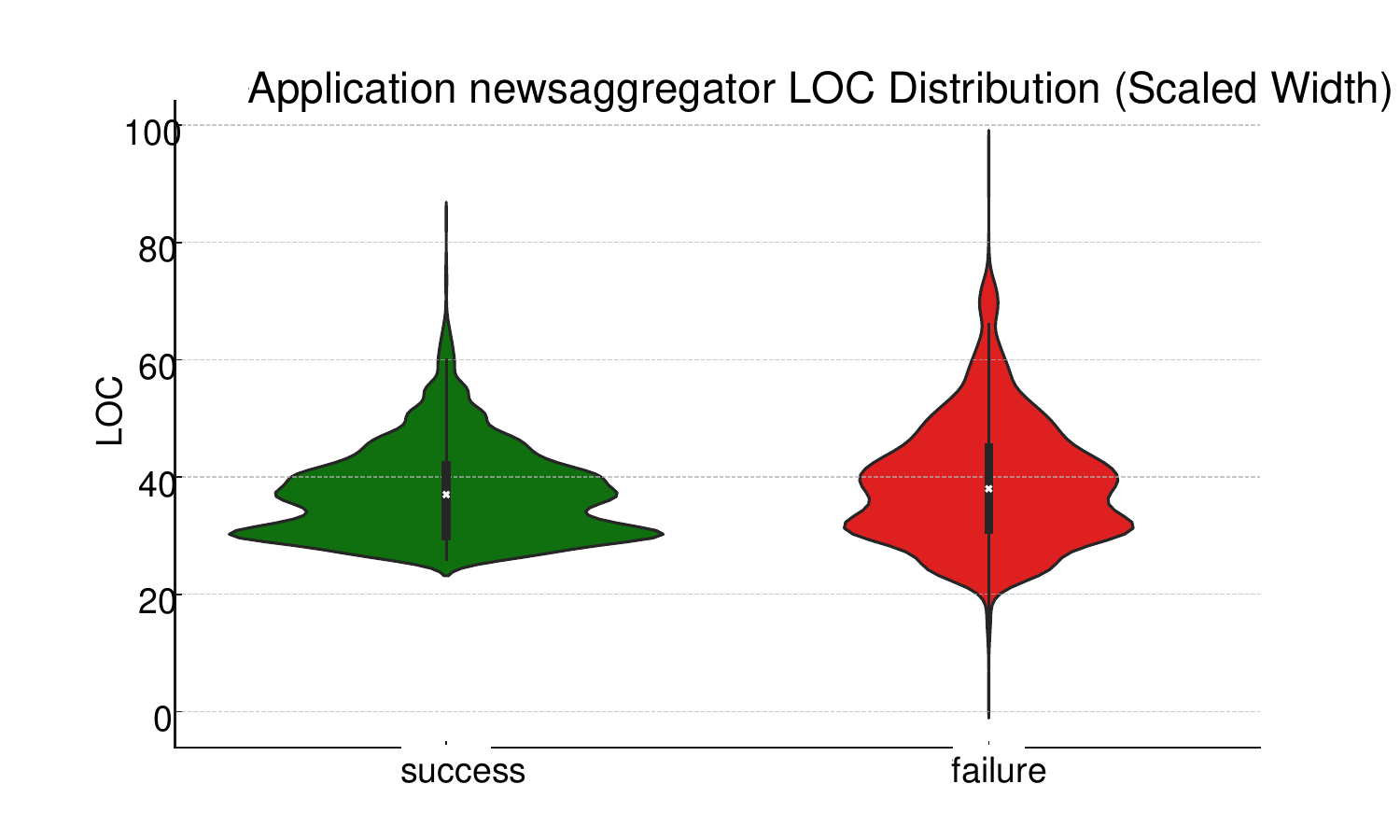}
        \caption{News aggregator (mean LOC = 37)}
    \end{subfigure}
    \begin{subfigure}{0.4\textwidth}
        \centering
        \includegraphics[width=\linewidth]{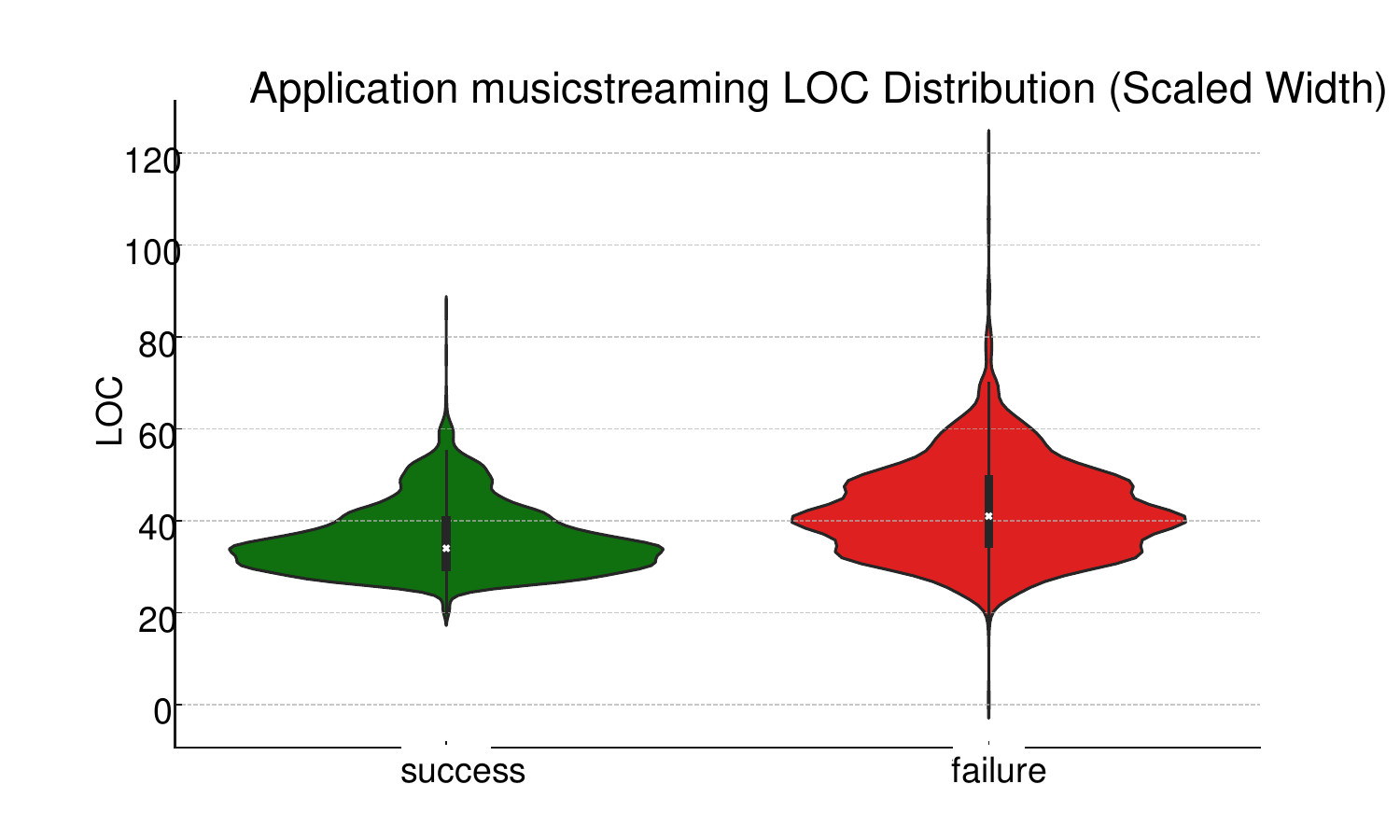}
        \caption{Music streaming (mean LOC = 37)}
    \end{subfigure}

    \begin{subfigure}{0.4\textwidth}
        \centering
        \includegraphics[width=\linewidth]{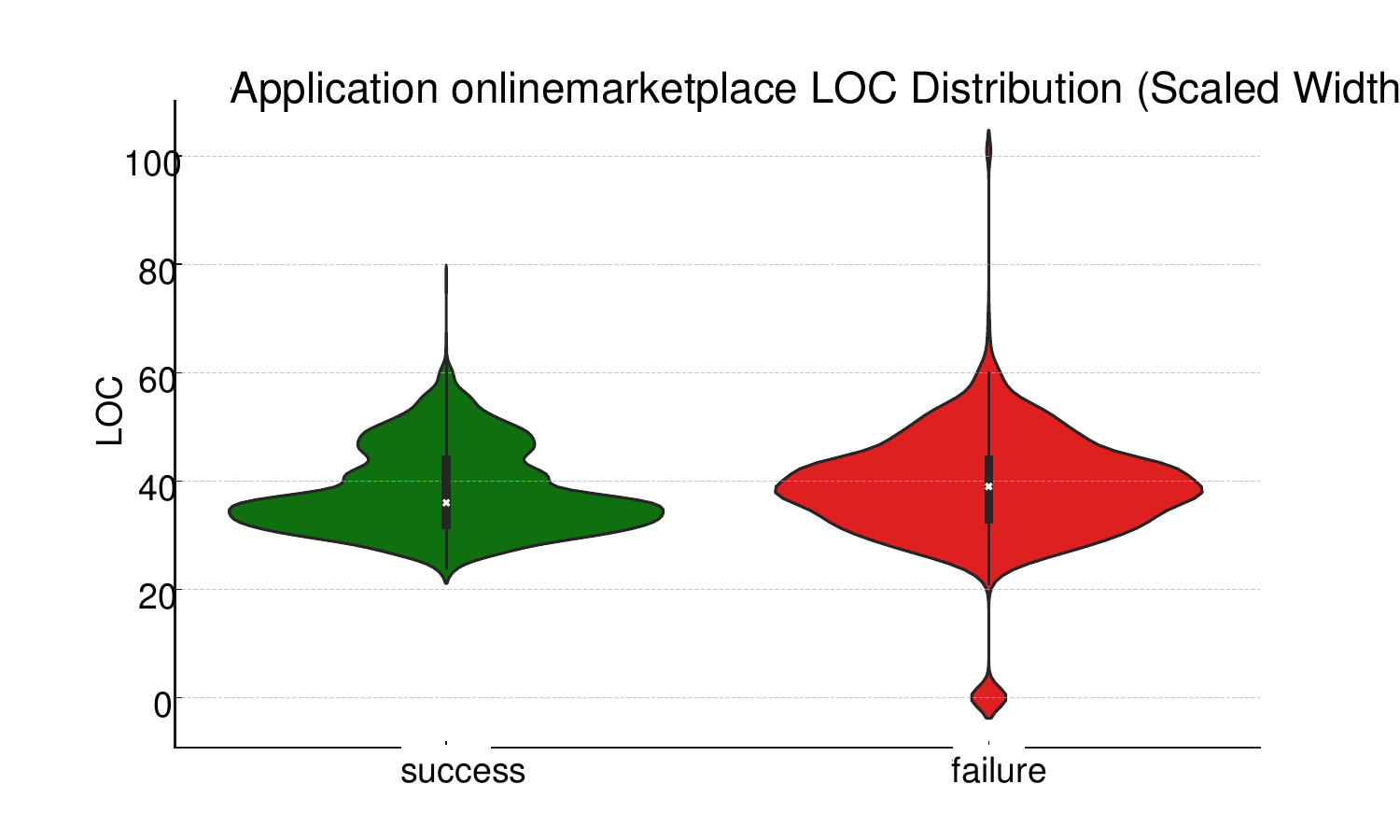}
        \caption{Online marketplace (mean LOC = 37)}
    \end{subfigure}
    \begin{subfigure}{0.4\textwidth}
        \centering
        \includegraphics[width=\linewidth]{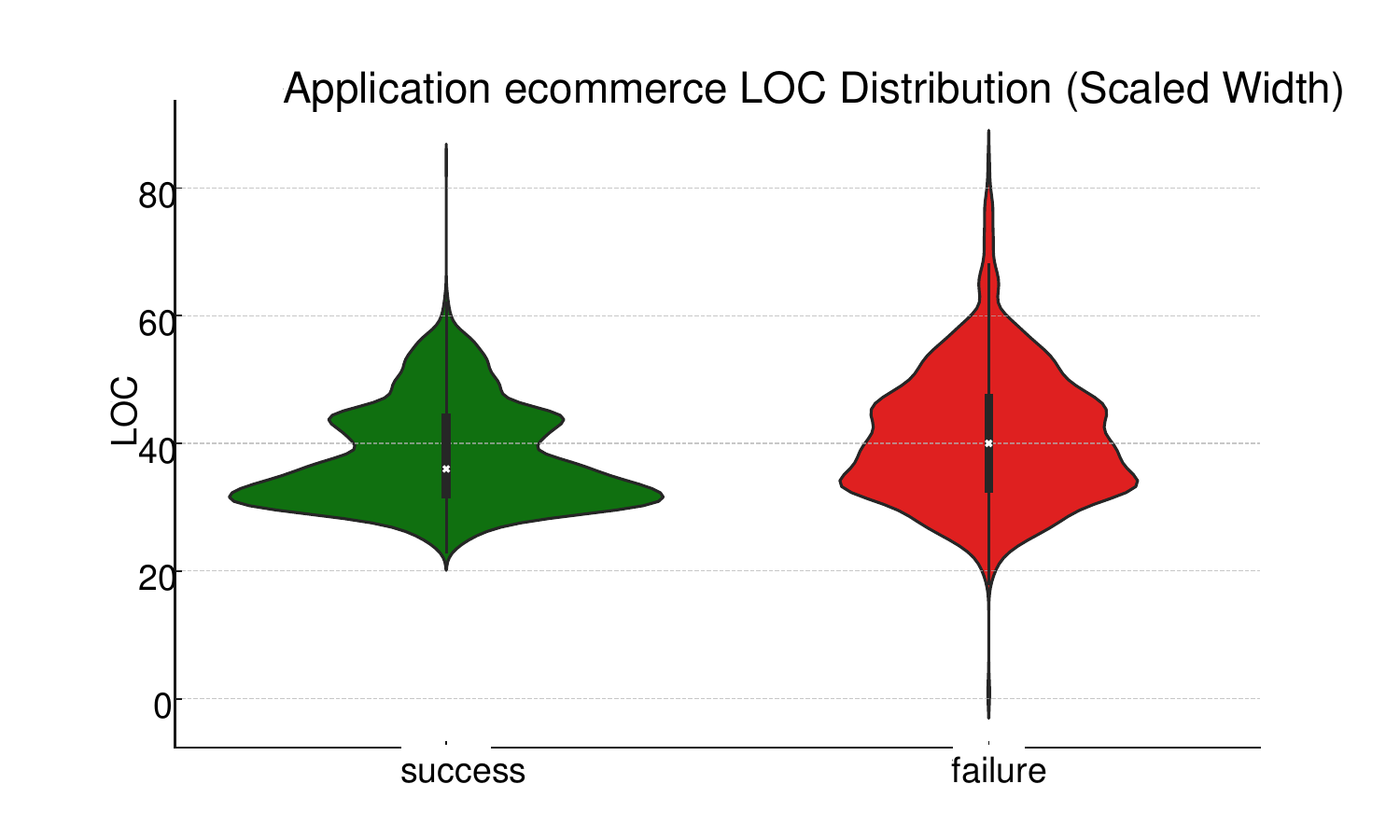}
        \caption{E-commerce (mean LOC = 37)}
    \end{subfigure}

    \begin{subfigure}{0.4\textwidth}
        \centering
        \includegraphics[width=\linewidth]{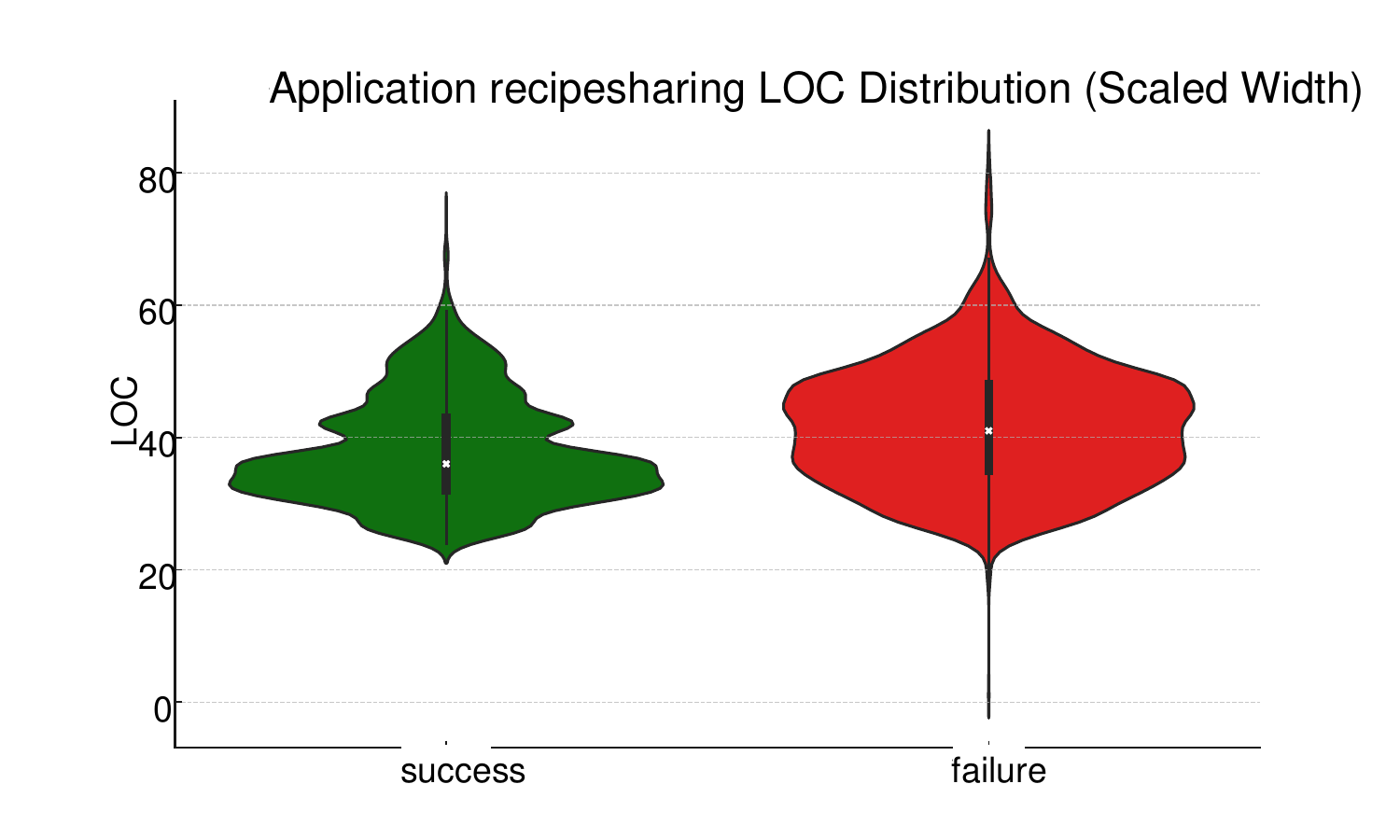}
        \caption{Recipe sharing (mean LOC = 38)}
    \end{subfigure}
    \begin{subfigure}{0.4\textwidth}
        \centering
        \includegraphics[width=\linewidth]{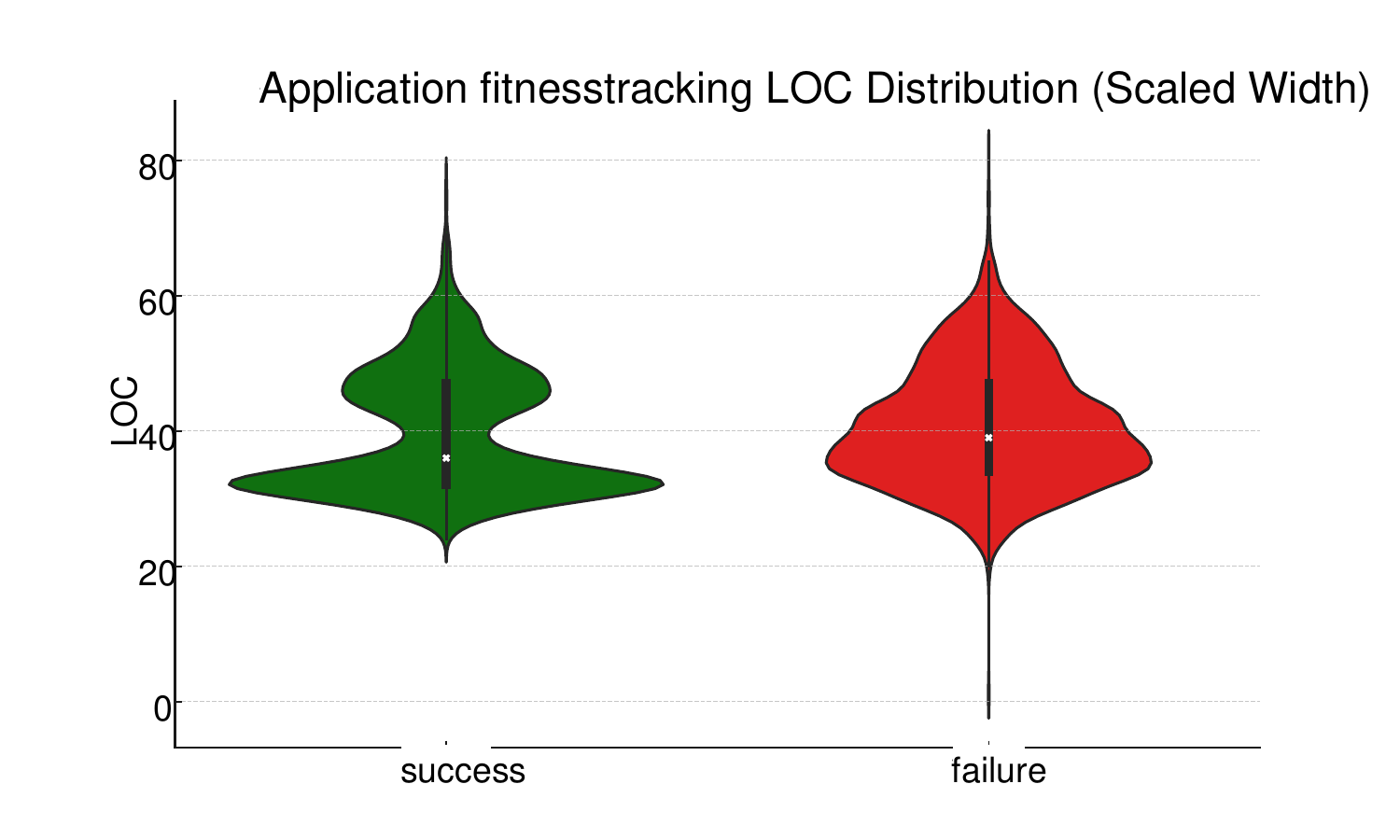}
        \caption{Fitness tracking (mean LOC = 38)}
    \end{subfigure}

    \begin{subfigure}{0.4\textwidth}
        \centering
        \includegraphics[width=\linewidth]{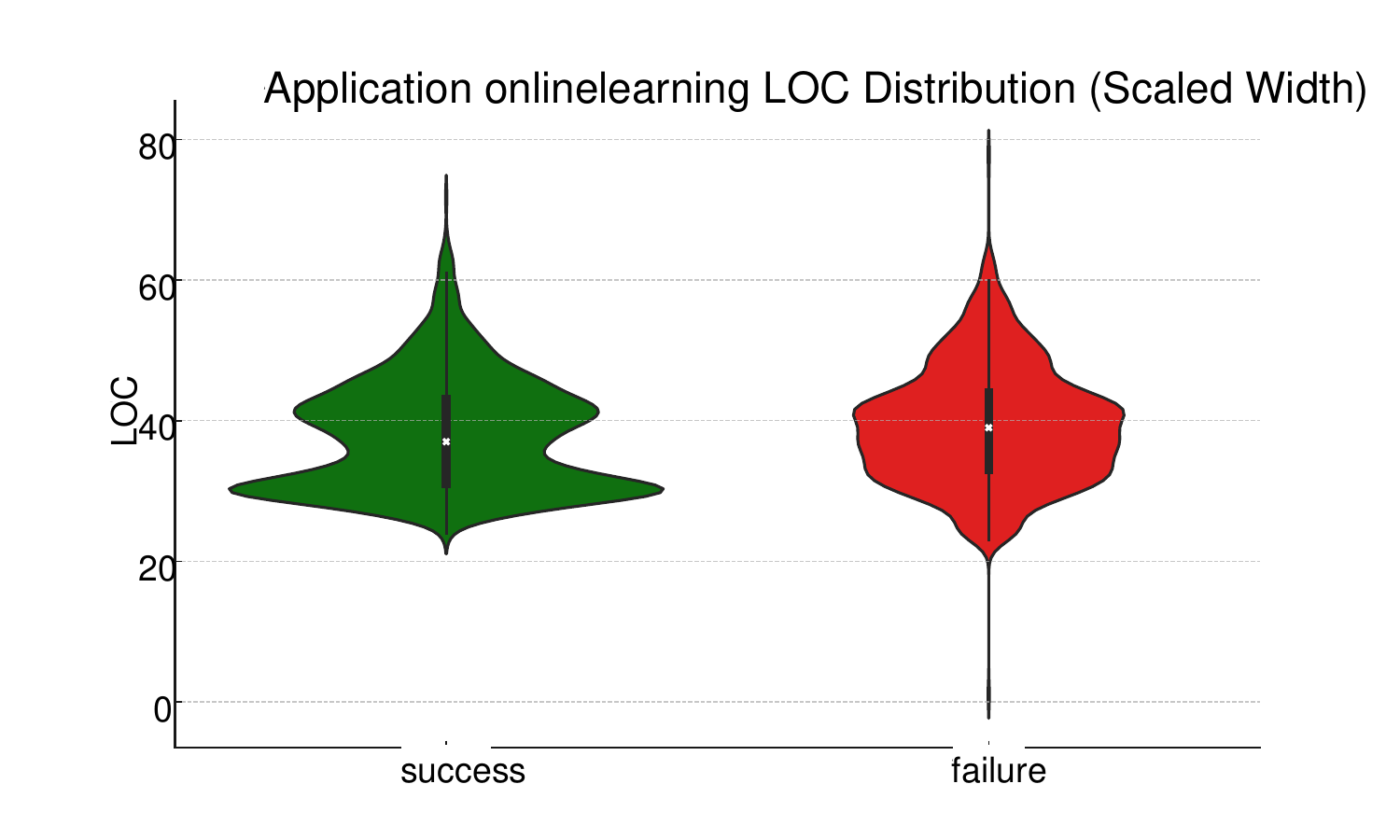}
        \caption{Online learning (mean LOC = 38)}
    \end{subfigure}
    \begin{subfigure}{0.4\textwidth}
        \centering
        \includegraphics[width=\linewidth]{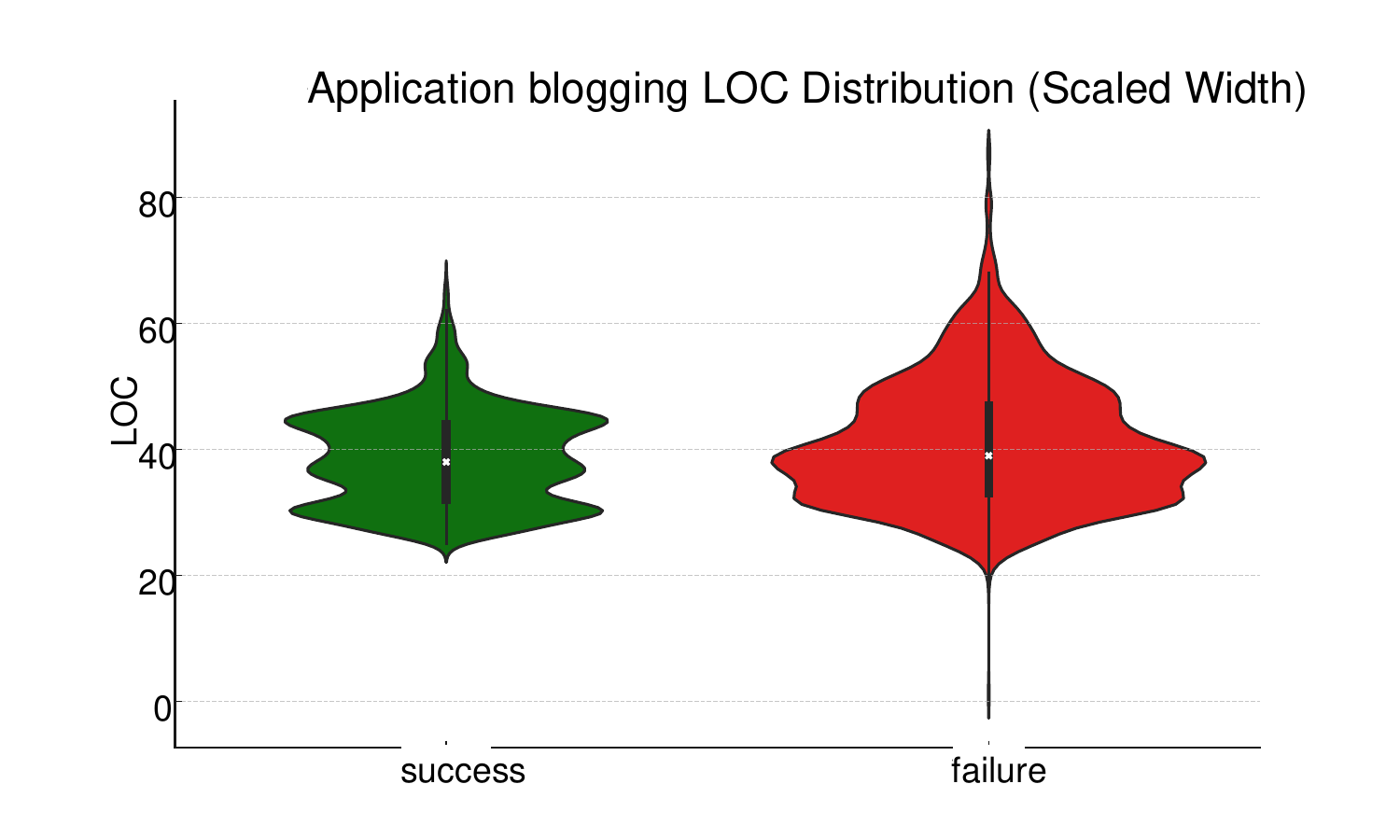}
        \caption{Blogging (mean LOC = 39)}
    \end{subfigure}

    \begin{subfigure}{0.4\textwidth}
        \centering
        \includegraphics[width=\linewidth]{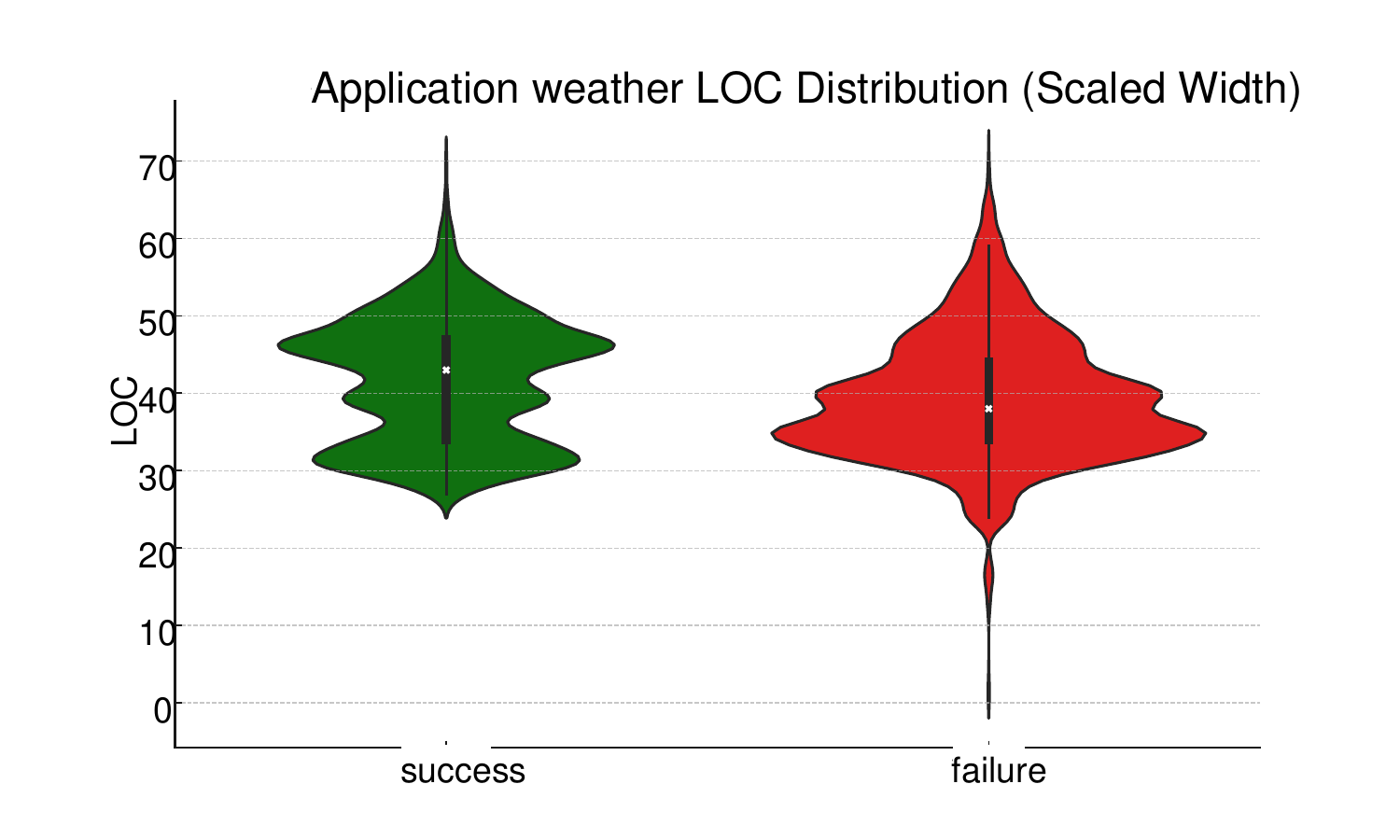}
        \caption{Weather (mean LOC = 40)}
    \end{subfigure}
    \begin{subfigure}{0.4\textwidth}
        \centering
        \includegraphics[width=\linewidth]{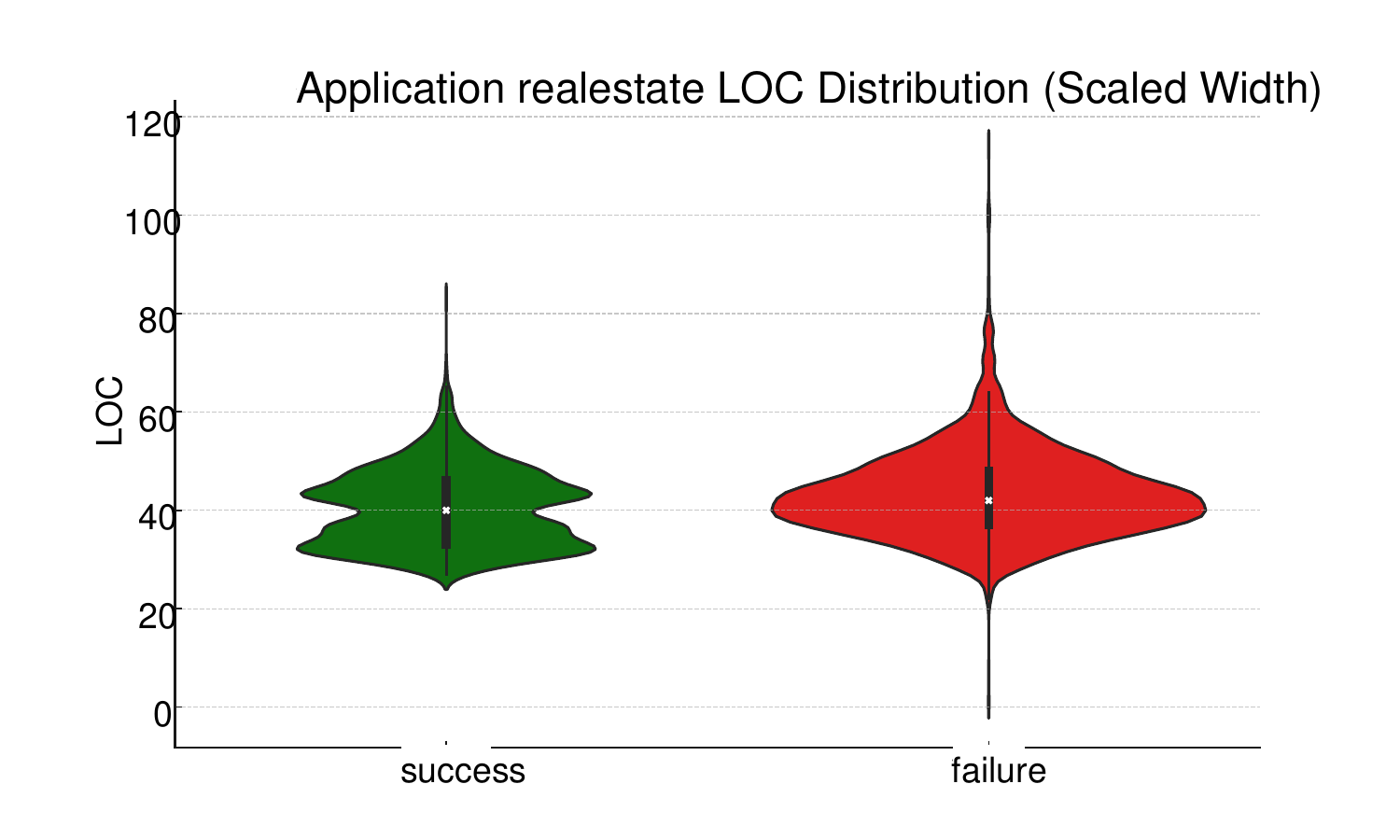}
        \caption{Real estate (mean LOC = 42)}
    \end{subfigure}
\end{figure}

\begin{figure}[h!]
    \ContinuedFloat
    \centering
    \begin{subfigure}{0.4\textwidth}
        \centering
        \includegraphics[width=\linewidth]{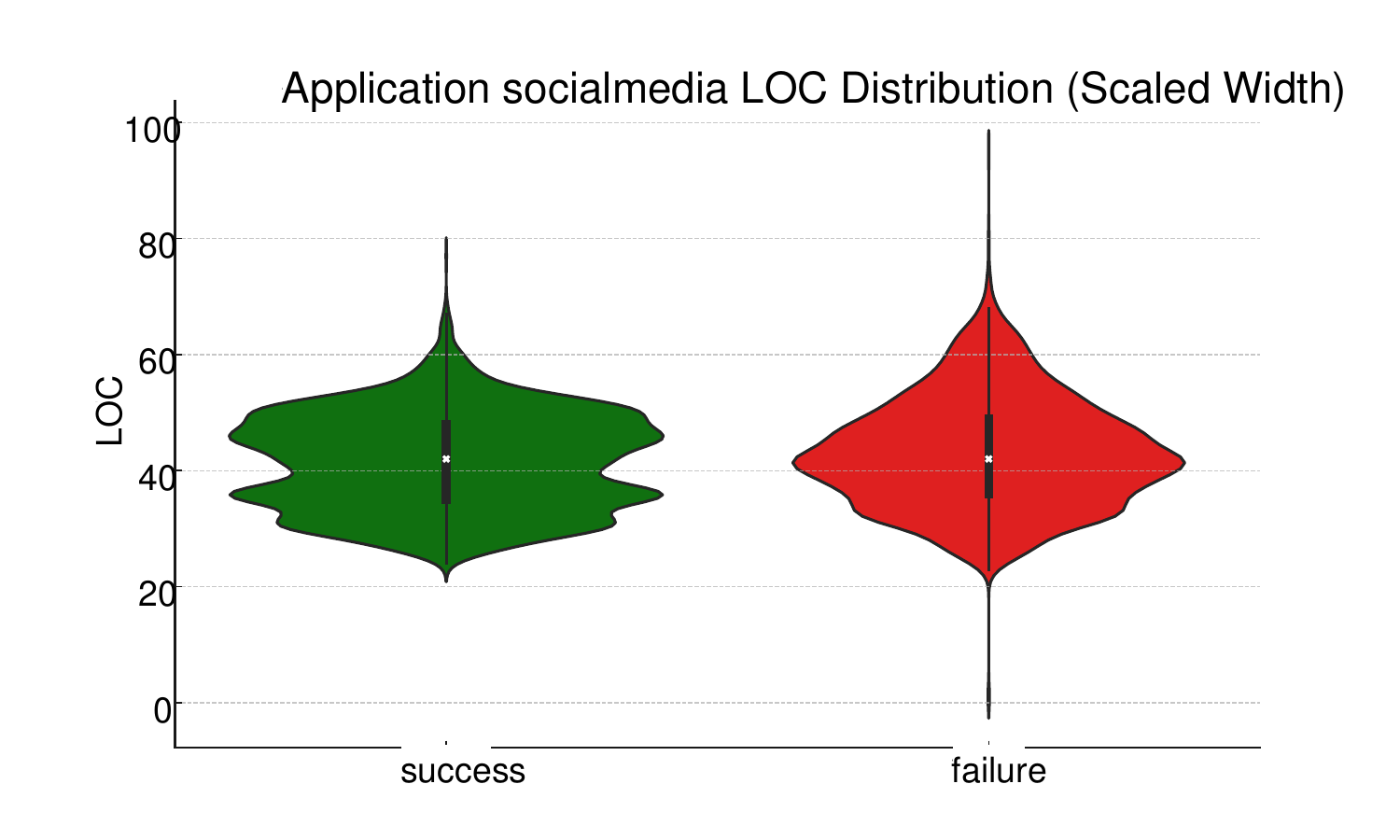}
        \caption{Social media (mean LOC = 42)}
    \end{subfigure}
    \begin{subfigure}{0.4\textwidth}
        \centering
        \includegraphics[width=\linewidth]{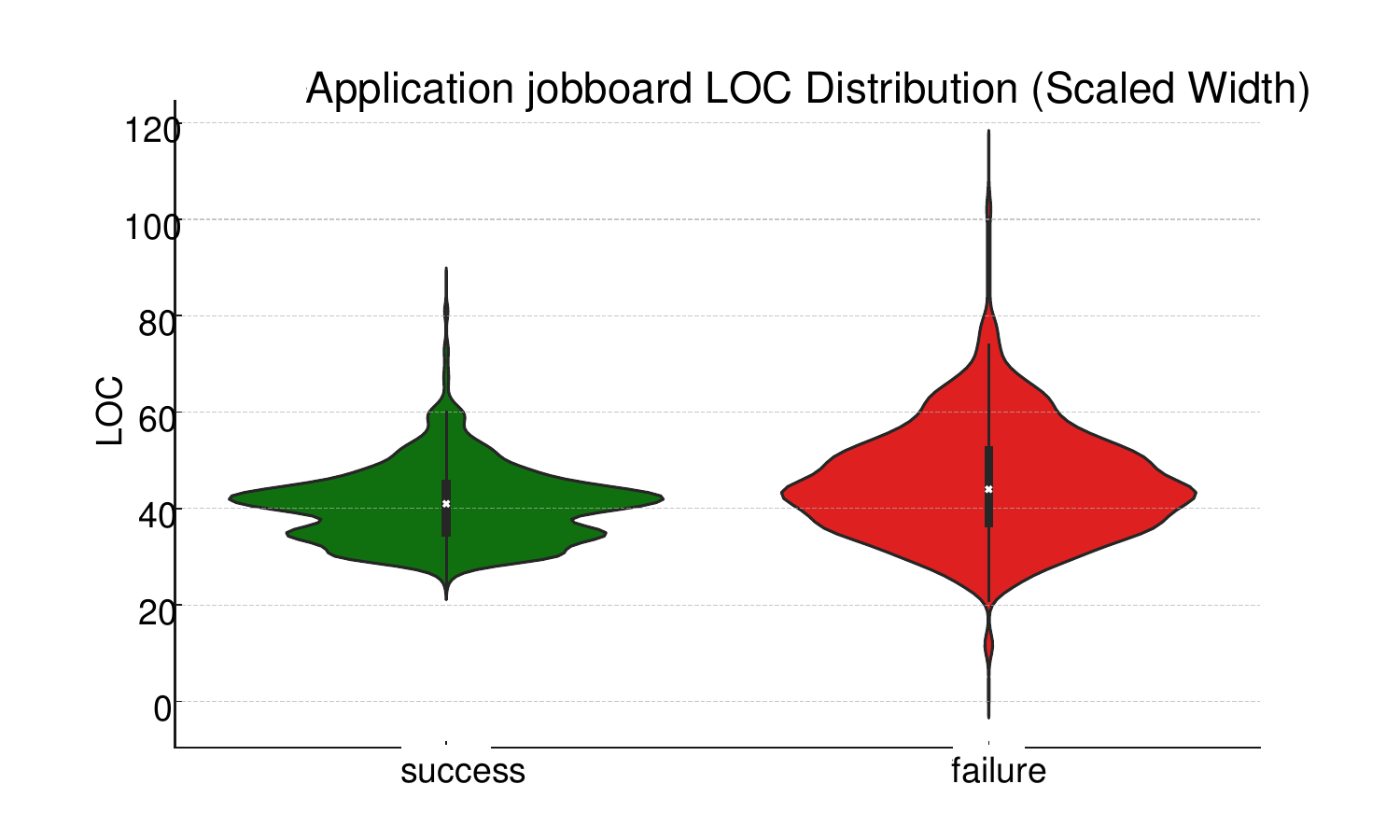}
        \caption{Job board (mean LOC = 42)}
    \end{subfigure}

    \begin{subfigure}{0.4\textwidth}
        \centering
        \includegraphics[width=\linewidth]{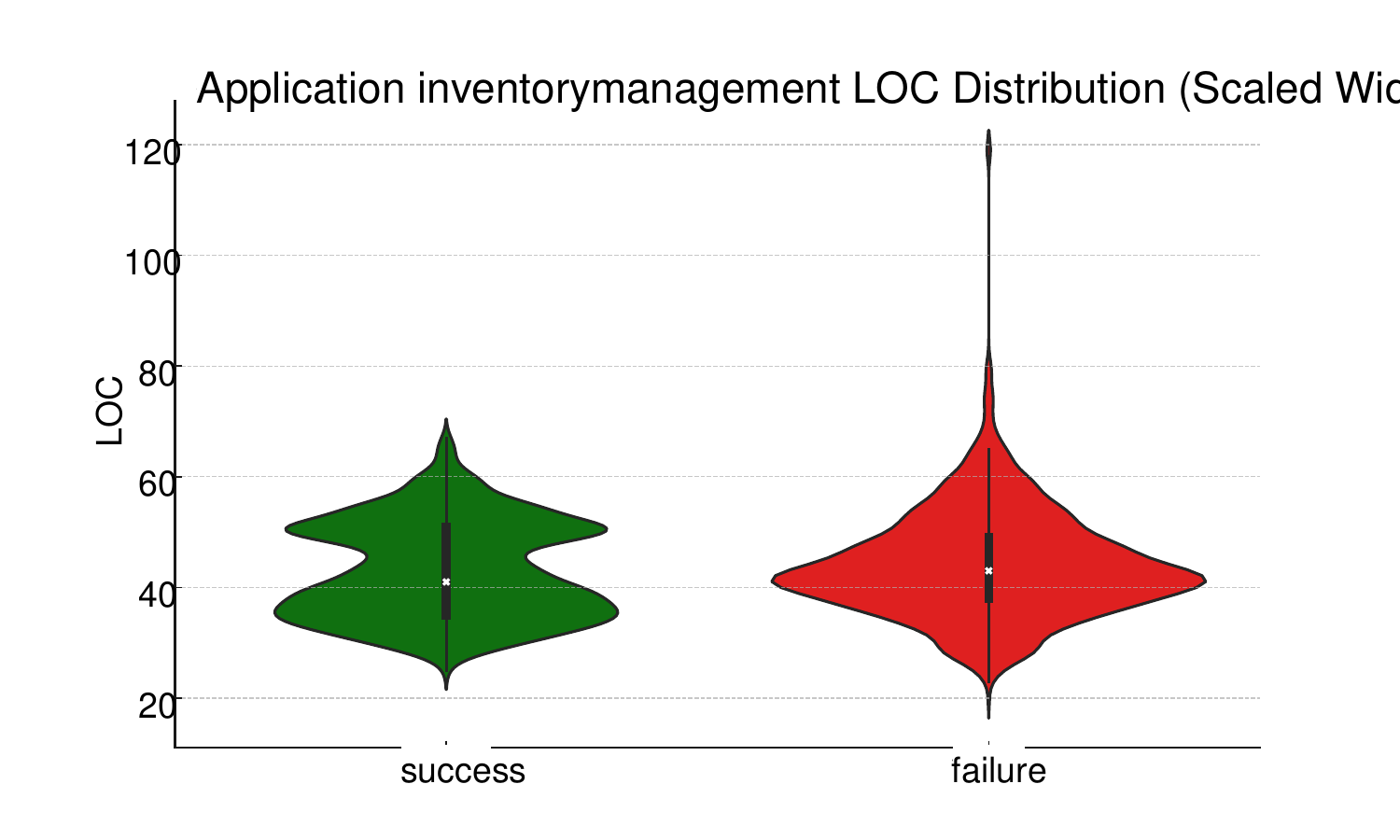}
        \caption{Inventory management (mean LOC = 42)}
    \end{subfigure}
    \begin{subfigure}{0.4\textwidth}
        \centering
        \includegraphics[width=\linewidth]{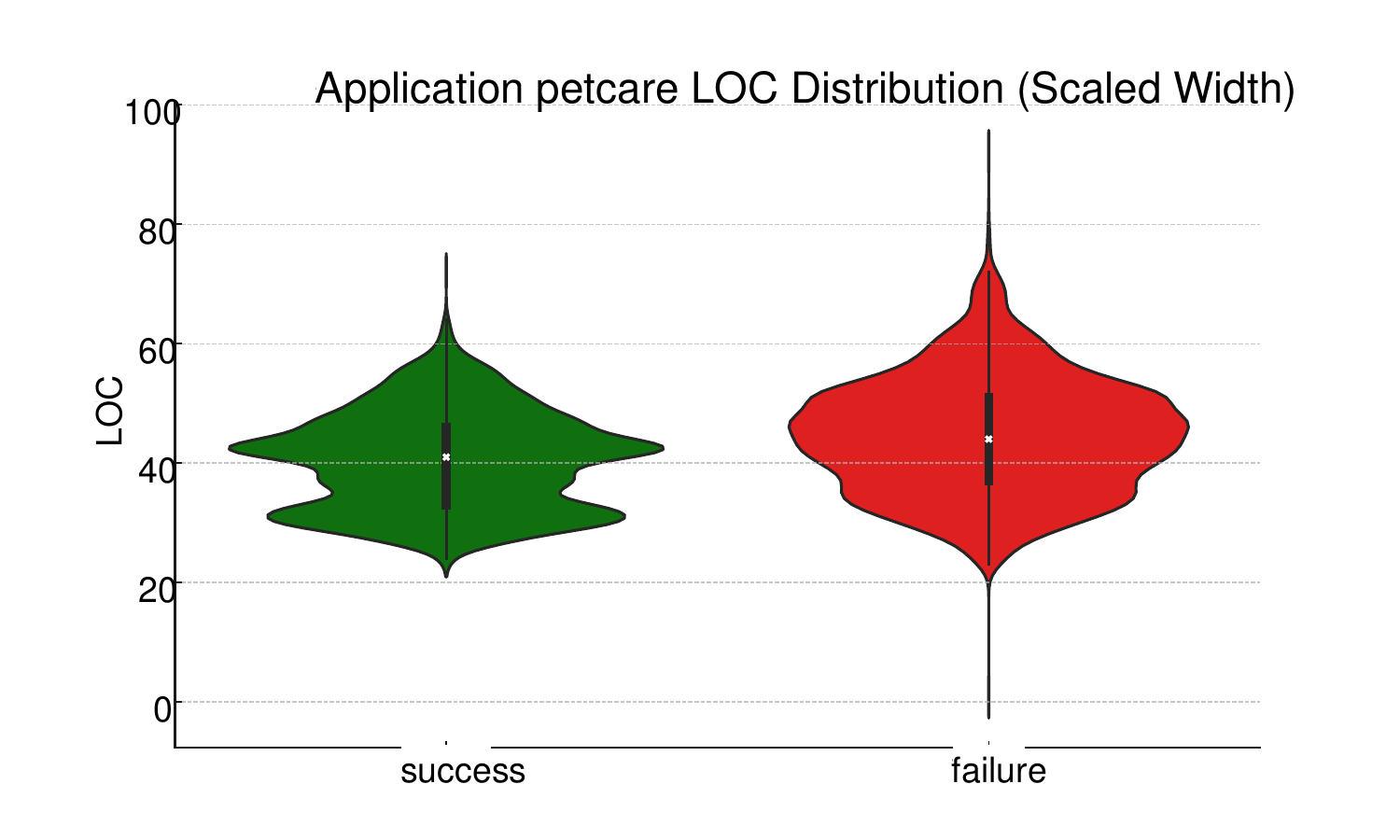}
        \caption{Pet care (mean LOC = 42)}
    \end{subfigure}

    \begin{subfigure}{0.4\textwidth}
        \centering
        \includegraphics[width=\linewidth]{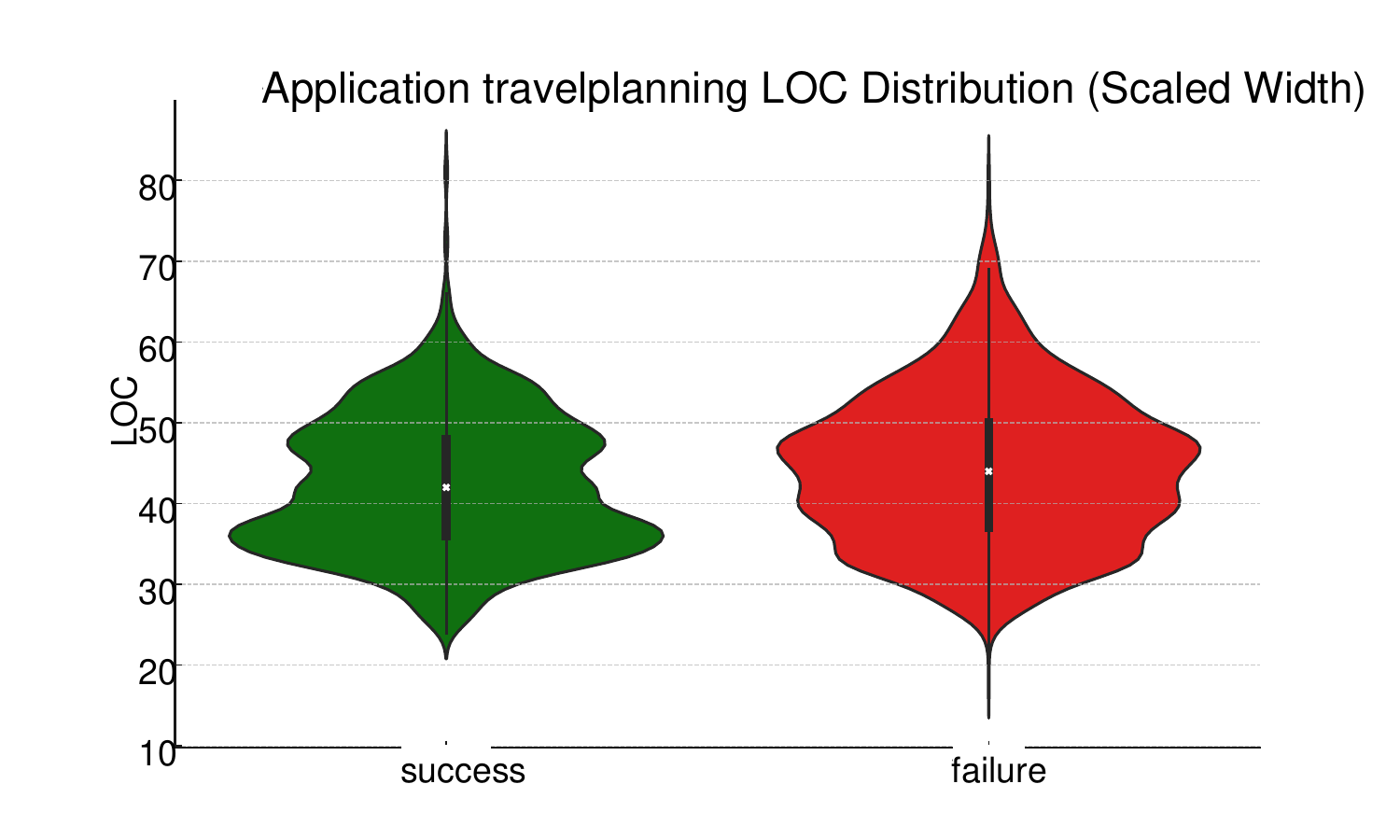}
        \caption{Travel planning (mean LOC = 42)}
    \end{subfigure}
    \begin{subfigure}{0.4\textwidth}
        \centering
        \includegraphics[width=\linewidth]{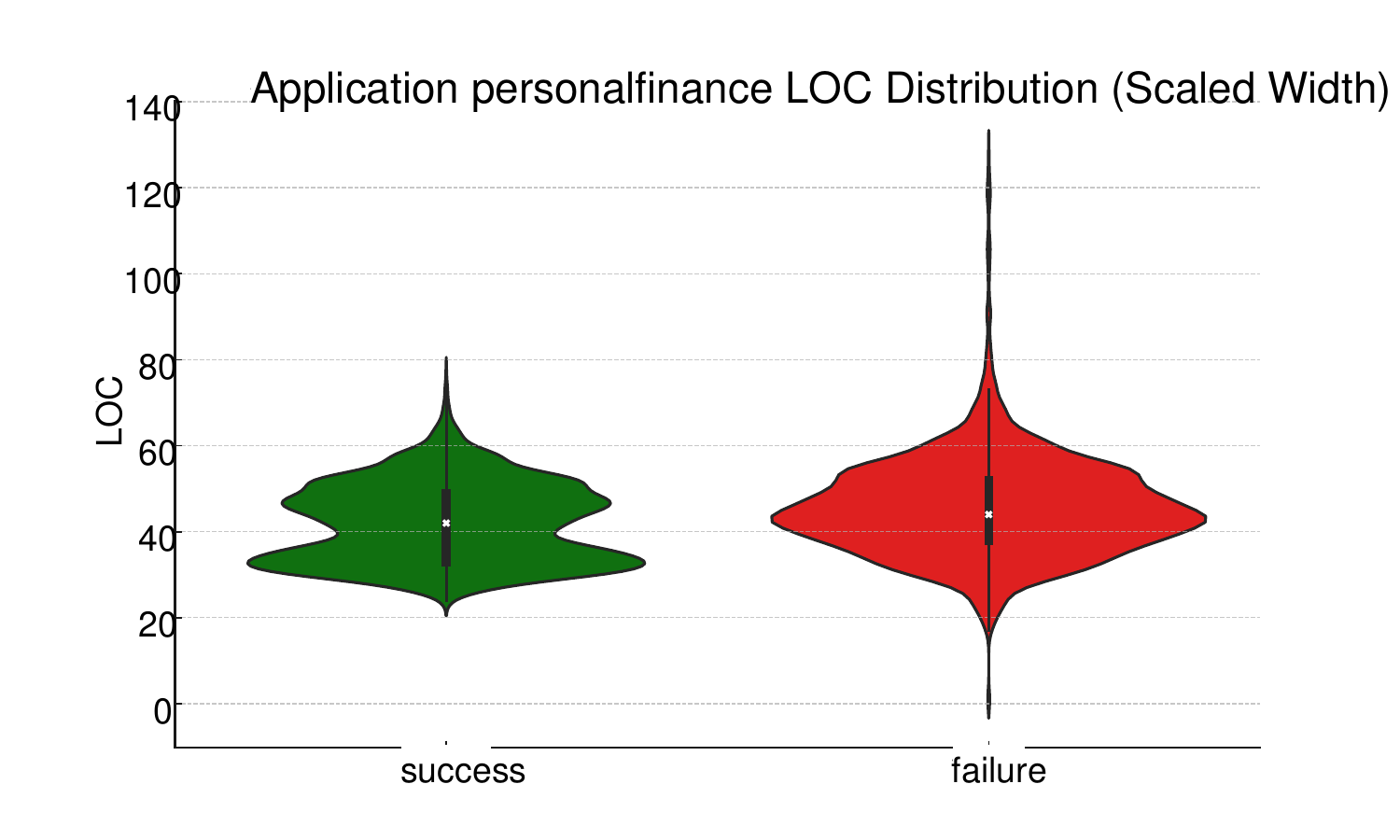}
        \caption{Personal finance (mean LOC = 43)}
    \end{subfigure}

    \begin{subfigure}{0.4\textwidth}
        \centering
        \includegraphics[width=\linewidth]{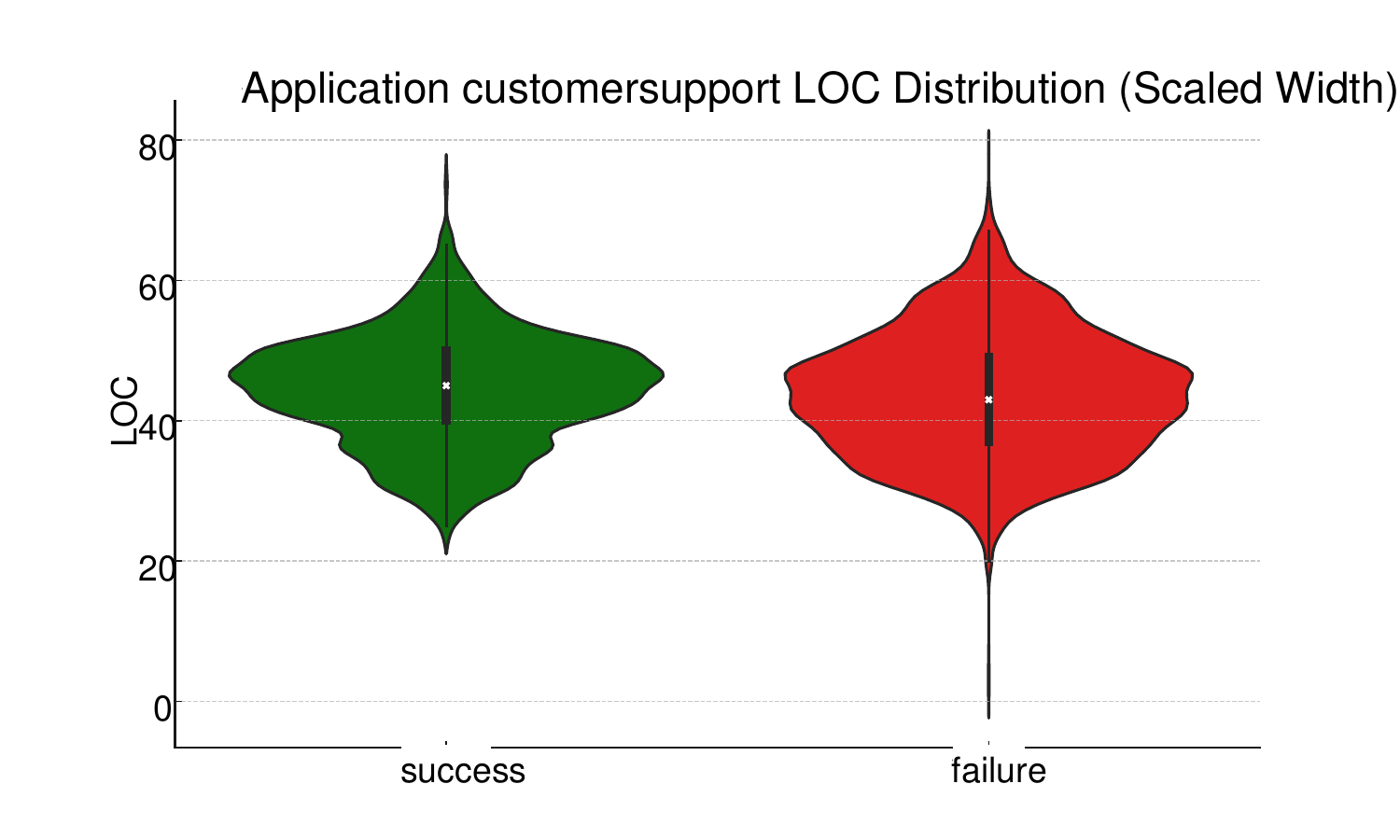}
        \caption{Customer support (mean LOC = 44)}
    \end{subfigure}
    \begin{subfigure}{0.4\textwidth}
        \centering
        \includegraphics[width=\linewidth]{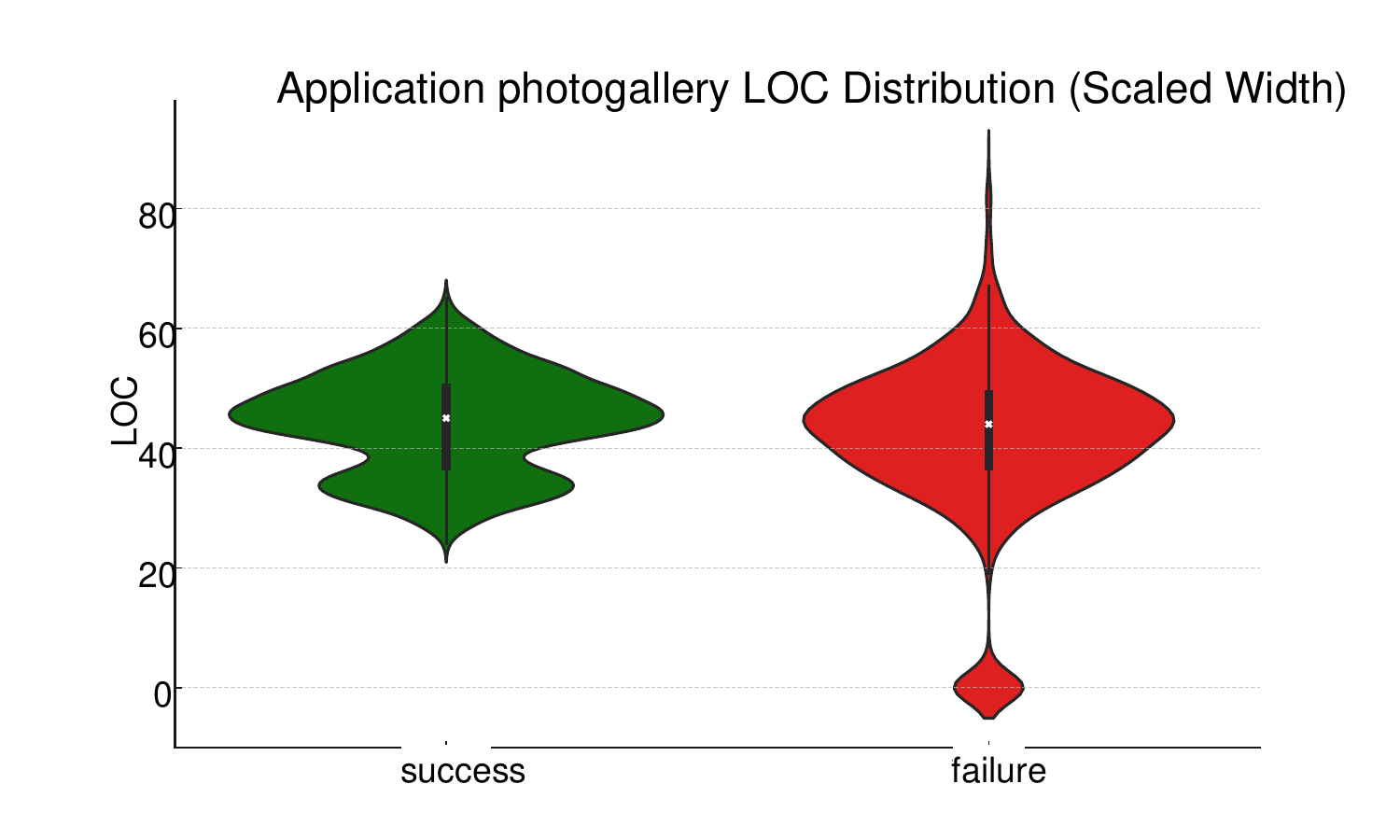}
        \caption{Photo gallery (mean LOC = 44)}
    \end{subfigure}

    \begin{subfigure}{0.4\textwidth}
        \centering
        \includegraphics[width=\linewidth]{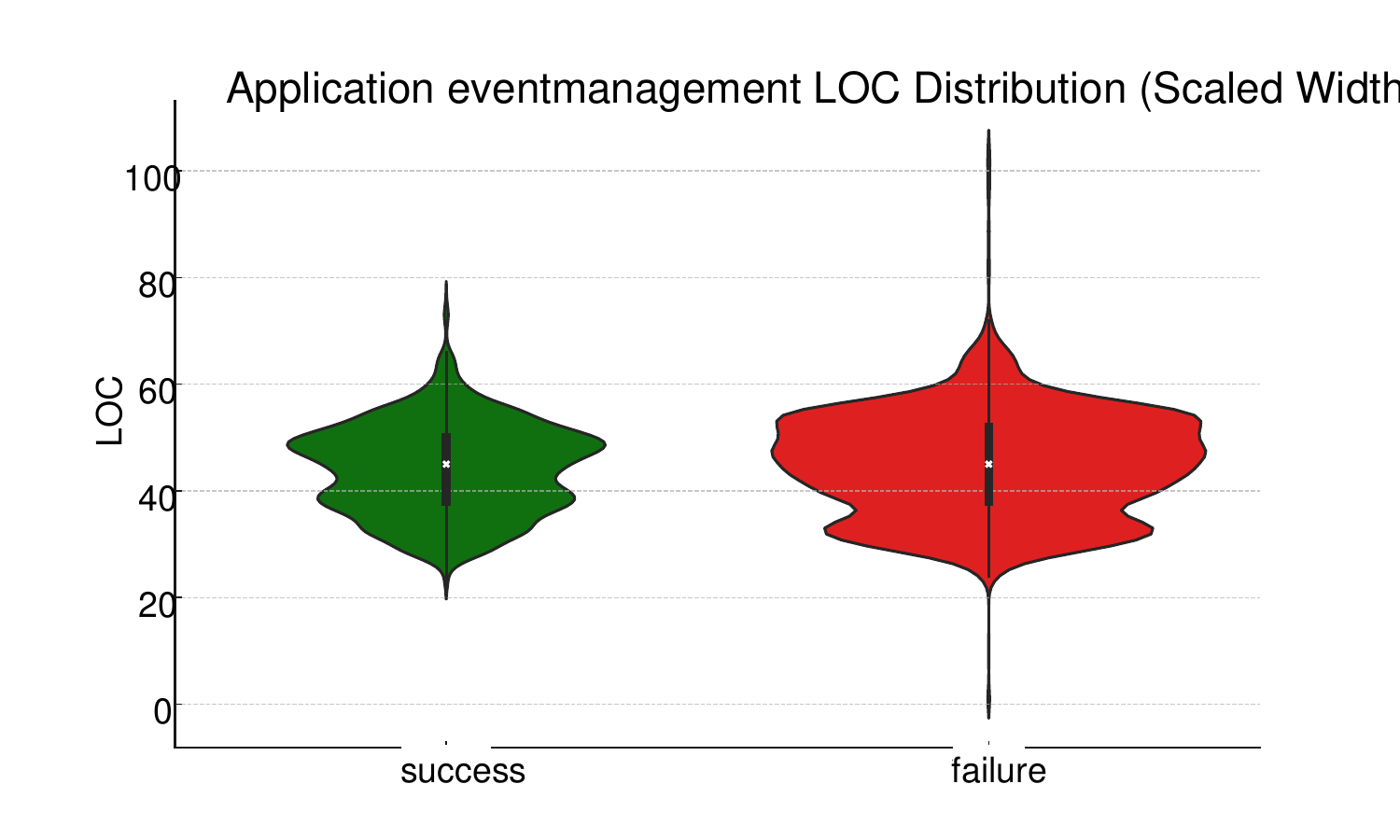}
        \caption{Event management (mean LOC = 45)}
    \end{subfigure}
    \begin{subfigure}{0.4\textwidth}
        \centering
        \includegraphics[width=\linewidth]{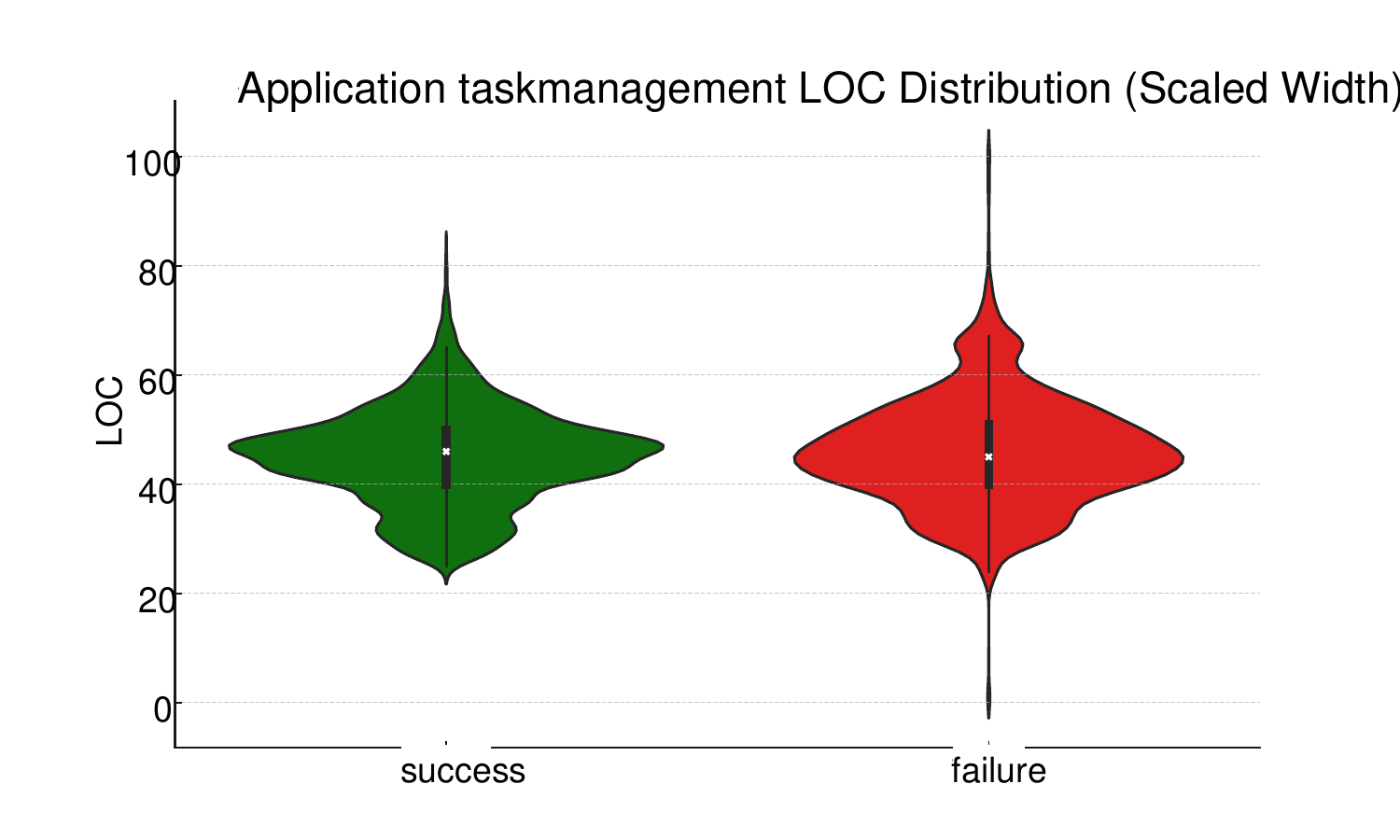}
        \caption{Task management (mean LOC = 46)}
    \end{subfigure}

    \caption{LOC Distribution by Application: Success vs Failure}
    \label{fig:loc_successfail_distribution_apps}
\end{figure}

Since each application assembles outputs from all models with full spectrum of performances, the success and failure data set are about the equal size. Similar to what we have observed in model-based sharding (Sec.~\ref{sec:loc_successfail}), the distribution pattern for success is equally or more complex than that for failure, summarized in Tab.~\ref{tab:loc_successfail_distribution_apps}.
\begin{table}[h!]
\caption{Summary of Fig.~\ref{fig:loc_successfail_distribution_apps}: unimodal vs multimodal}
\centering
\begin{tabular}{|c|c|c|}
\hline
& \textbf{UniModal Success} & \textbf{MultiModal Success} \\
\hline
\textbf{UniModal Failure} & (b) (q) (t) & (c) (d) (f) (g) (h) (j) (k) (l) (m) (n) (o) (p) \\
\hline
\textbf{MultiModal Failure} &  & (a) (e) (i) (r) (s) \\
\hline
\end{tabular}
\label{tab:loc_successfail_distribution_apps}
\end{table}
\section{Per-Application Error Analysis}
\begin{figure}[htbp]
    \centering
    \begin{subfigure}[b]{0.4\textwidth}
        \includegraphics[width=\textwidth]{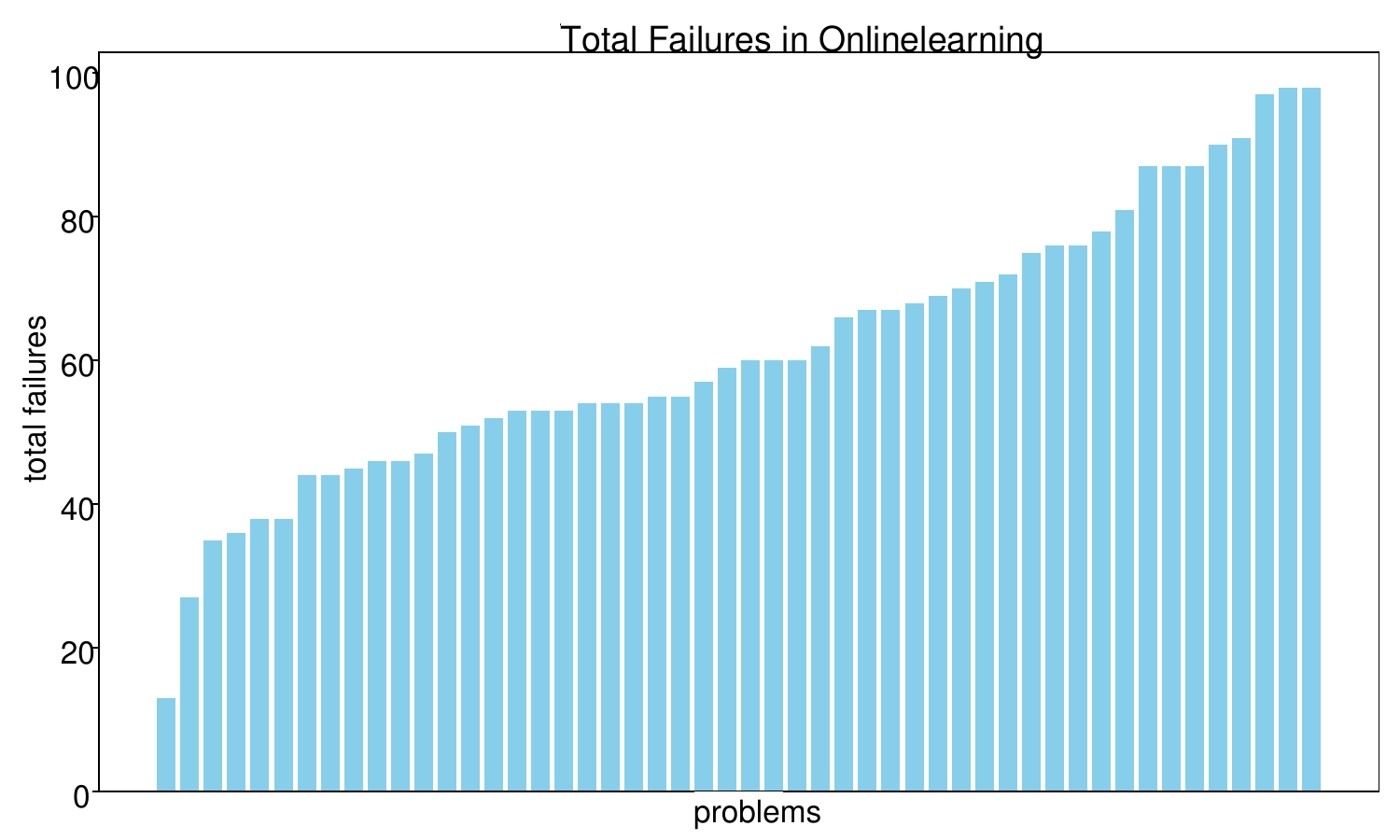}
        \caption{Online learning}
    \end{subfigure}
    \begin{subfigure}[b]{0.4\textwidth}
        \includegraphics[width=\textwidth]{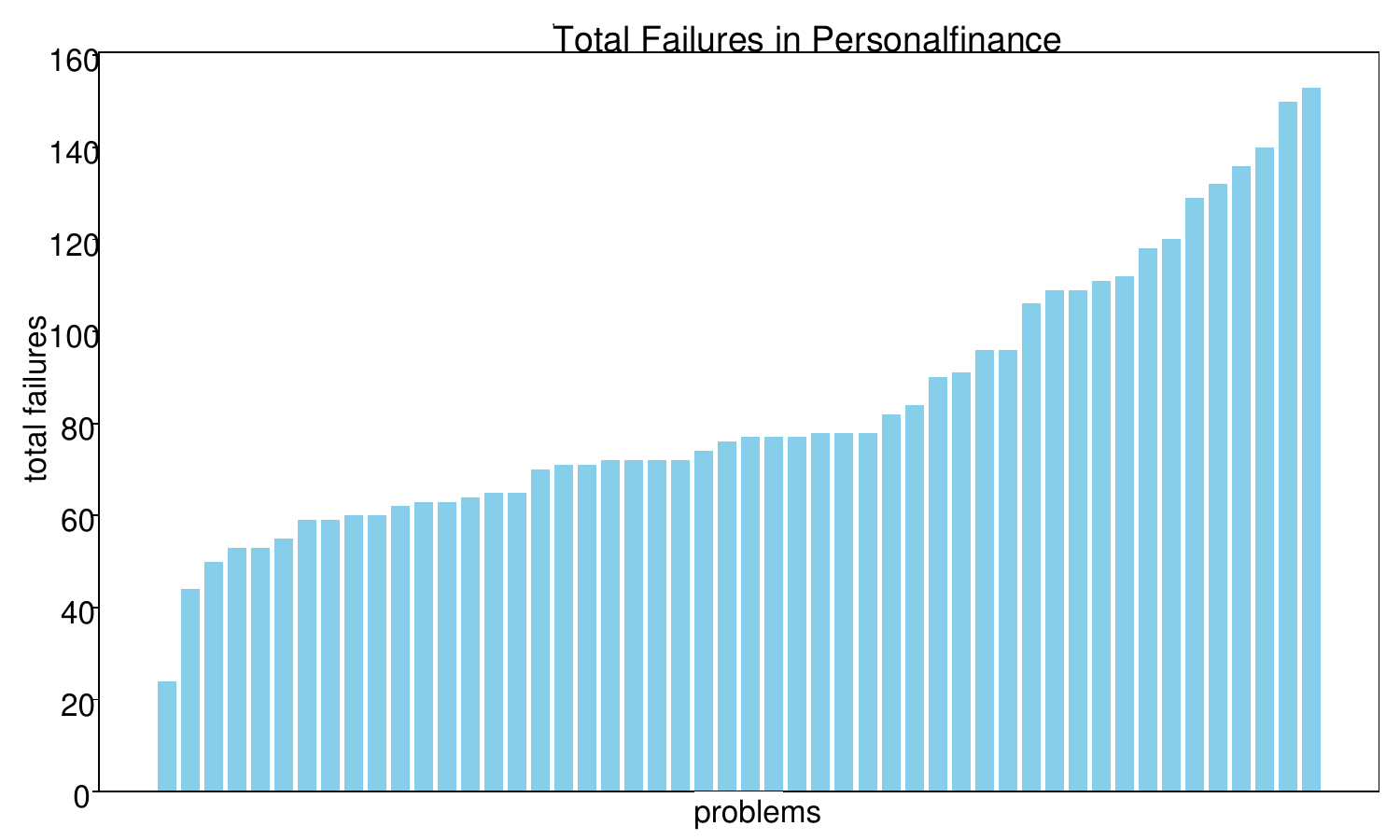}
        \caption{Personal finance}
    \end{subfigure}

    \begin{subfigure}[b]{0.4\textwidth}
        \includegraphics[width=\textwidth]{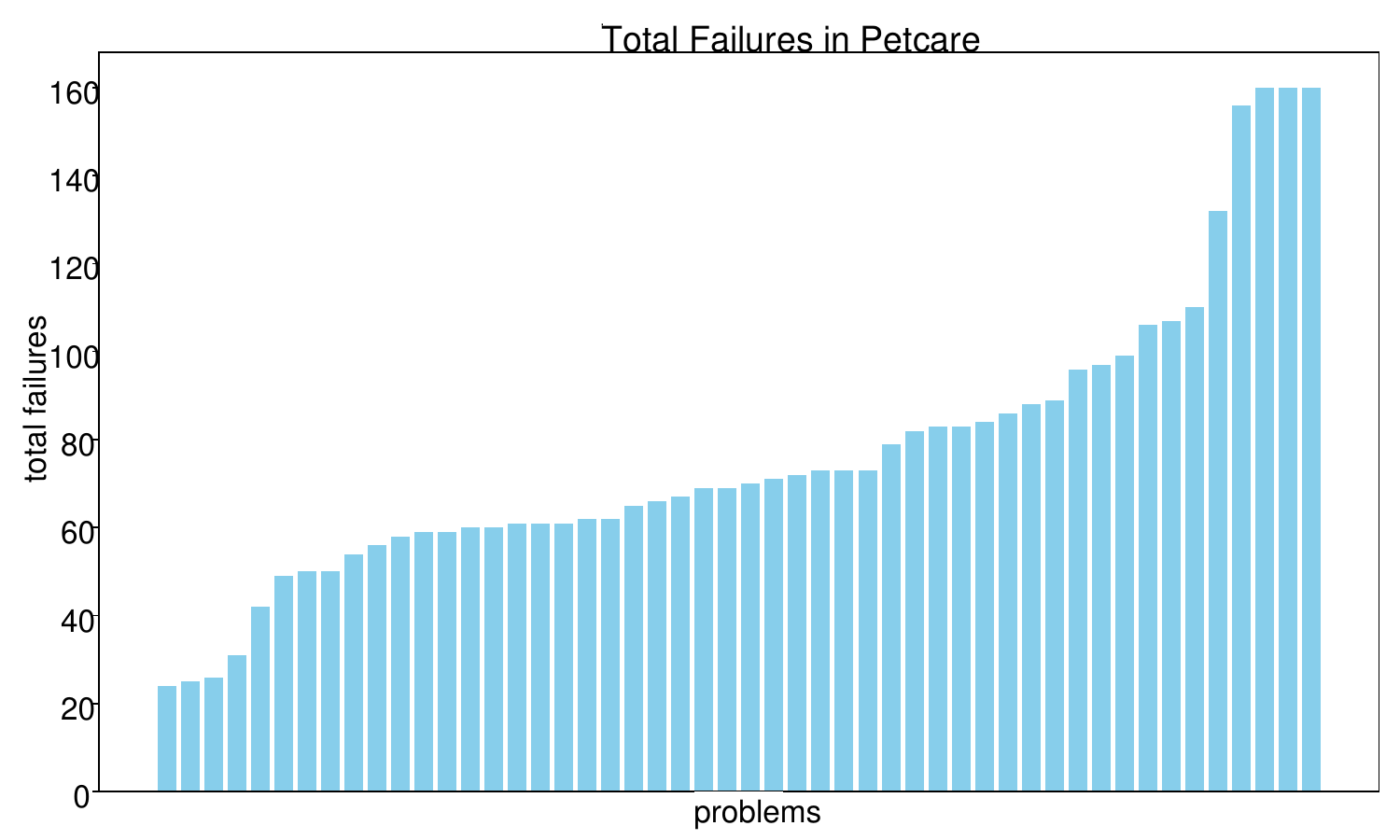}
        \caption{Pet care}
    \end{subfigure}
    \begin{subfigure}[b]{0.4\textwidth}
        \includegraphics[width=\textwidth]{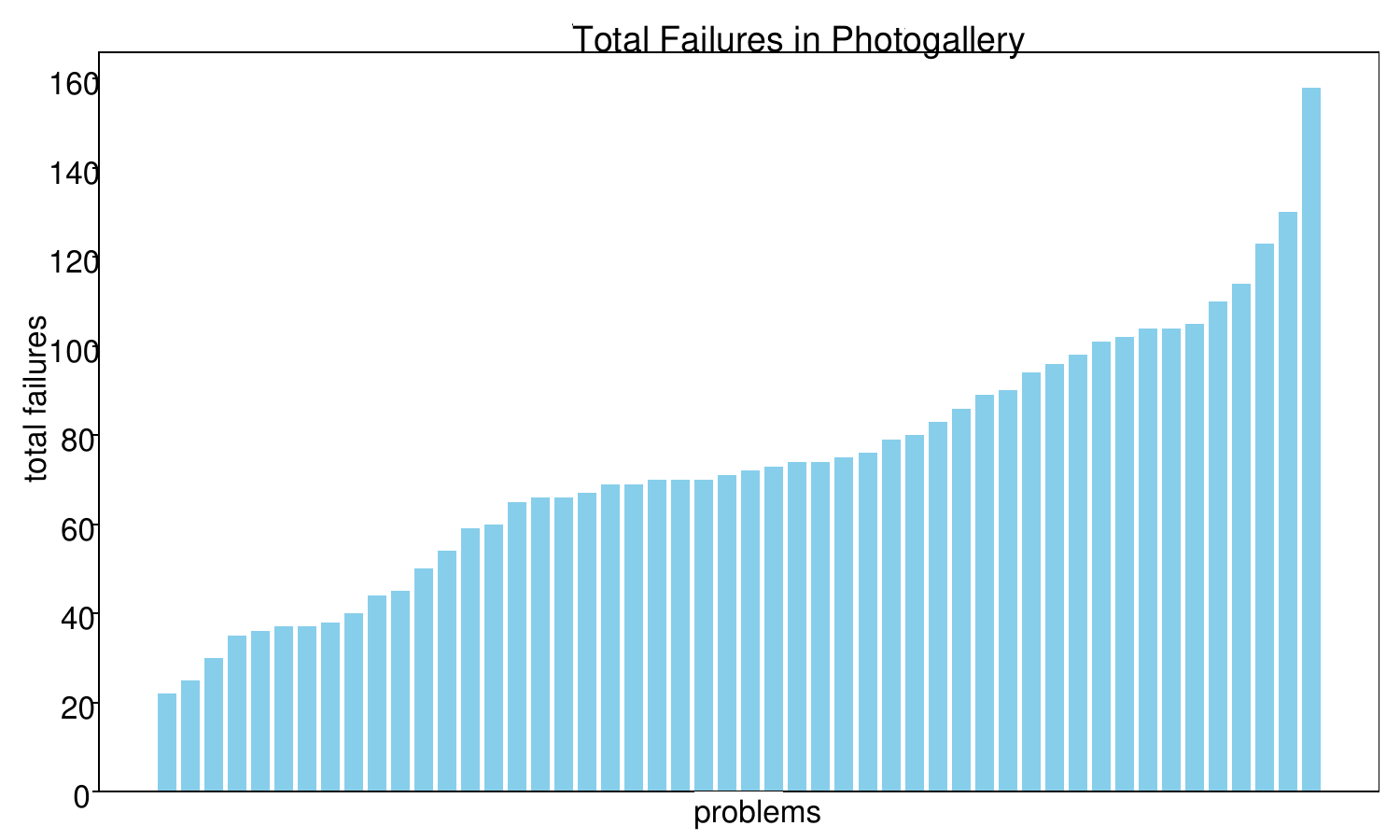}
        \caption{Photo gallery}
    \end{subfigure}

    \begin{subfigure}[b]{0.4\textwidth}
        \includegraphics[width=\textwidth]{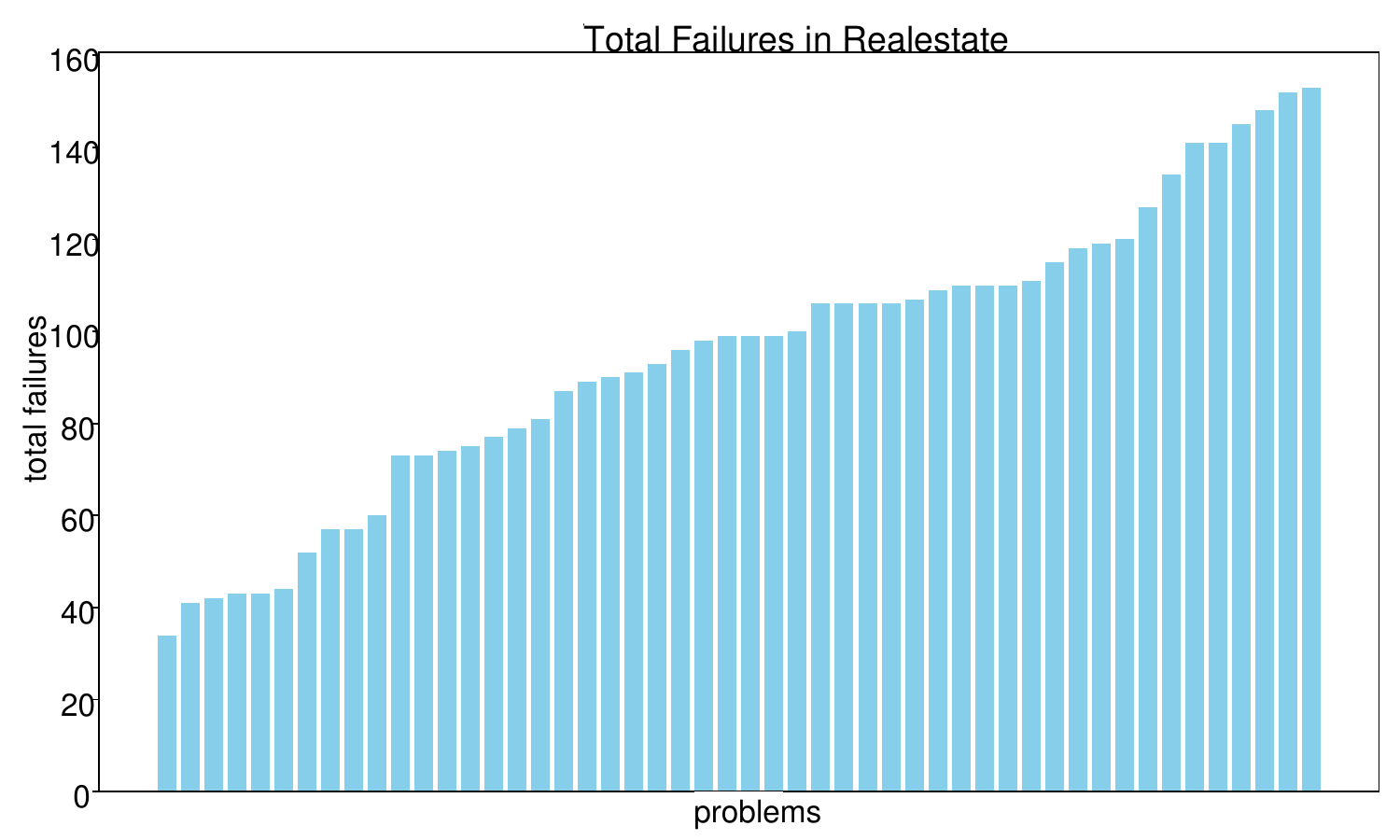}
        \caption{Real estate}
    \end{subfigure}
    \begin{subfigure}[b]{0.4\textwidth}
        \includegraphics[width=\textwidth]{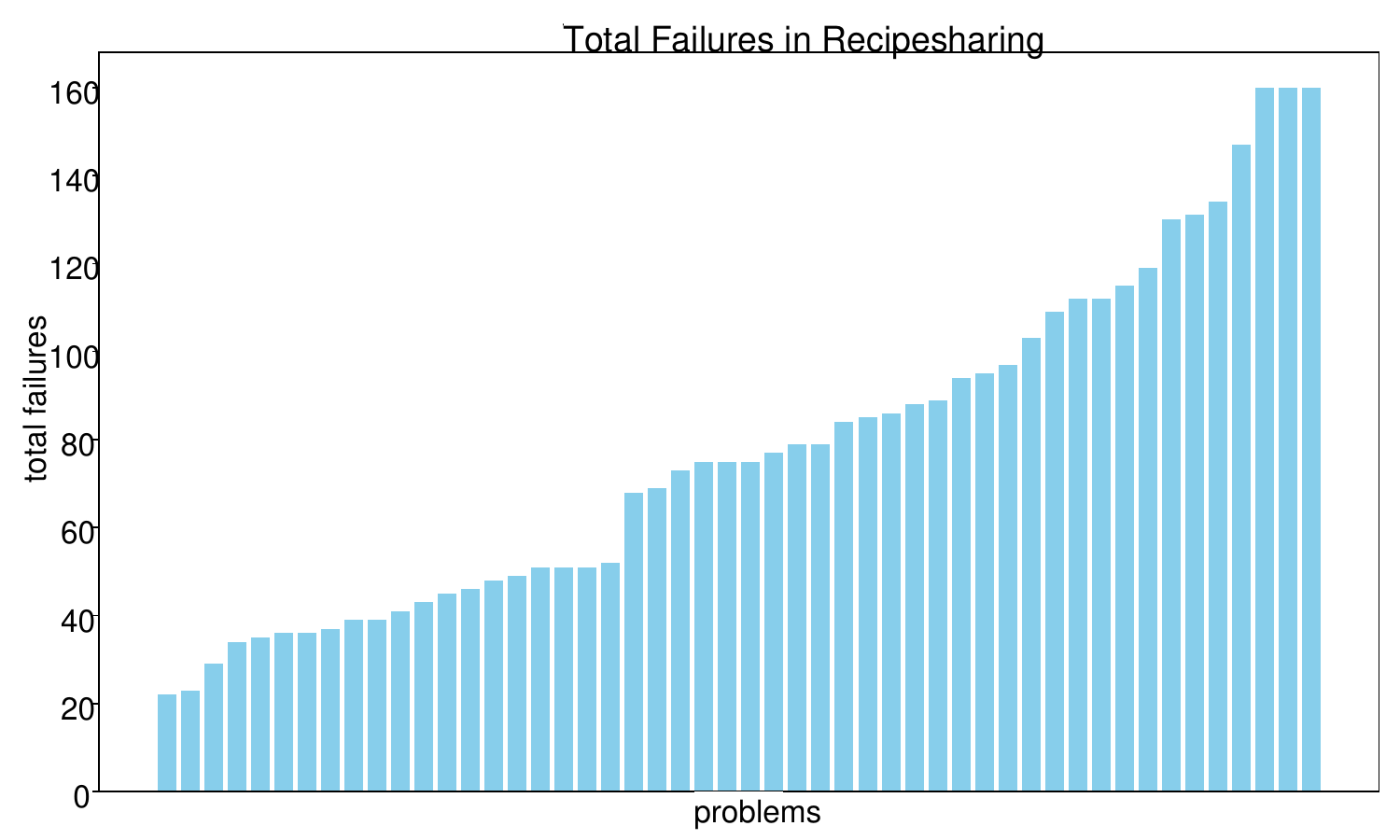}
        \caption{Recipe sharing}
    \end{subfigure}

    \begin{subfigure}[b]{0.4\textwidth}
        \includegraphics[width=\textwidth]{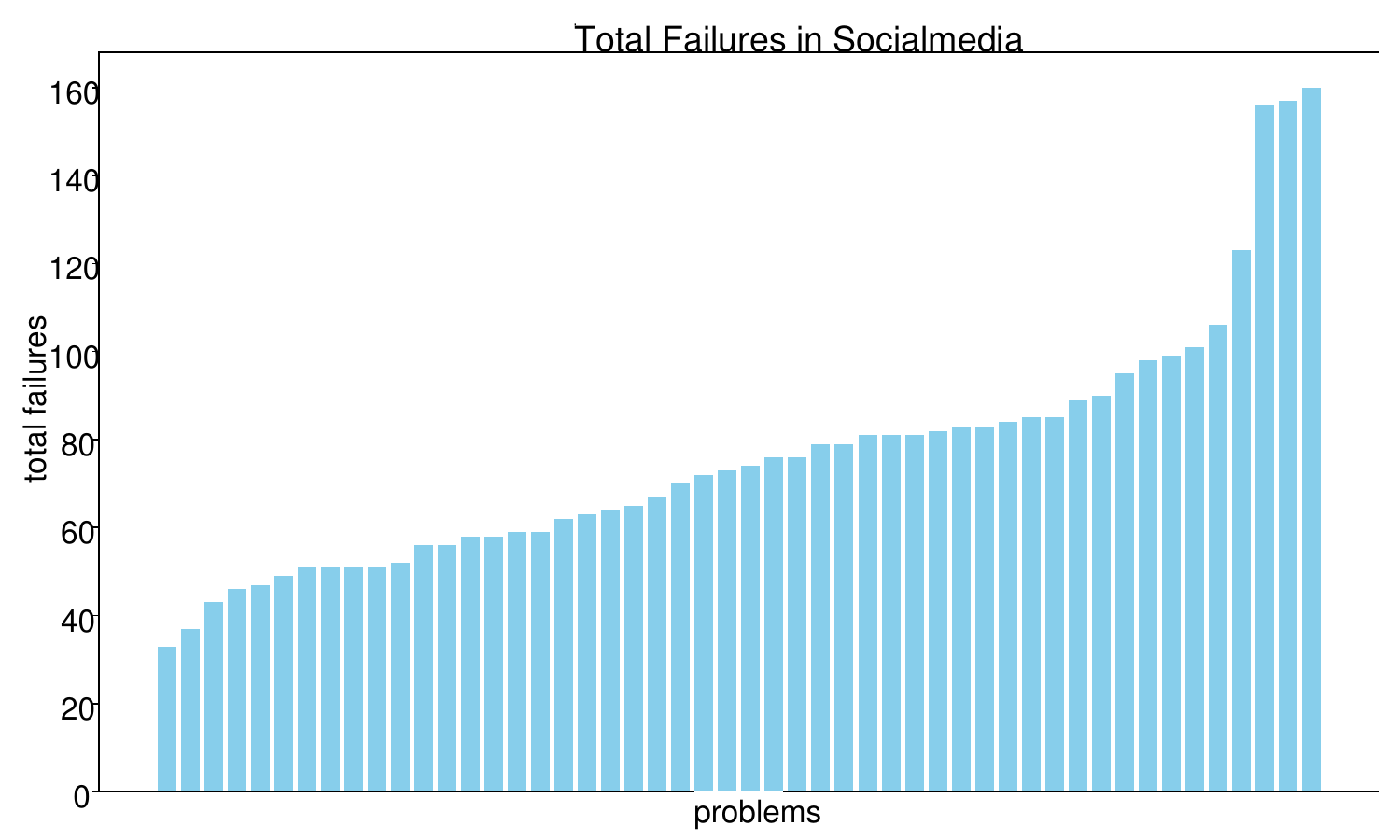}
        \caption{Social media}
    \end{subfigure}
    \begin{subfigure}[b]{0.4\textwidth}
        \includegraphics[width=\textwidth]{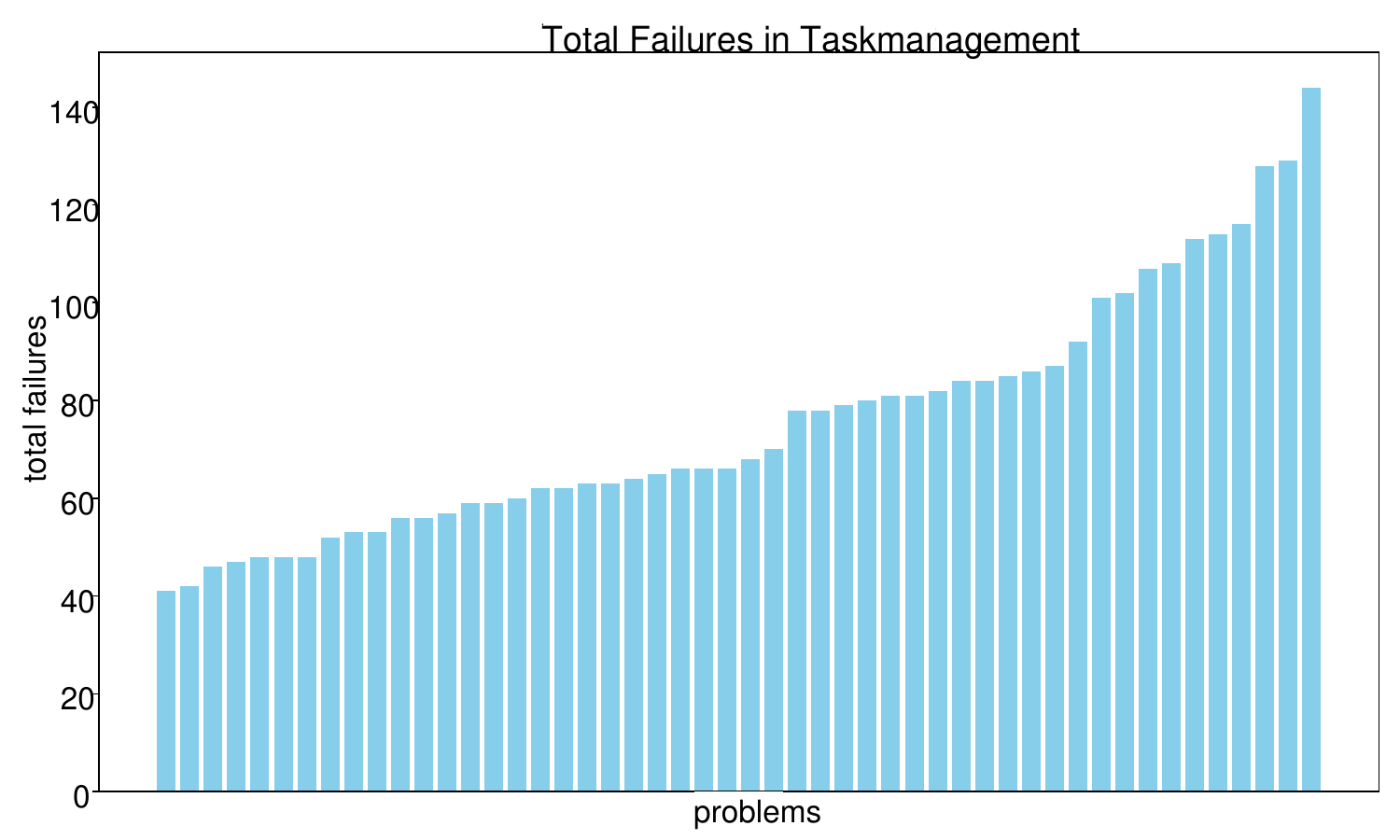}
        \caption{Task management}
    \end{subfigure}

    \begin{subfigure}[b]{0.4\textwidth}
        \includegraphics[width=\textwidth]{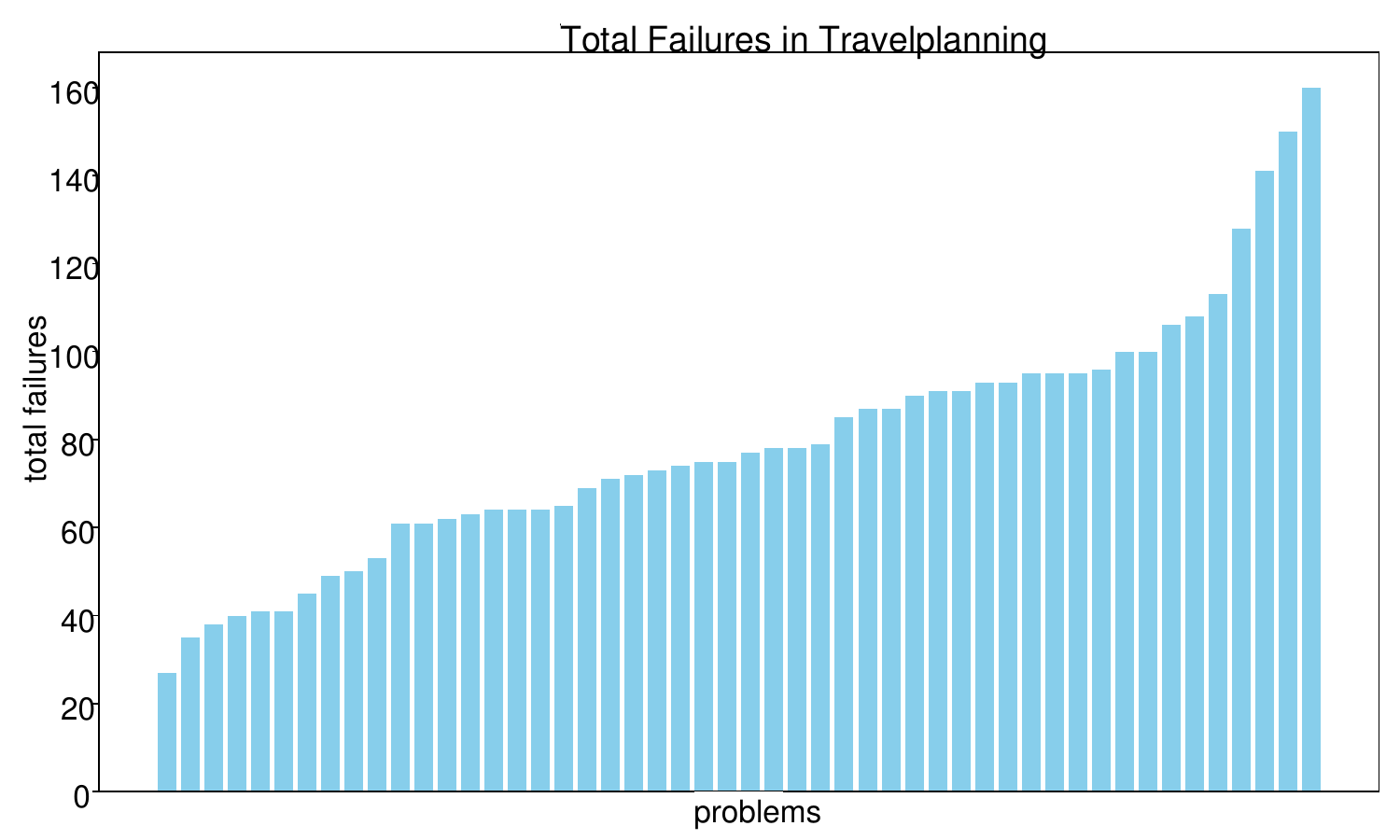}
        \caption{Travel planning}
    \end{subfigure}
    \begin{subfigure}[b]{0.4\textwidth}
        \includegraphics[width=\textwidth]{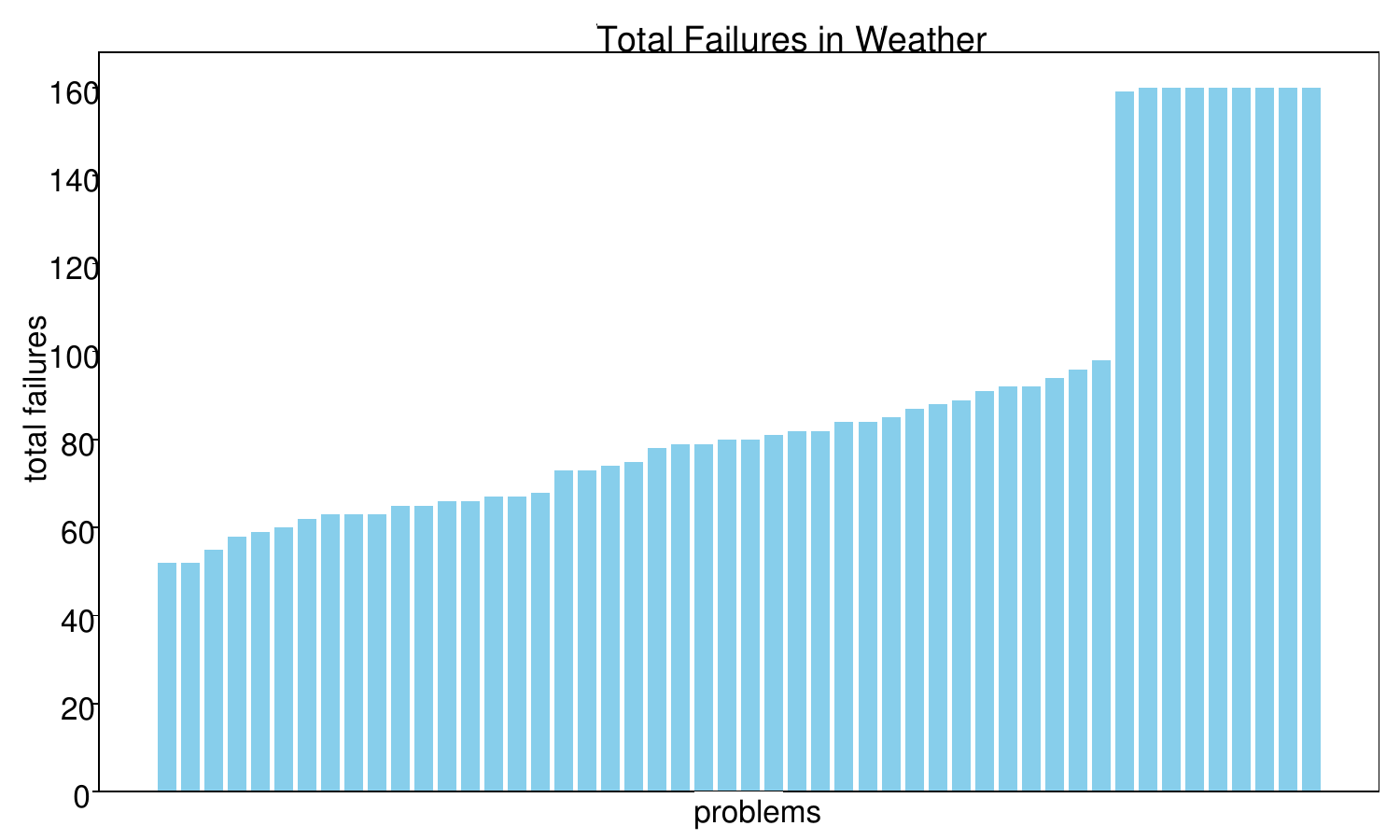}
        \caption{Weather}
    \end{subfigure}
 
    \caption{Failures per problem by application}
    \label{fig:failures_apps}
\end{figure}

Fig.~\ref{fig:failures_apps} shows the failure pattern broken down by applications. 
\begin{enumerate}
\item\textbf{Consistency Across Applications}: All applications exhibit the same general shape—a large concentration of easier problems on the left side and a few harder problems on the right side. This consistency suggests that across different domains, there are always a few particularly challenging problems that models struggle with.

\item\textbf{Variations in Skewness}: Some applications, such as Fitness Tracking and Music Streaming, show a more pronounced skew with a sharp rise in failure rates for a few problems, indicating a steeper difficulty curve. Others have a more gradual increase, indicating a more even distribution of problem difficulty.

\item\textbf{Extreme Difficulty in Certain Applications}: Applications like Customer Support and Pet Care have a sharper increase towards the right, implying that these domains have a subset of problems that are especially challenging.

\item\textbf{Easier Applications}: In applications like Weather and Photo Gallery, the overall number of failures seems lower compared to other appli cations, suggesting that the problems in these areas were generally easier.
\end{enumerate}
\begin{figure}[htbp]
    \centering
    \begin{subfigure}[b]{0.4\textwidth}
        \includegraphics[width=\textwidth]{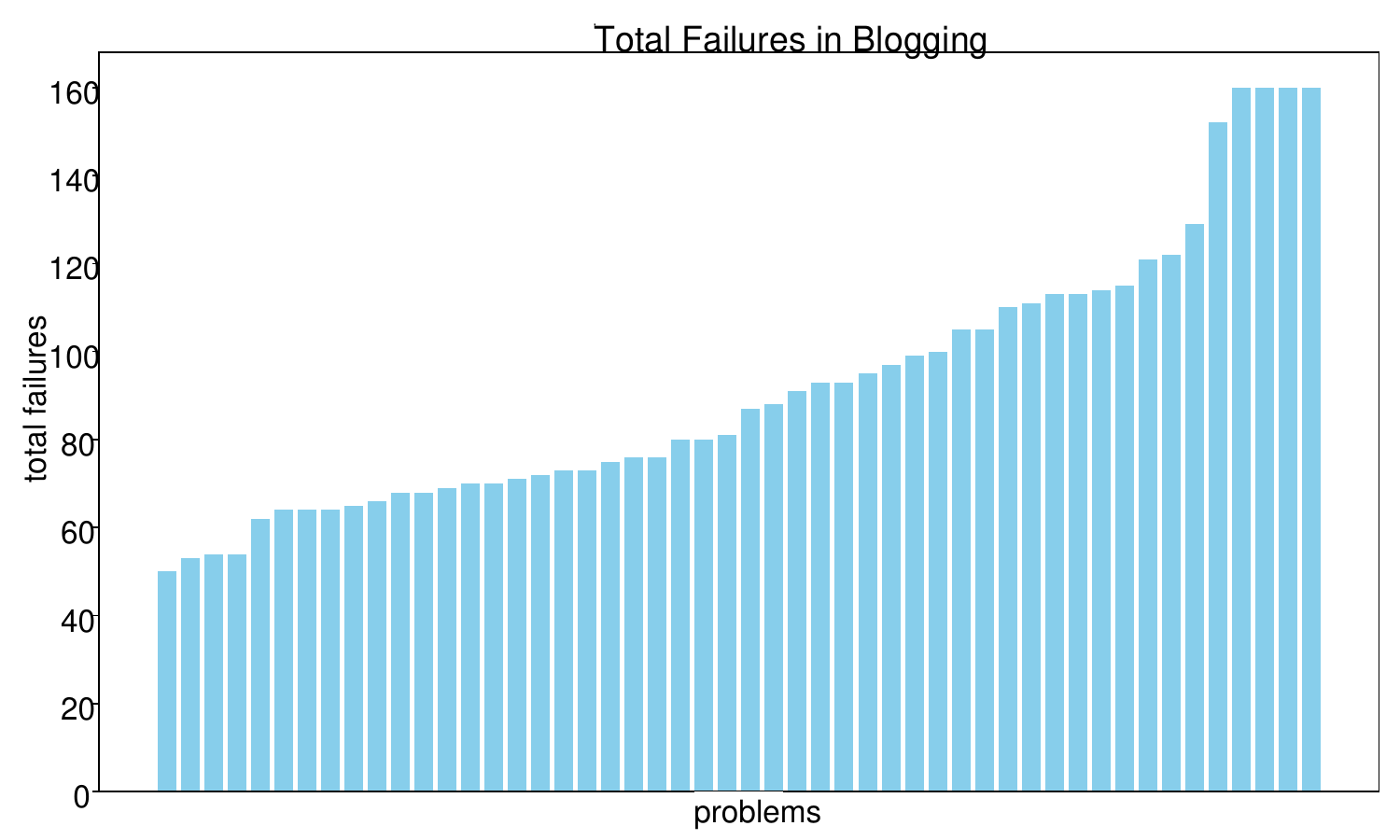}
        \caption{Blogging}
    \end{subfigure}
    \begin{subfigure}[b]{0.4\textwidth}
        \includegraphics[width=\textwidth]{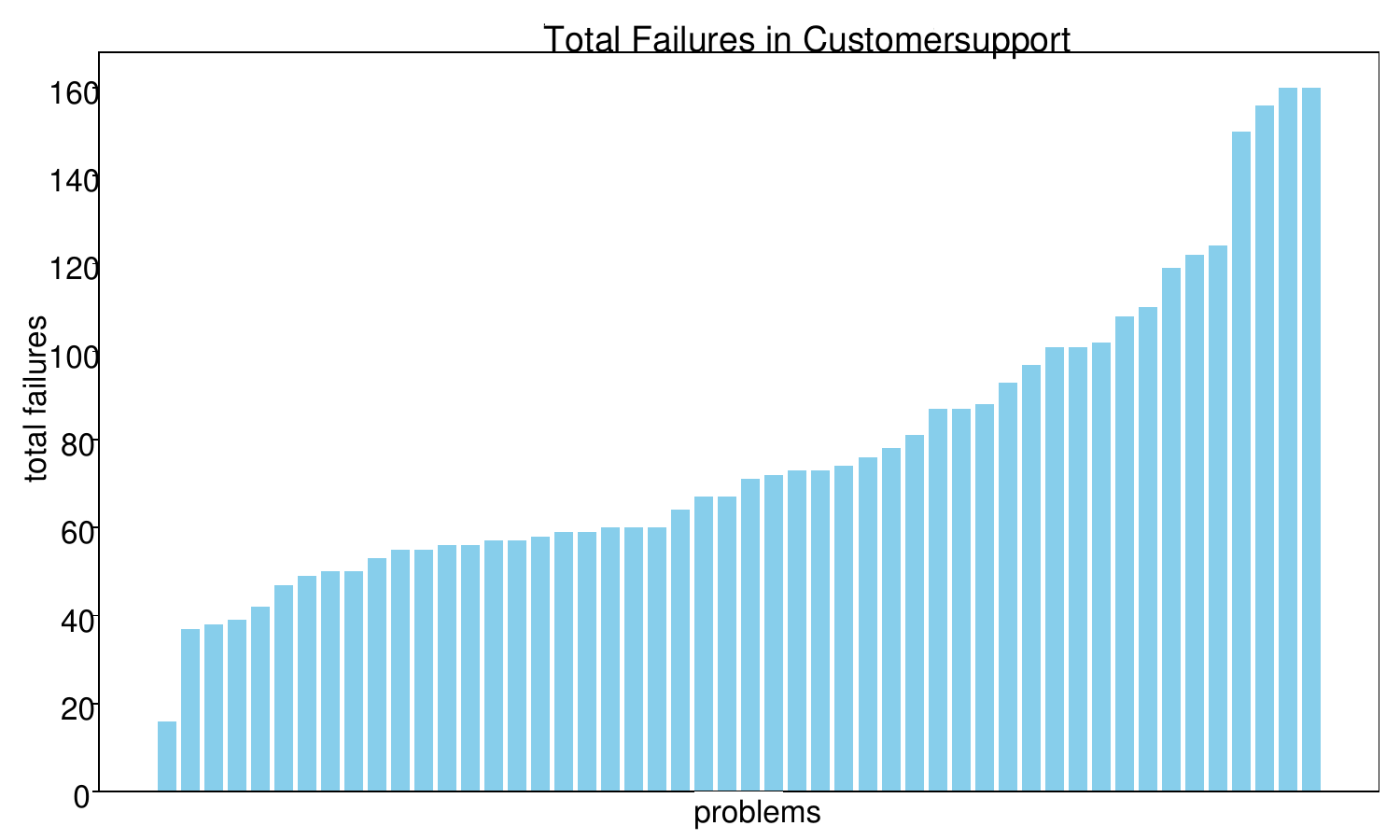}
        \caption{Customer support}
    \end{subfigure}

    \begin{subfigure}[b]{0.4\textwidth}
        \includegraphics[width=\textwidth]{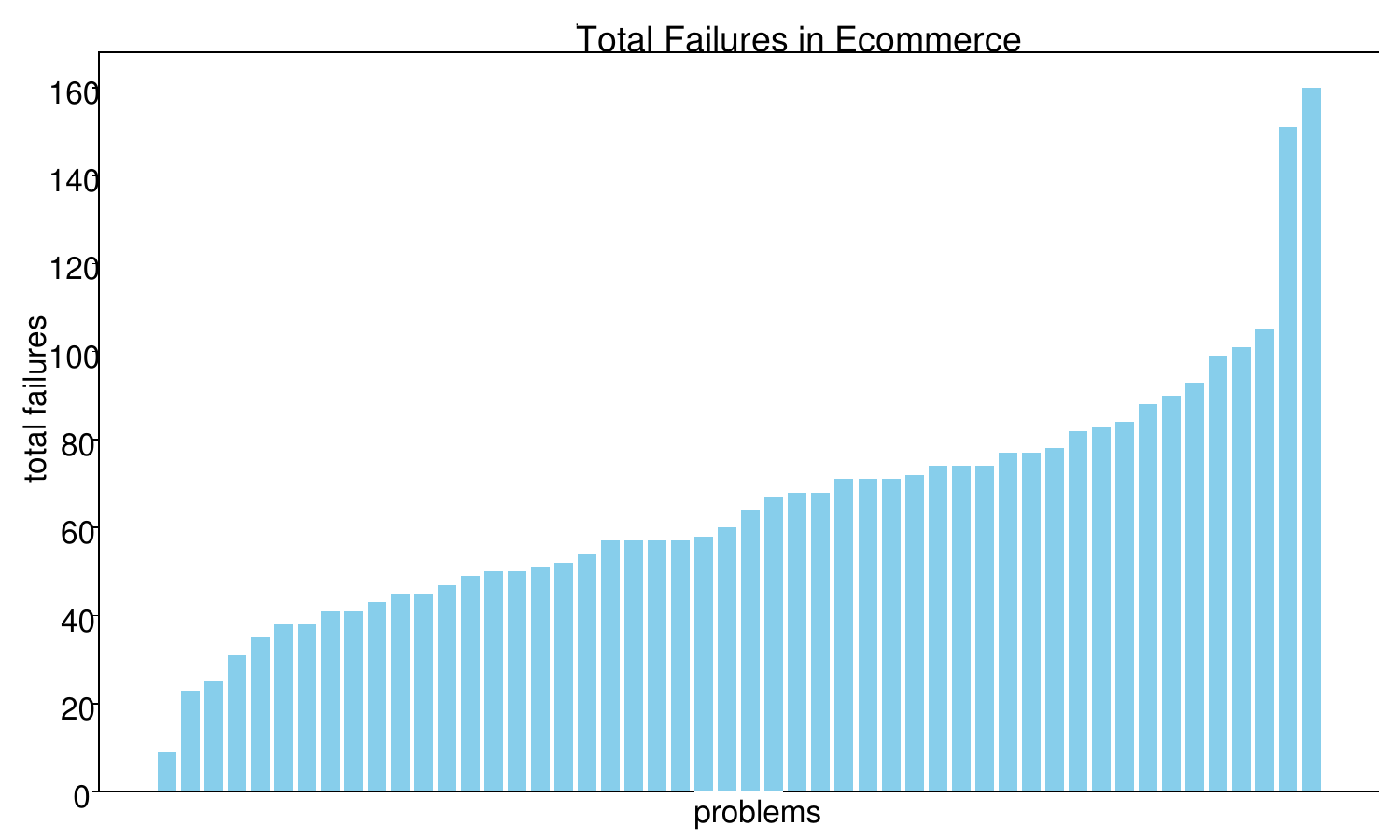}
        \caption{E-commerce}
    \end{subfigure}
    \begin{subfigure}[b]{0.4\textwidth}
        \includegraphics[width=\textwidth]{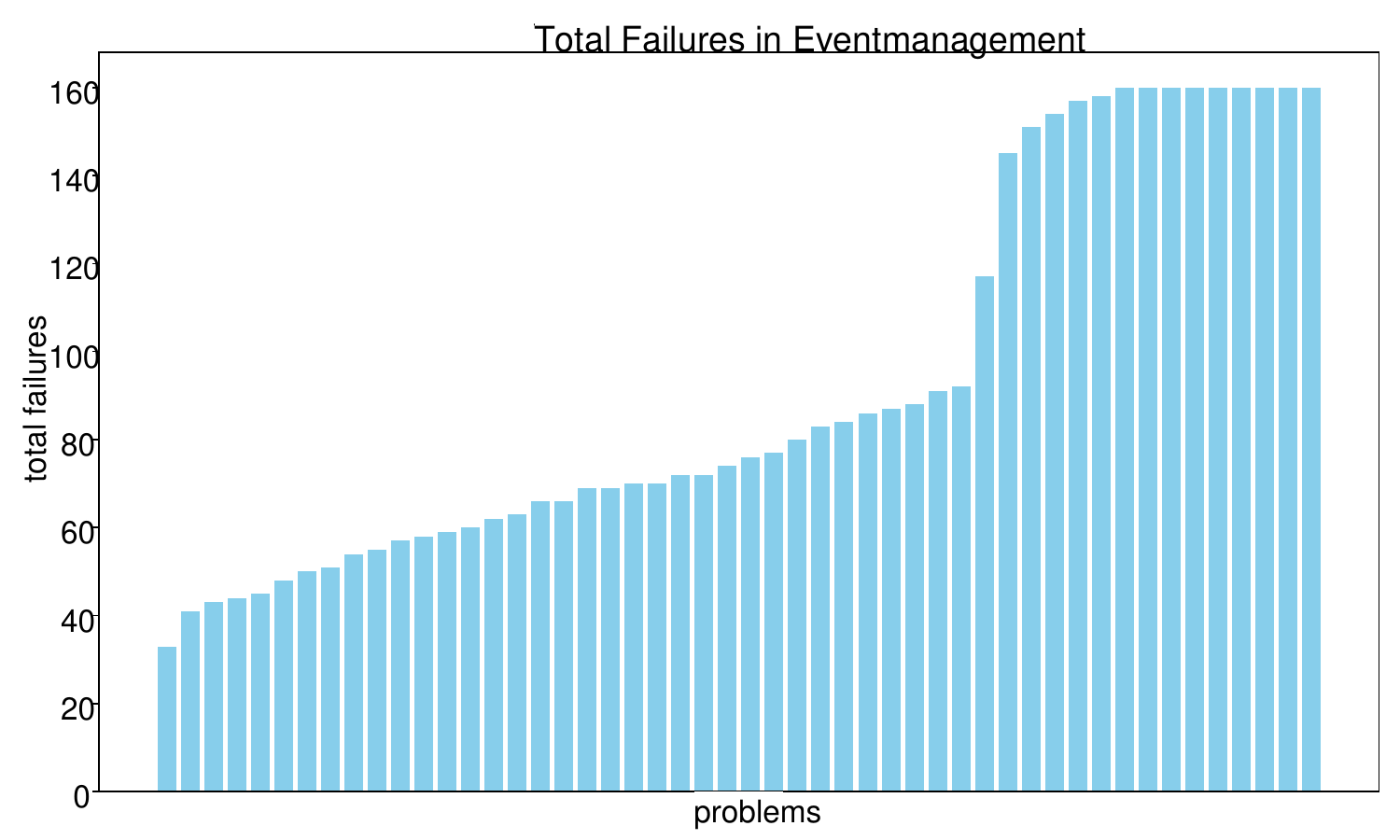}
        \caption{Event management}
    \end{subfigure}

    \begin{subfigure}[b]{0.4\textwidth}
        \includegraphics[width=\textwidth]{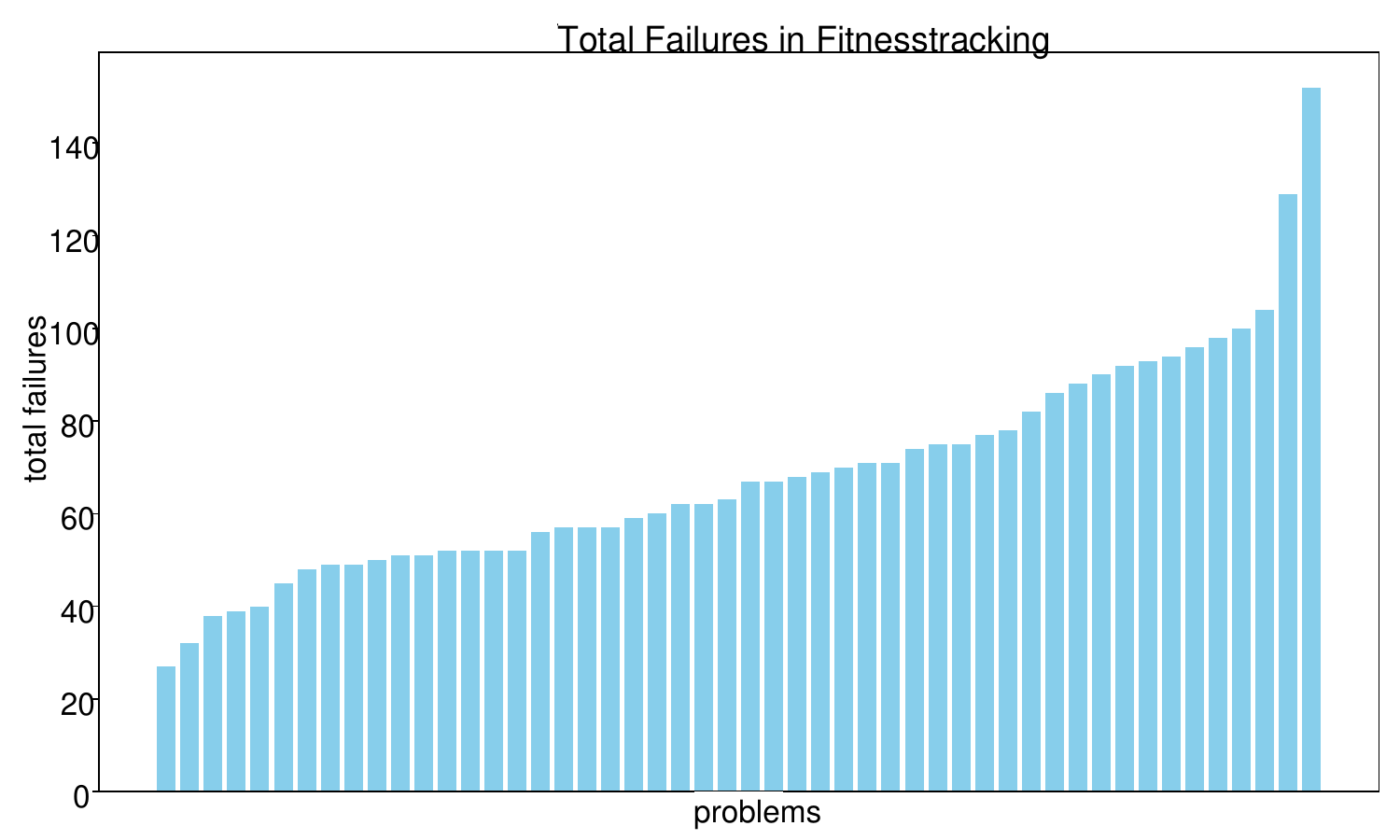}
        \caption{Fitness tracking}
    \end{subfigure}
    \begin{subfigure}[b]{0.4\textwidth}
        \includegraphics[width=\textwidth]{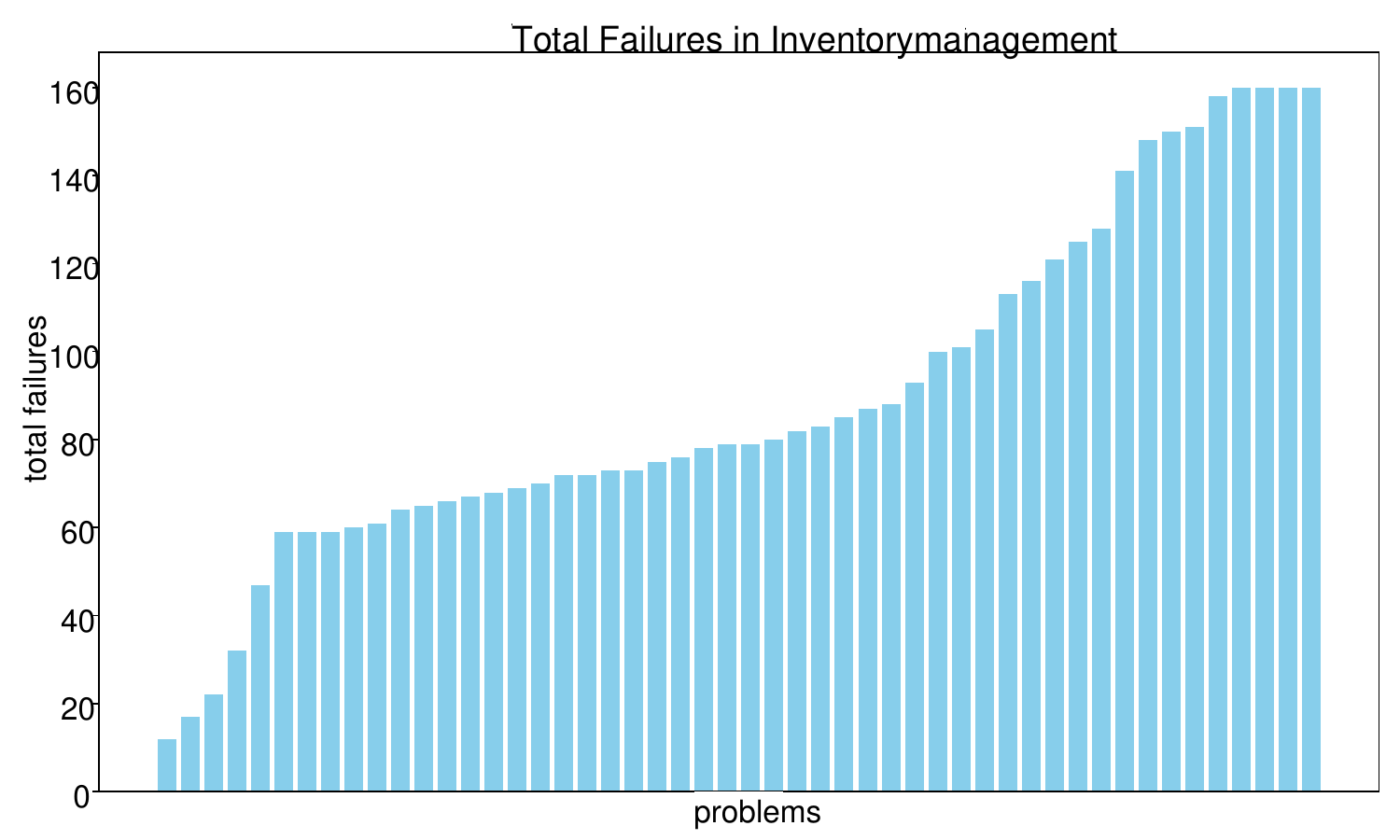}
        \caption{Inventory management}
    \end{subfigure}

    \begin{subfigure}[b]{0.4\textwidth}
        \includegraphics[width=\textwidth]{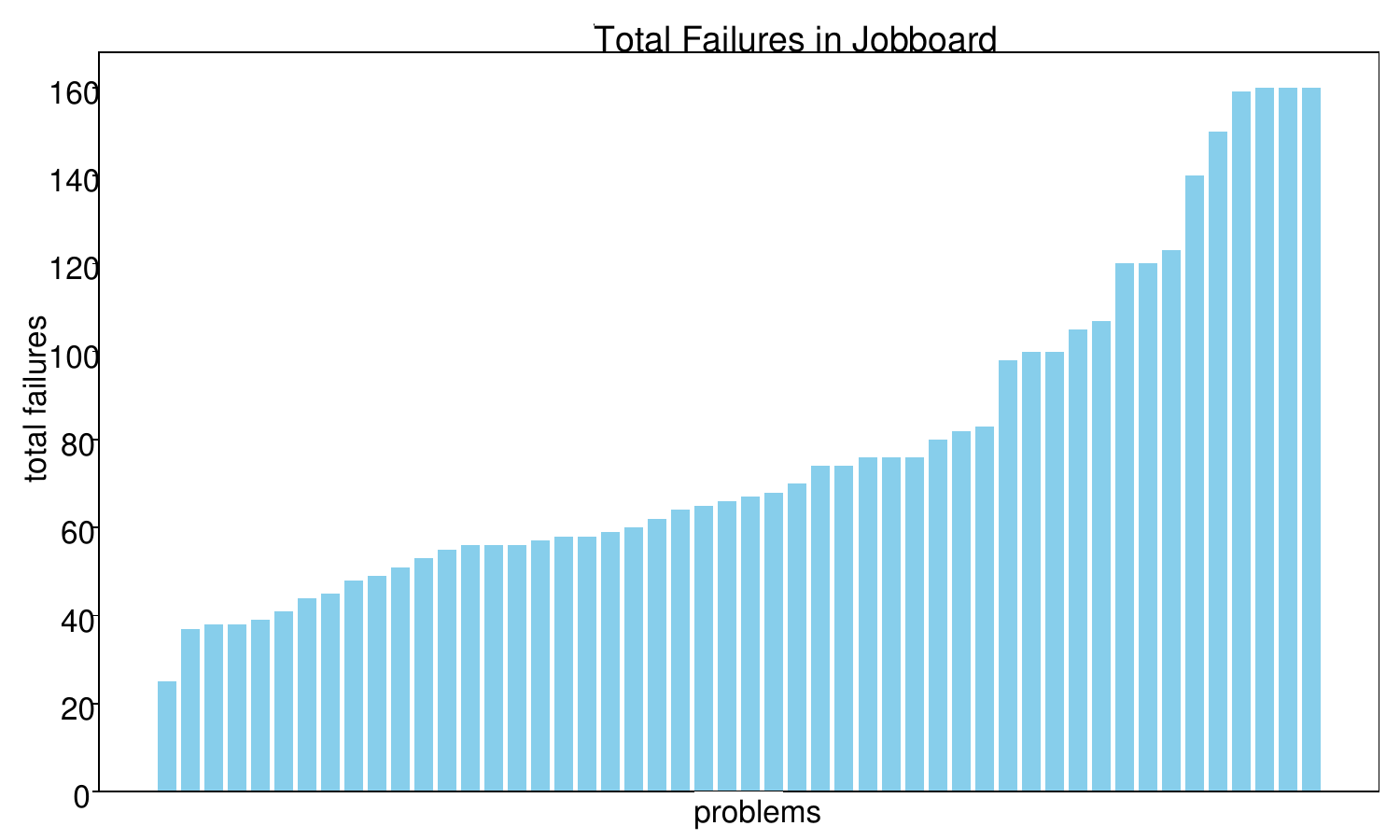}
        \caption{Job board}
    \end{subfigure}
    \begin{subfigure}[b]{0.4\textwidth}
        \includegraphics[width=\textwidth]{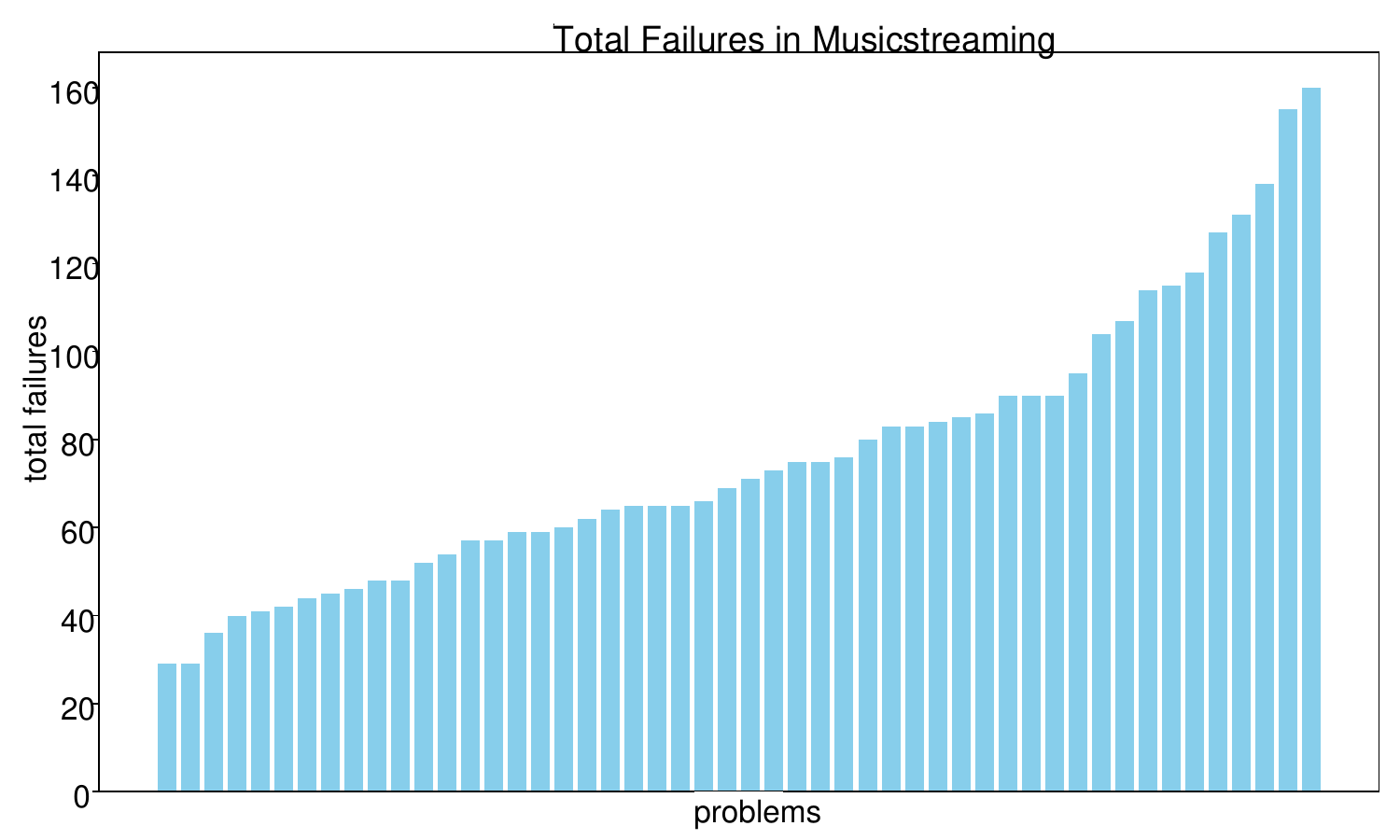}
        \caption{Music streaming}
    \end{subfigure}

    \begin{subfigure}[b]{0.4\textwidth}
        \includegraphics[width=\textwidth]{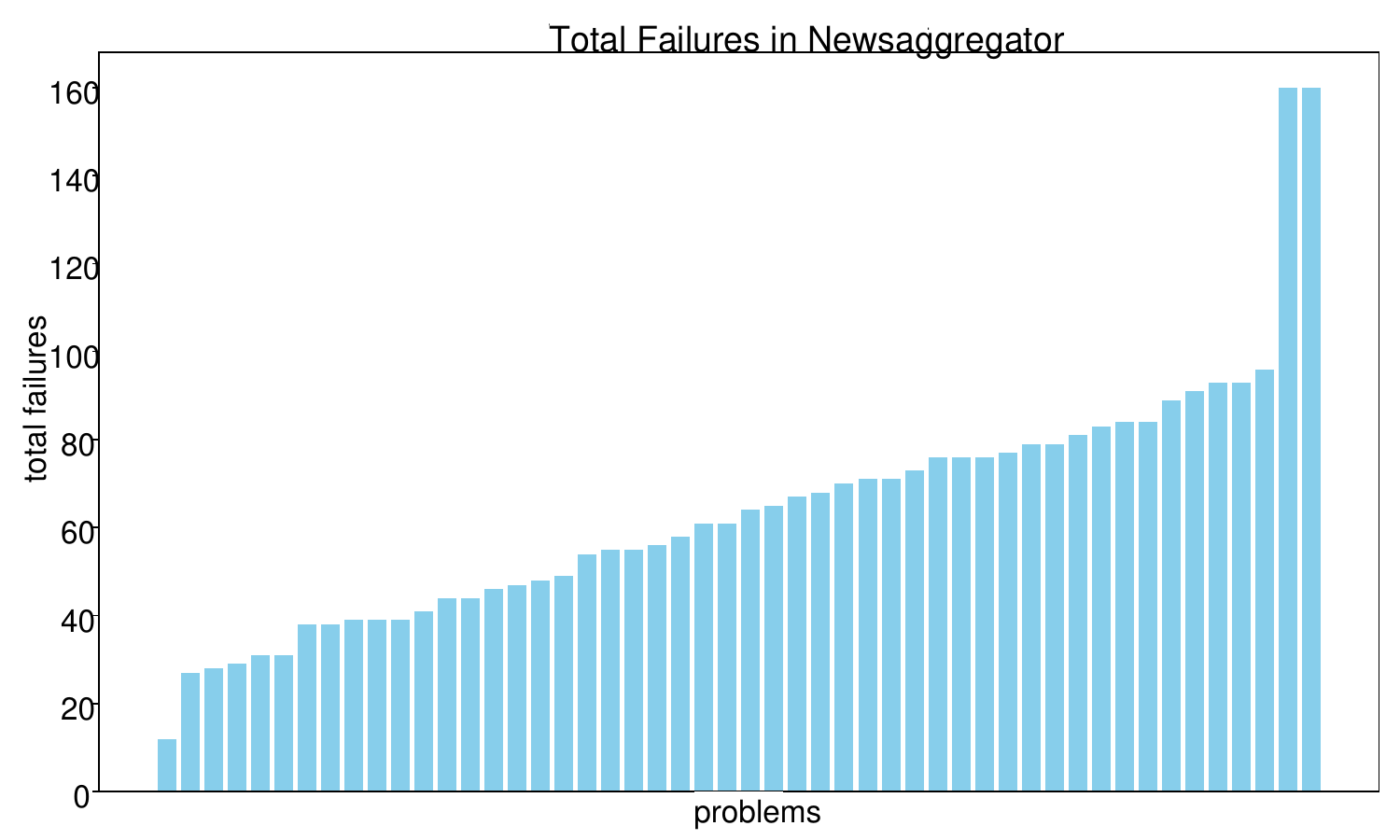}
        \caption{News aggregator}
    \end{subfigure}
    \begin{subfigure}[b]{0.4\textwidth}
        \includegraphics[width=\textwidth]{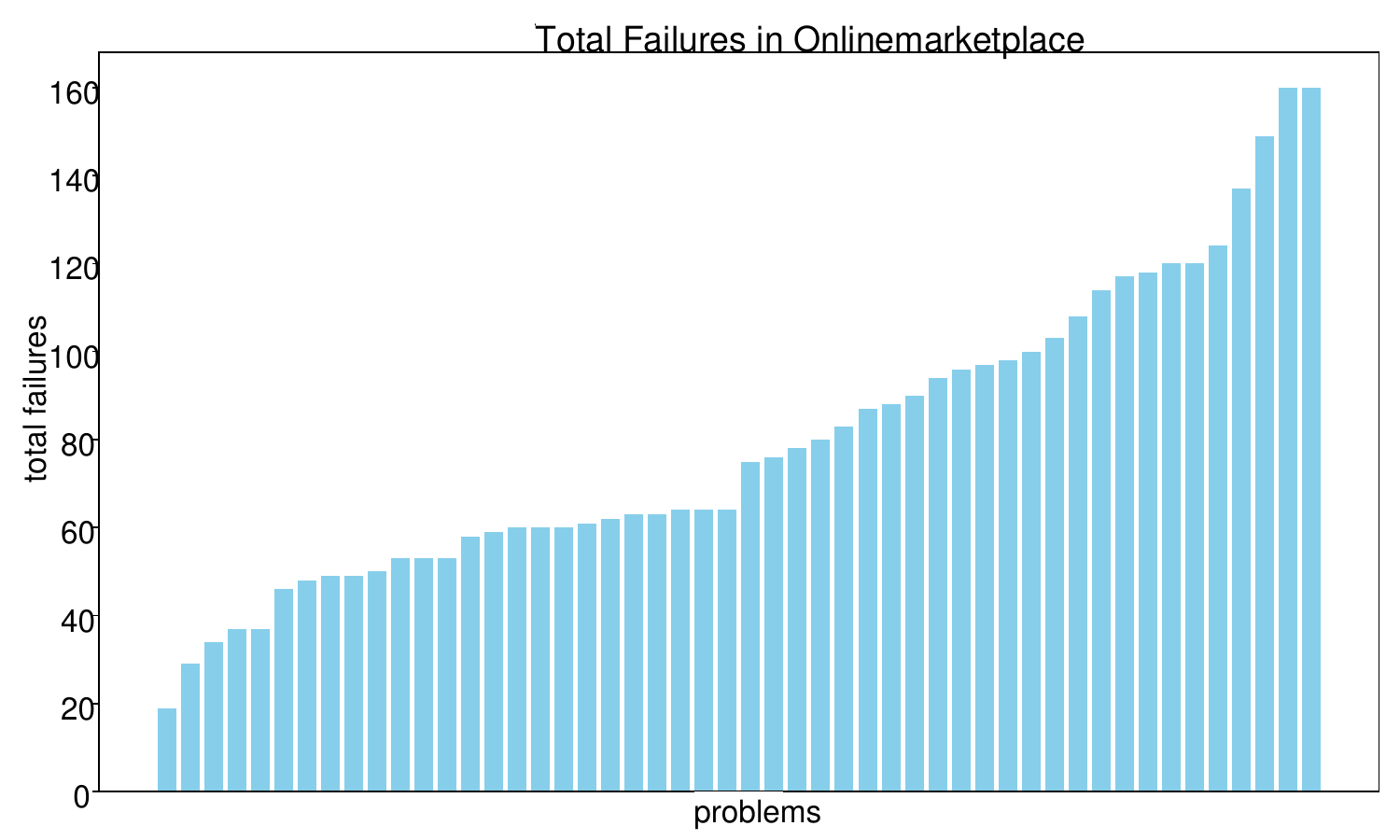}
        \caption{Online marketplace}
    \end{subfigure}
\end{figure}

\begin{figure}[h!]
    \centering
    \includegraphics[width=0.9\textwidth]{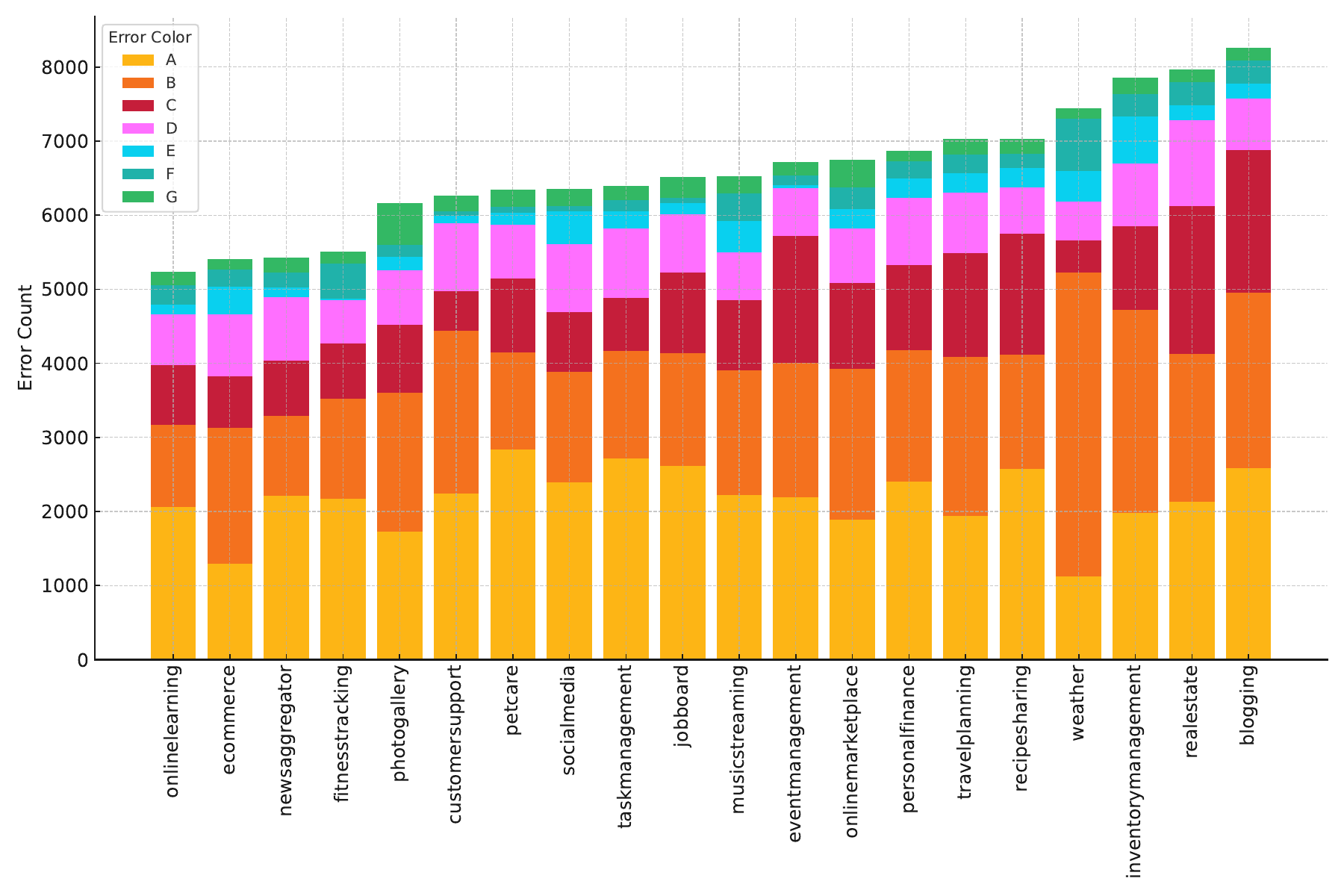}
    \caption{Errors by applications}
    \label{fig:errors_apps}
\end{figure}

Fig.~\ref{fig:errors_apps} shows error distribution by applications. Since each application assembles outputs from all models, the raw error counts are at the same scale for all applications. We do not find any distinctive patterns. There is neither special error nor special application. 
\section{Bias Analysis}
We conducted a preliminary investigation into potential biases within our benchmark, focusing on language bias, cultural inclusivity, and implicit assumptions. To this end, we searched the codebase for gendered terms, stereotypical language, and regional references using an automated analysis script. Additionally, we examined API endpoints and user-facing messages for exclusionary patterns or implicit biases. Our investigation did not identify any instances of such biases in WebApp1K.

While these findings are encouraging, we recognize the limitations of automated analysis and the potential for more nuanced biases that may require further investigation. We welcome additional guidance or suggestions for extending this analysis to ensure a comprehensive evaluation of fairness within our benchmark.
\end{document}